\documentclass[aps,twocolumn,preprintnumbers,amsmath,amssymb,nofootinbib,longbibliography,10pt]{revtex4-1}
\usepackage{graphicx}
\usepackage{dcolumn}
\usepackage[tight]{subfigure}
\usepackage{amsmath}
\usepackage{amssymb}
\usepackage{verbatim}
\usepackage{color}
\usepackage{bm} 
\usepackage{natbib}
\usepackage{xspace}
\usepackage{mathtools}
\usepackage{physics}
\usepackage{hyperref} 
\hypersetup{colorlinks=true,linktoc=all,linkcolor=blue,breaklinks=true,citecolor=blue,urlcolor=blue}
\usepackage{etoolbox}
\usepackage{epstopdf}

\newcommand{\Pibar}{\overline{\Pi}} 

\begin{document}
\newrobustcmd{\suppmattitle}{}
\title{
Finite-frequency noise of interacting single-electron emitters: spectroscopy with higher noise harmonics
}
\author{Niklas Dittmann$^{(1,2,3)}$}
\author{Janine Splettstoesser$^{(1)}$}
\affiliation{
  (1) Department of Microtechnology and Nanoscience (MC2), Chalmers University of Technology, SE-41296 G{\"o}teborg, Sweden\\
  (2) Institute for Theory of Statistical Physics, RWTH Aachen, 52056 Aachen, Germany\\
      and JARA -- Fundamentals of Future Information Technology\\
  (3) Peter-Gr\"unberg Institut and Institute for Advanced Simulation, Forschungszentrum J\"ulich, D-52425 J\"ulich, Germany\\
}


\begin{abstract}
We derive the symmetrized current-noise spectrum of a quantum dot, which is weakly tunnel-coupled to an electron reservoir and driven by a slow time-dependent gate voltage. 
This setup can be operated as an on-demand emitter of single electrons into a mesoscopic conductor.
By extending a real-time diagrammatic technique which is perturbative in the tunnel coupling, we obtain the time-resolved finite-frequency noise as well as its decomposition into noise harmonics
in the presence of both strong Coulomb interaction and slow time-dependent driving.
We investigate the noise over a large range of frequencies and point out where the interplay of Coulomb interaction and driving leads to unique signatures in finite-frequency noise spectra, in particular in the first harmonic.
Besides that, we employ the first noise harmonic as a spectroscopic tool to access individual fluctuation processes.
We discuss how the inverse noise frequency sets a time scale for fluctuations, which competes with time scales of the quantum-dot relaxation dynamics as well as the driving.
\end{abstract}

\maketitle

\section{Introduction}

\begin{figure}[b]
\begin{center}
\includegraphics[width=0.98\columnwidth]{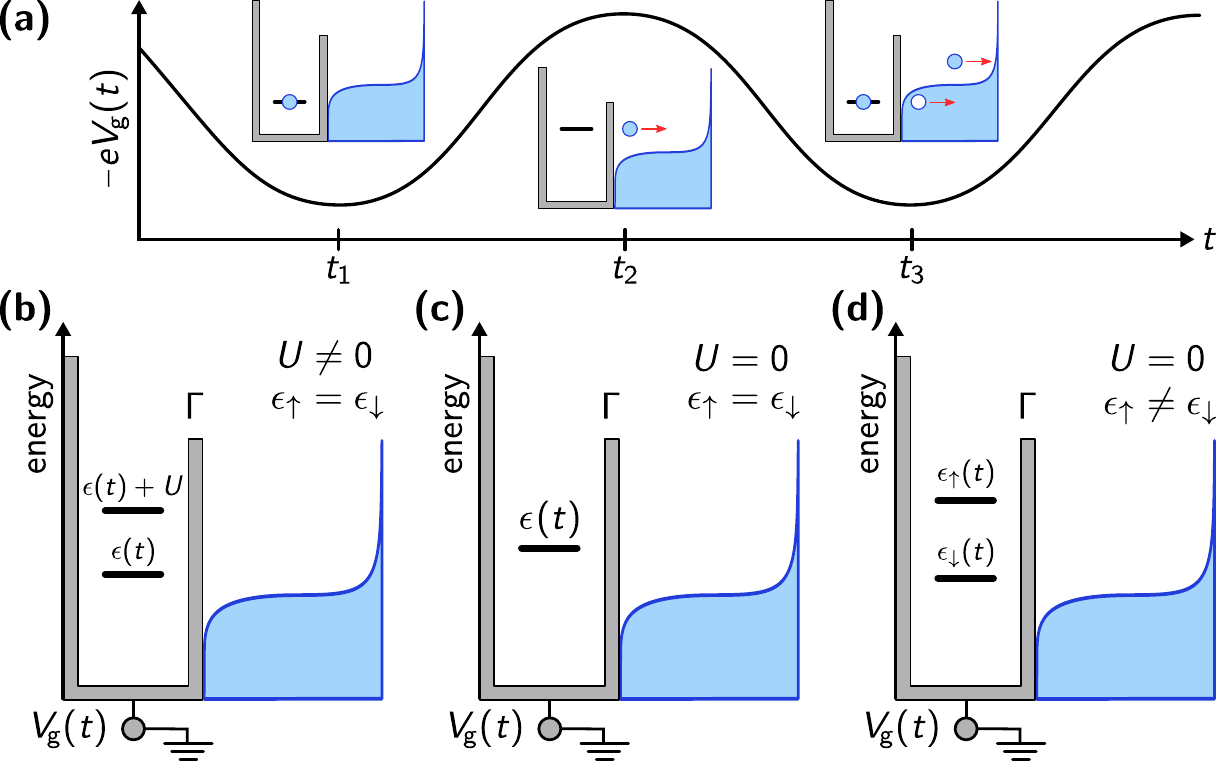}
\end{center}\vspace{-5mm}
	\caption{
	(a)~Working principle of the single-electron emitter: A~harmonic gate voltage, $V_\mathrm{g}(t)$, results in the periodic emission of single electrons and holes into the reservoir at crossings
	between the dot's and the reservoir's electrochemical potentials.
	(b)~Energy landscape of a spin-degenerate quantum dot with on-site Coulomb interaction $U$, tunnel coupled to an electron reservoir with coupling strength~$\Gamma$. 
	Similar setups with vanishing Coulomb interaction are shown in (c) for a spin-degenerate and in (d) for a spin-split energy level.
	}
	\label{fig_motivation}
\end{figure}

Quantum dots driven by time-dependent electric fields can be used to emit single particles into an electronic conductor in a controlled way~\cite{Feve07,Blumenthal07,Bauerle18}. 
Setups of this type, generating ultra-precise charge currents, are for example designed for metrological purposes~\cite{Giblin12,Pekola13}.
In the last years, it has been shown that these on-demand single-electron emitters also allow for novel studies in single-electron transport, 
such as in the emerging field of electron quantum-optics~\cite{Freulon15,Marguerite16}. 
Here, coherence properties on the level of the single-particle wave function can be accessed and exploited. 
The conceptually simplest implementation of a single electron source~\cite{Feve07} is particularly appropriate for this kind of application: 
a quantum dot---taking the role of a mesoscopic capacitor~\cite{Buttiker1993Jun,Pretre1996Sep}---which is coupled to a single electronic contact 
and charged/discharged by an AC-modulation of an applied gate voltage, see Fig.~\ref{fig_motivation}.

For the characterization of charge transport through quantum dots, a central quantity is provided by the current noise~\cite{Blanter00,Clerk10,Landauer98}.
In general, the noise constitutes a measure for thermal and quantum fluctuations and it reveals signatures of correlations between charge carriers, originating, e.\,g., 
from the Pauli exclusion principle or Coulomb interaction. 
In on-demand single-electron emitters realized by a purely ac-driven quantum dot in contact with a single electronic contact, 
the zero-frequency noise averaged over one driving period always equals zero due to charge-current conservation.
The finite-frequency noise, however, has been detected in single-electron emitters in the quantum Hall regime~\cite{Mahe10,Parmentier12,Marguerite16} and fluctuations in the emission process could be identified.
Moreover, the frequency dependence of the noise spectrum can be related to energy emission and absorption processes, 
which has been demonstrated in transport through nanoscale devices in the stationary regime as well~\cite{Onac06,Basset12,Ubbelohde12}. 
Theoretically, finite-frequency noise of quantum dots has been analyzed mostly for the stationary regime~\cite{Kack03,Engel04,Braun06,Rothstein09,Gabdank11,Marcos11,Orth12,Muller13,Moca14,Jin15,Droste15,Zamoum16,Stadler17,Crepieux17} 
and the study of time-dependent setups~\cite{Moskalets07,Moskalets09,Moskalets13,Zhao13} has so far been limited to systems where Coulomb interaction seems to play no major role.

In this paper, we study the finite-frequency noise of a slowly time-dependently driven quantum dot with possibly strong on-site Coulomb interaction, weakly tunnel-coupled to a single electron reservoir, as shown in Fig.~\ref{fig_motivation}~(b).
The working principle of this single-electron emitter is illustrated in Fig.~\ref{fig_motivation}~(a). 
A time-periodic gate voltage leads to a time-dependent modulation of the quantum-dot levels. 
This results in crossings of the addition energies of the quantum dot with the Fermi energy of the reservoir, making tunneling processes out of or into the quantum dot energetically accessible. 
Consequently, single electrons are emitted and absorbed (corresponding to an emitted hole). 
The finite-frequency noise contains information not only on the precision of the source, but also on temporal delays in the described emission process~\cite{Mahe10,Parmentier12} 
and has recently even been used for quantum-state tomography of the emitted particles~\cite{Marguerite2017Oct}.
Our work shows that Coulomb interaction can have a significant impact on such noise spectra.
To unambiguously identify interaction-induced features in the noise, we compare our results to a noninteracting  quantum dot, both in the spin-degenerate case as shown in Fig.~\ref{fig_motivation}~(c), 
as well as in the presence of a strong magnetic field, see Fig.~\ref{fig_motivation}~(d).

For our finite-frequency noise calculations, we extend a real-time diagrammatic technique~\cite{Schoeller94,Koenig96b}, which is based on a perturbative expansion in the tunnel coupling between quantum dot and reservoir.
For stationary systems, this method has, e.\,g.,~been applied to study the finite-frequency noise of a single-electron transistor~\cite{Kack03}, a quantum-dot spin valve~\cite{Braun06} and 
a quantum dot coupled to normal and superconducting contacts~\cite{Droste15}. 
For systems with slow periodic time dependence, Ref.~\cite{Riwar13} analyzes the noise of adiabatic quantum pumps, however restricted to the (long-time) zero-frequency noise, 
which vanishes when the dot is in contact with a single reservoir.
The present work extends the latter approach to finite noise frequencies. 
In our calculations of the finite-frequency noise power, we distinguish a high, an intermediate and a low noise-frequency regime, 
\mbox{$\omega \gg \Gamma$}, \mbox{$\omega \sim \Gamma$} and \mbox{$\omega \ll \Gamma$}, 
with noise frequency $\omega$ and tunnel-coupling strength $\Gamma$, 
and we discuss appropriate approximation schemes for these three regimes. 

Importantly, in contrast to previous works which studied the stationary regime, here also the time scale of the driving has to be treated with care. 
We investigate the noise for slowly driven system parameters, focusing on the instantaneous contribution to the noise, where the system is considered to always follow the driving.
In addition, we analyze the first-order correction, namely the adiabatic response, which takes into account the lag with respect to the time-dependent drive.
One generic consequence of the driving, independent of these approximation schemes, is the fact that the finite-frequency noise power also depends on time. 
We show in our work that the study of the zeroth and first harmonic of this time-dependent function is particularly insightful---a quantity that has attracted little attention so far~\cite{Marguerite2017Oct}.

The paper is structured as follows. 
In Sec.~\ref{sec_model} we specify the model of the driven quantum dot and carefully define all noise quantities studied in this paper. 
We also discuss general properties of the time-resolved finite-frequency noise and its harmonics.  
The derivation of our approach to access the finite-frequency noise is outlined in Sec.~\ref{sec_methods}, where technical details are shifted to the Apps.~\ref{app_method_details}-\ref{app_HFscheme}. 
This technical section is embedded in the manuscript such that it can be skipped by readers more interested in the specific results, which are presented in the subsequent sections.
In Secs.~\ref{sec_results_HF_general}-\ref{sec_results_HF_spectra_harmonics} we analyze finite-frequency noise spectra and in particular the noise harmonics of the quantum dot for high noise frequencies and slow driving, and
in Sec.~\ref{sec_results_resummedHF} we extend these results to faster driving schemes.
Noise spectra in the low noise-frequency regime are discussed in Sec.~\ref{sec_results_LF}. 
The crossover between these two regimes, i.\,e., noise spectra for intermediate frequencies, are given in Sec.~\ref{sec_results_crossover}.
We conclude in Sec.~\ref{sec_conclusion}.

\section{Model and general properties of the noise spectrum}
\label{sec_model}

\subsection{Quantum dot with time-periodic gate voltage}
\label{sec_model_1}

In this work, we investigate fluctuations in the dynamical charge current which is emitted from a single-electron source.
The model we study here is motivated by on-demand electron emitters~\cite{Feve07,Blumenthal07} based on quantum dots.
Crucial for the quantized charge emission in these setups is the discrete level structure and possibly the on-site Coulomb interaction together with the application of a tailored time-dependent gate voltage, 
such as shown in Fig.~\ref{fig_motivation} and described in the introduction.

We assume the quantum-dot levels to be well separated in energy such that only a single level participates in the transport process. 
The quantum dot can therefore be modeled by the Hamiltonian
\begin{align}
\label{eq_hamiltonian}
 H_\text{dot}(t) &= \sum_\sigma \epsilon_\sigma(t) d^\dagger_\sigma d_\sigma + U d^\dagger_\uparrow d_\uparrow d^\dagger_\downarrow d_\downarrow.
\end{align}
Annihilation (creation) operators of quantum-dot states are denoted as $d_\sigma$ ($d^\dagger_\sigma$) with spin index \mbox{$\sigma=\,\uparrow,\downarrow$}. 
The single-particle energy is given by \mbox{$\epsilon_\sigma(t)=\bar{\epsilon}_\sigma + \epsilon_\mathrm{g}(t)$}. 
The mean value $\bar \epsilon_\sigma$ contains an optional Zeeman splitting. 
The time-dependent gate voltage, which we consider to be a harmonic drive with frequency $\Omega$ and amplitude $\delta\epsilon$, results in the additional term \mbox{$\epsilon_\mathrm{g}(t)=\delta \epsilon \cos(\Omega t)$} 
(we furthermore define $T$ as the period of the gate-voltage drive with \mbox{$\Omega = \frac{2\pi}{T}$}).
The driving shifts quantum-dot addition energies above and below the Fermi energy, which in turn causes periodic current pulses, see Fig.~\ref{fig_motivation}~(a).
For the spin-symmetric case, without Zeeman splitting, we use the simplified notation \mbox{$\epsilon_\sigma(t)=\epsilon(t)$}. 
Besides the single-particle physics, we include an on-site Coulomb interaction~$U$, which can possibly be large.  
The eigenstates of the \textit{decoupled} quantum-dot Hamiltonian,~$H_\mathrm{dot}(t)$,  are $\big\{\!\ket{0},\ket{\uparrow},\ket{\downarrow},\ket{\mathrm 2}\!\big\}$ for dot occupations with zero,
one (with spin $\uparrow,\downarrow$) and two electrons, respectively.

The quantum dot is tunnel coupled to a single electronic reservoir with Hamiltonian \mbox{$H_\text{res} = \sum_{\sigma k} \epsilon_k c^\dagger_{\sigma k} c_{\sigma k}$}, 
where operators for annihilation (creation) of reservoir states  with orbital quantum number~$k$ are  $c_{\sigma k}$ ($c^\dagger_{\sigma k}$) and the single-particle energy $\epsilon_k$ is spin independent. 
A Fermi function \mbox{$f(\epsilon)=\left[1+\exp(\beta\epsilon)\right]^{-1}$} characterizes the occupation of this reservoir with the inverse temperature $\beta$ 
and the electrochemical potential of the reservoir \mbox{$\mu=0$}, taken as reference energy. 
Throughout this work, $k_\mathrm{B}$, $\hbar$ and $e$ are set to one.
The tunnel-coupling between dot and reservoir is included by
\begin{align}
\label{eq_hamiltonian_tun}
 H_\text{T}   &= \sum_{\sigma k} \left( \gamma d^\dagger_\sigma c_{\sigma k}  + \gamma^* c^\dagger_{\sigma k} d_\sigma \right),
\end{align}
where the coupling~$\gamma$ is considered as energy and spin independent. 
This parameter quantifies the tunnel-coupling strength \mbox{$\Gamma = 2\pi \nu_0 |\gamma|^2$}, with $\nu_0$ being the density-of-states at the Fermi energy.  
We consider systems which operate in the regime of weak reservoir-dot coupling, $\beta\Gamma\ll1$, where the broadening of dot energy levels caused by the coupling is small.

We illustrate our model for an interacting, spin-degenerate quantum dot in Fig.~\ref{fig_motivation}~(b) as well as for a noninteracting
quantum dot without and with spin-splitting in Fig.~\ref{fig_motivation}~(c) and (d). 
This basic setup has been realized with quantum dots in a 2D electron gas, where the interaction is thought to be essentially screened by a large metallic gate~\cite{Feve07}, 
as well as in temporarily created quantum dots~\cite{Blumenthal07}, where a strong interaction separates the dot addition energies.

\subsection{Finite-frequency noise and noise harmonics}
\label{sec_noisedef}
As discussed in the introduction, our main interest is the current noise, namely  fluctuations in the charge current emitted by the time-dependently driven quantum dot.

The current operator, measuring the charge current into the tunnel-coupled reservoir, is given by \mbox{$I = - i \sum_{\sigma k} \big(\gamma d^\dagger_\sigma c_{\sigma k} - \gamma^* c^\dagger_{\sigma k}d_\sigma\big)$}.
Using this expression we define the current-fluctuation operator, \mbox{$\delta I = I - \expval{I}$}, together with its symmetrized two-time correlator  \mbox{$\mathcal{C}(t,\tau)=\expval{\left\{ \delta I(t), \delta I(t+\tau)\right\}}$}.  
The time $t$ is taken as the reference measurement time and $\tau$ is the time difference between two current measurements at $t$ and $t'=t+\tau$. 
Here, curly brackets denote an anti-commutator and the time dependence of operators is treated following the Heisenberg picture.
As usual, the finite-frequency current noise spectrum is obtained by a Fourier transform of this correlator with respect to the \textit{time difference},
\begin{align}
\label{eq_noise_tfr}
S(t;\omega) =  \int d\tau e^{i \omega \tau} \mathcal{C}(t,\tau).
\end{align}
However, in contrast to standard treatments, this finite-frequency noise still depends on the time~$t$. 
This is caused by the time-dependent driving of the quantum dot, breaking time-translational invariance. 
We here refer to the quantity defined in Eq.~(\ref{eq_noise_tfr}) as the \textit{time-resolved finite-frequency noise}. 
The study of this quantity gives us, to some extend, an intuitive understanding of the effect of time-dependent driving and the response times on the noise spectrum, as demonstrated in Sec.~\ref{sec_results_HF}. 
However, it is at the same time hard to disentangle various effects governing this quantity and it is also expected to be difficult to measure in a realistic experiment.
In addition to the time-resolved finite-frequency noise we therefore promote the study of the symmetric \emph{current-noise harmonics},
\begin{align}
\label{eq_noise}
\mathcal S(n;\omega) &= \int_0^T \frac{dt}{T} \int d\tau e^{i n\Omega t + i \omega \tau} \mathcal{C}(t,\tau),
\end{align}
with index~$n$. 
Studying these noise harmonics, in particular the case \mbox{$n=1$}, is a main focus of this paper.
Thereby, we investigate temporal correlations between the driving signal and the noise spectrum.
Equation~\eqref{eq_noise} also defines the more standardly studied \emph{time-averaged noise spectrum}, \mbox{$\mathcal S(n=0;\omega)$}. 
This quantity has, e.\,g.,~proven to be helpful to characterize the precision of single-electron emitters \cite{Mahe10,Parmentier12}.
Here, we show that information on fluctuation processes, which is hard to extract from these time-averaged noise spectra, can be accessed by analyzing the first noise harmonic.

The main quantities investigated in this paper are listed in Tab.~\ref{tab_quant}.

\begin{table}[t]
 \begin{tabular}{|l|l|l|}
 \hline
 quantity & brief description & reference\\
 \hline
 \hline
 $S(t;\omega)$ & time-resolved finite-frequency noise & Eq.~\eqref{eq_noise_tfr}\\
 $\mathcal S(n=0;\omega)$ & time-averaged noise spectrum & Eq.~\eqref{eq_noise} \\
 $\mathcal S(n\neq0;\omega)$ & $n$'th noise harmonic & Eq.~\eqref{eq_noise}\\
 $\tilde S(t;\omega)$ & auxiliary function for diagrammatic & Eq.~\eqref{eq_auxiliarynoise}\\
 & noise calculations & \\
 \hline
 $\boldsymbol P(t)$ & quantum-dot occupation vector & Sec.~\ref{sec_rtdpt}\\
 $\boldsymbol F^{(i)}(t)$ & instantaneous fluctuation vector & Eqs.~\eqref{eq_flvector},\\
 & & \eqref{eq_noise_LF_for_flvector_i}, \eqref{eq_noise_crossover_for_flvector_i}\\
 \hline
 \end{tabular}
 \caption{Main quantities analyzed in this paper.}
 \label{tab_quant}
\end{table}

\subsection{Expansion for slow gate-voltage driving}
\label{sec_model_adexp}

In subsequent sections we analyze both the noise harmonics [Eq.~\eqref{eq_noise}] and the time-resolved finite-frequency noise [Eq.~\eqref{eq_noise_tfr}] of the quantum dot, while the latter is operated as an on-demand electron emitter.
As explained in Sec.~\ref{sec_methods}, we therefore extend a real-time diagrammatic perturbative approach for weakly coupled quantum dots. 
For this purpose, as well as for an expansion in small driving frequencies which is detailed below, it turns out to be helpful to rewrite the noise harmonics as
\begin{align}
\label{eq_noise2}
\mathcal S(n;\omega) &= \int_0^T \frac{dt}{T} e^{in\Omega t} \left[ \tilde{S}\left(t;\omega\right) + \tilde{S}\left(t;n\Omega-\omega\right) \right],
\end{align}
with the \textit{auxiliary function},
\begin{align}
\label{eq_auxiliarynoise}
\tilde{S}\left(t;\omega\right) &= 
   \begin{aligned}[t] 
       &\int_{-\infty}^t\! dt' e^{i \omega (t'-t)} \Big[\expval{\left\{I(t), I(t')\right\}} \\ & - 2 \expval{I(t)} \expval{I(t')} \Big].
   \end{aligned}
\end{align}
See App.~\ref{app_method_details} for the derivation of this expression. The advantage of this rewriting is the treatment of the reference time $t$, in a way that all other times, $t'=t+\tau$, lie in the past with respect to it.

The single-electron emission is achieved by a slow time-dependent gate-voltage driving, namely $\delta\epsilon\,\Omega\beta/\Gamma\ll1$. 
This condition ensures that the system has enough time for electron emissions/absorptions to occur during each level crossing caused by the drive.
This justifies an expansion of our noise expression in Eq.~\eqref{eq_noise2} in terms of the small parameter $\delta\epsilon\,\Omega\beta/\Gamma$, see e.\,g.~Refs.~\cite{Splettstoesser06,Riwar13}.
In the next paragraph, we outline the main idea of this expansion, which should suffice to follow our discussions of results in Secs.~\ref{sec_results_HF}-\ref{sec_results_crossover}.
For technical aspects we refer to Sec.~\ref{sec_methods}, where we explain all steps in our noise derivations in detail.

In order to obtain the noise-harmonics expression in Eq.~\eqref{eq_noise2} for slow driving, we expand the auxiliary function as
\mbox{$\tilde{S}(t;\omega) \rightarrow \tilde{S}^\mathrm{(i)}(t;\omega) + \tilde{S}^\mathrm{(a)}(t;\omega) + \dots$} .
The first term in this series is the \textit{instantaneous} contribution, marked with the superscript~(i).
It describes a time evolution of a system which always follows its instantaneous stationary state. It thus corresponds to the auxiliary function derived for a stationary quantum dot with parameters frozen at time $t$.
The second term in the series above takes into account a small retarded response of the system with respect to the time-dependent driving.
Therefore, we call terms of this type, indicated with the superscript~(a), the \textit{adiabatic response}.
By inserting the expansion of the auxiliary function into Eq.~\eqref{eq_noise2} and expanding rigorously up to first order in the small-driving parameter, we obtain the instantaneous noise and its adiabatic response,
\begin{subequations}
\label{eq_noise_adexp}
\begin{align}
\label{eq_noise_adexp_i}
\mathcal S^\mathrm{(i)}(n;\omega) &=  
\int_0^T \frac{dt}{T} e^{i n \Omega t}\Big[ \tilde{S}^\mathrm{(i)}(t;\omega) + \tilde{S}^\mathrm{(i)}(t;-\omega) \Big]\\
\label{eq_noise_adexp_a}
\mathcal S^\mathrm{(a)}(n;\omega) &=  
  \begin{aligned}[t]
    &\int_0^T\! \frac{dt}{T} e^{i n \Omega t} \Big[ \tilde{S}^\mathrm{(a)}(t;\omega) + \tilde{S}^\mathrm{(a)}(t;-\omega) \\ &+ \Omega n \, \partial_x \tilde{S}^\mathrm{(i)}(t;x)\big|_{x=-\omega} \Big].
  \end{aligned}
\end{align}
\end{subequations}
As pointed out before, the technical calculations of the expressions $\tilde S^\mathrm{(i)}(t;\omega)$ and $\tilde S^\mathrm{(a)}(t;\omega)$ are presented in Sec.~\ref{sec_methods}.
Prior to these derivations, we now discuss some general properties of the instantaneous and adiabatic-response contributions to the noise harmonics [Eqs.~\eqref{eq_noise_adexp}].

\subsubsection{Instantaneous finite-frequency noise}
\label{sec_generalprop_FDT}

The instantaneous contribution to the noise can in many respects be understood as the noise of a stationary equilibrium system (due to the presence of only a single reservoir).
The time $t$ merely enters as a parameter, and the instantaneous noise therefore inherits a number of properties that are known for these kind of non-time-dependently driven systems.

First of all, the symmetrized and time-averaged instantaneous finite-frequency noise, Eq.~\eqref{eq_noise_adexp_i} with \mbox{$n=0$}, is always real. 
The technical reason is that the auxiliary function, Eq.~\eqref{eq_auxiliarynoise}, fulfills \mbox{$\tilde S(t,\omega) = \tilde S^*(t,-\omega)$}, which also holds for the instantaneous and adiabatic-response parts separately.
Furthermore, because we consider a cosine driving of the gate voltage, we find that the instantaneous part of the auxiliary function, $\tilde S^\mathrm{(i)}(t,\omega)$, is an even function in~$t$. 
We conclude that all noise harmonics in instantaneous order, Eq.~\eqref{eq_noise_adexp_i} with \mbox{$n\neq0$}, are real quantities. 
The same is then true for the time-resolved finite-frequency noise, Eq.~\eqref{eq_noise_tfr}, in instantaneous order.

The instantaneous part of the time-resolved finite-frequency noise, being the one of an equilibrium system with parameters frozen at the reference time $t$, is furthermore expected to fulfill a fluctuation-dissipation theorem. 
Here, we give its explicit shape and show an interesting extension for the first noise harmonic. 
More specifically, the instantaneous contribution to the noise fulfills a fluctuation-dissipation theorem \cite{Callen51} at every fixed level position $\epsilon_\sigma(t)$, where $t$ serves as a parametrization,
\begin{align}
 \label{eq_fdt_eq}
 S^\text{(i)}(t;\omega) &= 2 \omega \coth\left(\frac{\beta\omega}{2}\right)\mathrm{Re}\, G(\left\{\epsilon_\sigma(t)\right\};\omega).
\end{align}
It connects the equilibrium noise spectrum at time $t$, \mbox{$S^\text{(i)}(t;\omega)$}, 
to the finite-frequency linear-response conductance, \mbox{$G(\left\{\epsilon_\sigma(t)\right\};\omega) = \frac{\partial I(\omega)}{\partial \epsilon_\mathrm{g}(\omega)}\Big|_{\left\{\epsilon_\sigma(t)\right\}}$}, 
evaluated for the system being in equilibrium at the level positions given by $\left\{\epsilon_\sigma(t)\right\}$.

We now use the expression given in Eq.~(\ref{eq_fdt_eq}), in order to derive extensions of the fluctuation-dissipation theorem for zeroth and first noise harmonics.
For \mbox{$n=0$}, we find
\begin{align}
 \label{eq_fdt_i0}
 \mathcal S^\mathrm{(i)}(0;\omega) &= \int_0^T \frac{dt}{T} 2 \omega \coth\left(\frac{\beta\omega}{2}\right)\mathrm{Re}\, G(\left\{\epsilon_\sigma(t)\right\};\omega).
\end{align}
In the same way, we derive the fluctuation-dissipation theorem for the first noise harmonic, \mbox{$n=1$}.
For our cosine gate-voltage driving, \mbox{$\epsilon_\mathrm{g}(t)=\delta \epsilon \cos(\Omega t)$}, we make a parameter replacement using $\epsilon_\mathrm{g}$ instead of time, such that the relation reads
\begin{align}
 \label{eq_fdt_i1}
 \mathcal S^\mathrm{(i)}(1;\omega) &=
\begin{aligned}[t] 
 &\frac{2 \omega}{\pi} \coth\left(\frac{\beta\omega}{2}\right)  \\
 &\times \int_{-\delta \epsilon }^{\delta \epsilon } d\epsilon_\mathrm{g}\sqrt{1-\frac{\epsilon_\mathrm{g}^2}{\delta \epsilon^2}}
\ \mathrm{Re}\frac{\partial G\left(\epsilon_\mathrm{g};\omega\right)}{{\partial\epsilon_\mathrm{g}}}.
\end{aligned}
\end{align}
This shows that, while the zeroth noise harmonic is directly related to an average over the finite-frequency conductance of the system, the first harmonic reveals nonlinearities 
(namely first-order derivatives of the conductance, equivalent to second-order derivatives of the current).
In the limit in which the driving amplitude, $\delta \epsilon$, is smaller than the scale on which variations in the conductance occur, Eq.~\eqref{eq_fdt_i1} simplifies to
\begin{align}
 \label{eq_fdt_i1_LR}
 \mathcal S^\mathrm{(i)}(1;\omega) &\approx \delta \epsilon\, \omega \coth\left(\frac{\beta\omega}{2}\right) \mathrm{Re}\frac{\partial G\left(\epsilon_\mathrm{g};\omega\right)}{\partial\epsilon_\mathrm{g}}\Big|_{\epsilon_\mathrm{g}=0},
\end{align}
and the first noise harmonic is proportional to this nonlinearity.

In Secs.~\ref{sec_results_HF}-\ref{sec_results_crossover} we use these relations to interpret the instantaneous contribution to the finite-frequency noise.
Similar extensions of the fluctuation-dissipation theorem, connecting noise harmonics to derivatives of the finite-frequency conductance, also hold for \mbox{$n\geq 2$}.
In these cases, higher derivatives as well products of derivatives of different order would appear.
Importantly, no such extension of the fluctuation-dissipation theorem is expected to hold for the adiabatic-response contribution to the noise \cite{Riwar13}.

\subsubsection{Adiabatic-response contribution to the noise}
\label{sec_generalprop_vanishing_a}

We also want to collect a few general properties of the adiabatic-response contribution to the noise.
We first point out that the adiabatic-response correction of the zeroth noise harmonic always vanishes for our system.
This result is a consequence of having a single contact and single-parameter driving,~$\epsilon_\mathrm{g}(t)$, which leads to an integrand in Eq.~\eqref{eq_noise_adexp_a} which is linear in~$\epsilon_\mathrm{g}'(t)$ and otherwise depends on~$\epsilon_\mathrm{g}(t)$. 
Due to the periodicity, \mbox{$\epsilon_\mathrm{g}(0)=\epsilon_\mathrm{g}(T)$}, we derive \mbox{$\mathcal S^{\mathrm{(a)}}(0;\omega)=0$}. 
Importantly, non-vanishing contributions occur in the adiabatic response of higher noise harmonics and hence in the time-resolved finite-frequency noise.

Besides that, the adiabatic response of the auxiliary function, $\tilde S^\mathrm{(a)}(t,\omega)$, turns out to be an odd function in~$t$, because it contains a factor $\epsilon'_\mathrm{g}(t)$
in front of an otherwise even expression, see also Sec.~\ref{sec_methods}.
Since, however, the third term on the right-hand side~in Eq.~\eqref{eq_noise_adexp_a} remains an even function, the harmonics of the adiabatic-response contribution to the noise are generally complex valued.
When the latter term does not contribute, which, as we explain in Sec.~\ref{sec_results_HF}, is for instance the case for high noise frequencies, we find that $\mathcal S^\mathrm{(a)}(n=1,\omega)$ is purely imaginary.
Consequently, we also find that the adiabatic-response contribution to the time-resolved finite-frequency noise in Eq.~\eqref{eq_noise_tfr} is generally complex valued (and real in the high noise-frequency regime).
The reason that these contributions to the symmetrized noise can become complex is due to the lag of the system. 
In any non-zero order in the expansion in the small parameter $\delta\epsilon\,\Omega\beta/\Gamma$, the two constituents on the right-hand side~in Eq.~\eqref{eq_noise2} are not each others 
complex conjugates.\footnote{We note that $\tilde S(t;\omega)$ is an auxiliary function and should not be confused with unsymmetrized noise.}	

The general properties of the noise are summarized in Tab.~\ref{tab_ReIm}.

\begin{table}[t]
 \begin{tabular}{|l|l|}
 \hline
 quantity & general property\\
 \hline
 \hline
 $ S^\mathrm{(i)}(t;\omega)$ & real \\
 $ S^\mathrm{(a)}(t;\omega)$ & complex  (real for $\omega \gg \Gamma$)  \\ \hline
 $\mathcal S^\mathrm{(i)}(n;\omega)$ & real for zero and non-zero harmonics\\
 $\mathcal S^\mathrm{(a)}(n=0;\omega)$ & vanishes for one-parameter driving\\
 $\mathcal S^\mathrm{(a)}(n\neq0;\omega)$ & complex (imaginary for $\omega \gg \Gamma$ and $n=1$) \\
 \hline
 \end{tabular}
 \caption{Properties of the time-resolved finite-frequency noise and the noise harmonics for a cosine driving of the gate voltage, see Sec.~\ref{sec_model_adexp}.}
 \label{tab_ReIm}
\end{table}

\section{Derivation of the finite-frequency noise}
\label{sec_methods}

\subsection{Real-time diagrammatic technique for reduced density matrix and current}
\label{sec_rtdpt}
We now outline the derivation of finite-frequency noise and noise harmonics of the time-dependently driven, interacting quantum dot.
Further details are given in respective appendices.
To derive the finite-frequency current noise, we extend a non-equilibrium real-time diagrammatic technique, 
which is based on a perturbative expansion in the reservoir-dot coupling strength~$\Gamma$, see Refs.~\cite{Schoeller94,Koenig96b,Splettstoesser06}.
For weak coupling and high temperature, $\beta \Gamma \ll 1$, as considered here, the system dynamics is well described by leading terms in this series.

\begin{figure}[tb]
\begin{center}
\includegraphics[width=0.98\columnwidth]{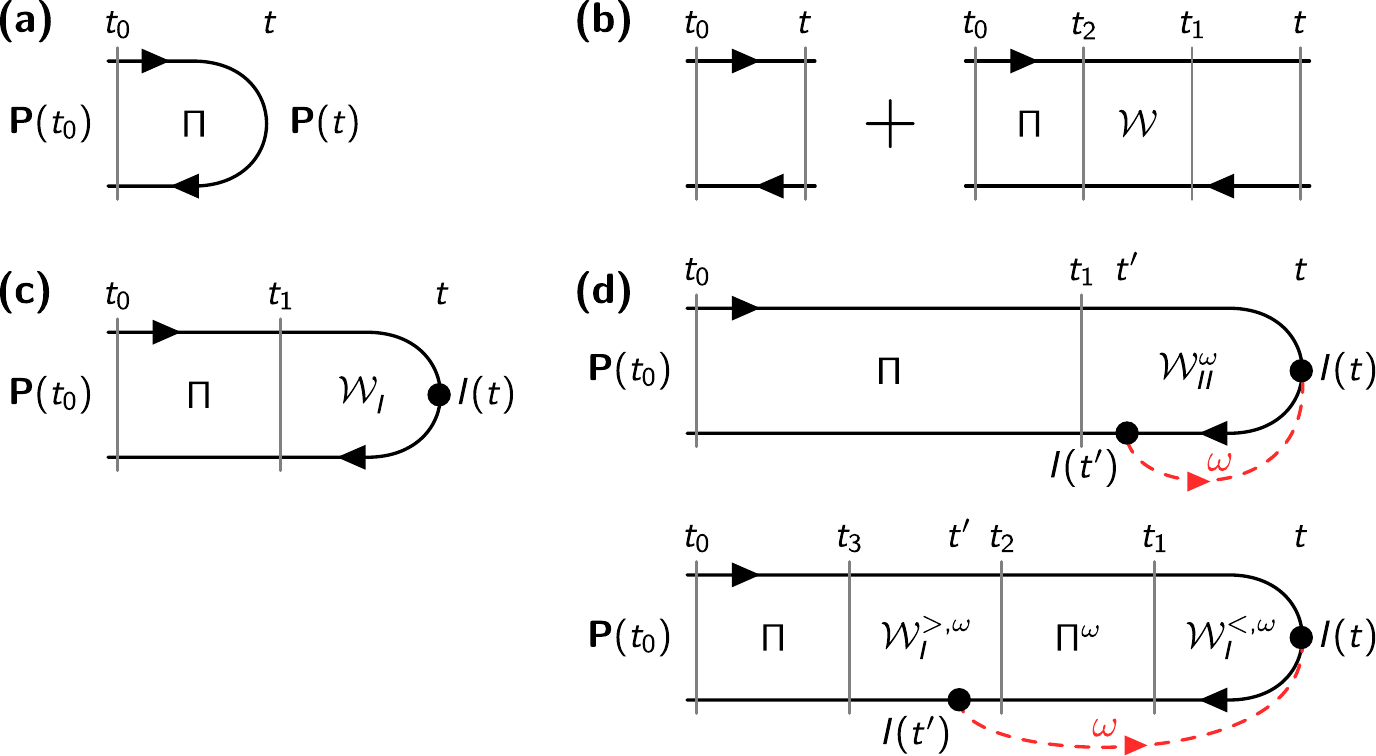}
\end{center}\vspace{-5mm}
	\caption{
	Time evolution on the Keldysh time contour sketched for (a)~the occupation vector (Eq.~\eqref{eq_occvec_propagation}); (b)~the propagator (Eq.~\eqref{eq_Pidef}); (c)~the current (Eq.~\eqref{eq_current});
	(d)~two possible contributions to the auxiliary function for the finite-frequency-noise calculation (Eq.~\eqref{eq_auxiliarynoise_blocks}).
	}
	\label{fig_keldysh}
\end{figure}

To describe the time evolution of the dot state, we consider its reduced density matrix, which we obtain by tracing out the reservoir degrees-of-freedom.
Because tunneling between quantum dot and reservoir conserves spin, the evolution of the diagonal part of the reduced density matrix decouples from the one for the off-diagonal part (coherences).
Therefore, for the calculation of the current and its finite-frequency noise, it is sufficient to consider the diagonal part only, which is given by the occupation probabilities of dot states, 
here collected into a vector, \mbox{$\boldsymbol P(t) = \big(P_0(t),P_\uparrow(t),P_\downarrow(t),P_\text{2}(t)\big)$}. 
The time-evolution of this probability vector is given by
\begin{align}
 \label{eq_occvec_propagation}
 \boldsymbol P(t) &= \Pi(t,t_0) \boldsymbol P(t_0),
\end{align}
where we assume that at an initial time, $t_0$, correlations between reservoir and quantum dot are absent. 
The propagator, $\Pi(t,t_0)$, takes tunneling to the reservoir into account.
It can be depicted on the Keldysh time contour, in which the forward (backward) part of the time evolution of the reduced density matrix is represented by a forward (backward) time line, see Fig.~\ref{fig_keldysh}~(a).

Treating the tunnel Hamiltonian, $H_\mathrm{T}$ in Eqs.~\eqref{eq_hamiltonian_tun}, perturbatively, is illustrated by the insertion of tunnel vertices on the forward and backward parts of the contour.
When tracing out the reservoir degrees-of-freedom, the tunnel vertices become pair-wise contracted (Wick's theorem), which in a diagrammatic language can be indicated by tunneling lines, see e.\,g.~Refs.~\cite{Schoeller94,Koenig96b,Splettstoesser06,Riwar13}.
The propagator, $\Pi(t,t_0)$, then fulfills the Dyson equation
\begin{align}
 \label{eq_Pidef}
 \Pi(t,t_0) &= \boldsymbol 1 + \int_{t_0}^t dt_1 \int_{t_0}^{t_1} dt_2 \, \mathcal W(t_1,t_2) \Pi(t_2,t_0),
\end{align}
where the kernel, $\mathcal W(t_1,t_2)$, is given by the sum of all irreducible diagrams on the Keldysh contour, i.\,e.,~all diagrams in which any vertical cut crosses a tunneling line.
This Dyson equation is sketched in Fig.~\ref{fig_keldysh}~(b), where the first term in the figure as well as the right-hand side~of the second term correspond to free parts of the contour.
Inserting Eq.~\eqref{eq_Pidef} into Eq.~\eqref{eq_occvec_propagation} and taking a time derivative, we derive the following kinetic equation~\cite{Schoeller94,Koenig96b} for the time evolution of the occupation vector:
\begin{align}
 \label{eq_master}
 \partial_t \boldsymbol P(t) &= \int_{-\infty}^t dt_1 \mathcal W(t,t_1)\boldsymbol P(t_1).
\end{align}
In Eq.~\eqref{eq_master}, we replaced the initial time $t_0\rightarrow-\infty$, assuming that it is far away from the measurement time~$t$. Similarly, an equation for the charge current can be written,
\begin{align}
 \label{eq_current}
 \expval{I(t)} &= \frac{\boldsymbol e^T}{2} \int_{-\infty}^t dt_1 \mathcal W_I(t,t_1) \boldsymbol P(t_1),
\end{align}
with \mbox{$\boldsymbol e^T=(1,1,1,1)$}. 
The additional current kernel, $\mathcal W_I(t,t_1)$, includes all irreducible diagrams in which an additional current vertex is placed at time $t$, see Fig.~\ref{fig_keldysh}~(c). 
Details on the calculation of kernels as well as explicit expressions are given in App.~\ref{app_adexp_kernels}.

\subsection{Real-time diagrammatic description of the finite-frequency noise}
\label{sec_auxnoise}

We now turn to the calculation of the  finite-frequency noise, Eq.~\eqref{eq_noise}, based on the auxiliary function defined in Eq.~\eqref{eq_auxiliarynoise}. 
This auxiliary function consists of correlation functions of two current operators at time $t$ and an earlier time $t'$, which have to be placed as external vertices along the Keldysh contour 
(at the turning point and, respectively, on the forward or backward contour). 
Illustrations of two possible configurations are shown in Fig.~\ref{fig_keldysh}~(d).
In total, we write the auxiliary function as
\begin{widetext}
\begin{eqnarray}
\label{eq_auxiliarynoise_blocks}
\tilde{S}(t;\omega) &=  &\lim_{t_0 \rightarrow -\infty} \frac{\boldsymbol e^T}{2} \Bigg( \int_{t_0}^{t} dt_1 \mathcal W_{II}(t,t_1;\omega)\boldsymbol P(t_1)+\int_{t_0}^{t} dt_1 \int_{t_0}^{t_1} dt_2 \int_{t_0}^{t_2} dt_3 \mathcal W_I^{<}(t,t_1;\omega)\Pi(t_1,t_2;\omega) \mathcal W_I^{>}(t_2,t_3;\omega)\boldsymbol P(t_3)\nonumber \\ 
& &
   - \int_{t_0}^t dt_1 \int_{t_0}^t dt_2 \int_{t_0}^{t_1} dt_3 e^{i \omega (t_1-t)} \mathcal W_I(t,t_2) \boldsymbol P(t_2)\otimes\boldsymbol e^T \mathcal W_I(t_1,t_3) \boldsymbol P(t_3) \Bigg).
\end{eqnarray}
\end{widetext}
The first term of this complex expression contains all diagrams in which the two current vertices are part of the same irreducible kernel, 
indicated by $\mathcal W_{II}(t,t_1;\omega)$, see the upper plot in Fig.~\ref{fig_keldysh}~(d). 
The exponential function in Eq.~\eqref{eq_auxiliarynoise} enters as an additional frequency line \cite{Braun06,Kack03} going from $t$ to $t'$ and carrying the noise frequency~$\omega$,
hence the frequency dependence of the kernel.

The lower sketch of Fig.~\ref{fig_keldysh}~(d) is an example for all those contributions in which the current vertices are part of two irreducible current kernels separated by a propagator. 
This is summarized in the second expression of Eq.~\eqref{eq_auxiliarynoise_blocks}. The first factor of this term, $\mathcal W_I^{<}(t,t_1;\omega)$, 
contains all irreducible diagrams which include a single current vertex and a frequency line, which enters the diagram from the left-hand side~and ends at the current vertex.
The third factor, $\mathcal W_I^{>}(t,t_1;\omega)$, is of similar nature, but here the frequency line begins at the current vertex and leaves the diagram to the right.
The kernels are separated by a propagator with an external frequency line, $\Pi(t,t_1;\omega)$, for which we write the modified Dyson equation
\begin{align}
 \label{eq_dysontime}
 \Pi(t,t_1;\omega) &= 
  \begin{aligned}[t]
    &\boldsymbol 1 \, e^{i \omega (t_1-t)} + \int_{t_1}^t dt_2 \int_{t_1}^{t_2} dt_3 \, e^{i \omega (t_2-t)} \\& \times \mathcal W(t_2,t_3;\omega) \Pi(t_3,t_1;\omega),
  \end{aligned}
\end{align}
with the kernel \mbox{$\mathcal W(t_2,t_3;\omega) = e^{i \omega (t_3-t_2)}\mathcal W(t_2,t_3)$}. 
Equation~\eqref{eq_dysontime} is given by Eq.~\eqref{eq_Pidef} multiplied with the factor $e^{i \omega (t_1-t)}$.
This factor takes into account the part of the frequency line between the two current vertices which runs over the propagator, see the lower sketch in Fig.~\ref{fig_keldysh}~(d).

Finally, the third term of Eq.~\eqref{eq_auxiliarynoise_blocks} stems from the second term of the auxiliary function in Eq.~\eqref{eq_auxiliarynoise}, consisting of a product of current expectation values, see Eq.~\eqref{eq_current}.
Importantly, the limit \mbox{$t_0 \rightarrow -\infty $} is not taken separately for the different terms in Eq.~\eqref{eq_auxiliarynoise_blocks}, because the terms do not converge independently.
This problem is caused by the propagator, $\Pi(t,t_1;\omega)$, which does not decay for large time differences, \mbox{$t \gg t_1$}.
To solve this issue, we split the propagator into a decaying, \emph{reduced} part~\cite{Thielmann03,Riwar13}, defined by
\begin{align}
\label{eq_Pibardef}
\Pibar(t,t_1;\omega) &= \Pi(t,t_1;\omega) - \boldsymbol P(t)\otimes \boldsymbol e^T e^{i\omega(t_1-t)},
\end{align}
and a non-decaying part.
The reduced propagator approaches zero for $t \gg t_1$, on a time scale given by the relaxation dynamics of the occupation vector, i.\,e.,~roughly on the scale $\Gamma^{-1}$.

\subsection{Expansion for slow gate-voltage driving}
\label{sec_adexp}

\begin{table*}
\begin{tabular}{|l|l|l|l|}
\hline
  abbr. & definition & brief description & further reading\\
 \hline
 \hline
 $(\mathrm{i})$ & instantaneous contribution &  instantaneous response of the system to the time- & Sec.~\ref{sec_adexp}, \\
 & &                                            dependent change of parameters  & Apps.~\ref{app_adexp_auxfunc}-\ref{app_adexp_reducedpropagator} \\
 \hline
 $(\mathrm{a})$ & adiabatic response &          small retarded response of the system to the time-& Sec.~\ref{sec_adexp},\\
 & &                                            dependent change of parameters  & Apps.~\ref{app_adexp_auxfunc}-\ref{app_adexp_reducedpropagator} \\
 \hline
 $(\mathrm{s})$ & resummed adiabatic expansion & sum of all orders in the frequency expansion, valid for  & Sec.~\ref{sec_results_resummedHF} \\
 & & driving frequencies not exceeding the tunneling rate & \\
\hline
\hline
\hline
  abbr. & definition & employed approximation scheme & further reading\\
 \hline
 \hline
 $(\mathrm{HF})$ & high noise frequencies (\mbox{$\omega \gg \Gamma$}) & order-by-order scheme of the $\Gamma$ expansion:  & Secs.~\ref{sec_gammaschemes} and \ref{sec_results_HF},\\
 & & observables are expanded in leading order in $\Gamma$ & App.~\ref{app_HFscheme} \\
 \hline
 none & all noise frequencies & when the noise-frequency regime is not specified, we & Secs.~\ref{sec_gammaschemes} and \ref{sec_results_crossover}\\
 & &  employ the crossover scheme of the $\Gamma$ expansion: & \\
 & &  only kernels are expanded in leading order in $\Gamma$ & \\
 \hline
 $(\mathrm{LF})$ & low noise frequencies  (\mbox{$\omega \ll \Gamma$}) & crossover scheme of the $\Gamma$ expansion with   & Secs.~\ref{sec_gammaschemes} and \ref{sec_results_LF}\\  
 & &   neglected frequency dependence of kernels& \\
 \hline
 \end{tabular}
 \caption{List of abbreviations, indicating the applied approximation scheme when used as a superscript.}
 \label{tab_super}
\end{table*}

We now come to the expansion for slow driving as introduced in Sec.~\ref{sec_model_adexp}.
To justify this expansion, we make an assumption concerning the time scale of the gate-voltage driving with respect to the response time of the system and the reservoirs, summarized in the condition $\delta\epsilon\Omega\beta/\Gamma\ll1$. 
In the following description, we closely follow the lines of Refs.~\cite{Splettstoesser06,Riwar13}. 
Thereby, we take into account the typical time scale for the support of kernels, which is given by the reservoir correlation time, $\beta$, as well as for changes in the occupation, $\boldsymbol P(t)$, given by~$\Gamma^{-1}$
(which sets the support of the reduced propagator).

We start by setting up the slow-driving expansion for the kinetic equation~\eqref{eq_master}, which determines the evolution of the occupation vector, $\boldsymbol{P}(t)$.
Therefore, we first expand the occupation vector $\boldsymbol{P}(t_1)$ in the integrand on the rhs.~of Eq.~\eqref{eq_master} around the reference time~$t$. 
In addition, an expansion of the kernel, describing a time evolution governed by time-dependently driven parameters, is performed,
\mbox{$\mathcal W(t,t_1) \rightarrow \mathcal W^\mathrm{(i)}_t(t-t_1)+ \mathcal W^\mathrm{(a)}_t(t-t_1)+\dots$}. 
See Refs.~\cite{Splettstoesser06,Riwar13} for details. 
Here, the superscript~(i) indicates that the kernel in lowest order describes a system that \textit{instantaneously} follows the time-dependent driving.
In other words, parameters are frozen at time~$t$. 
In contrast, the superscript~(a) refers to the \textit{adiabatic response}, taking into account the finite lag of the system with respect to the gate-voltage driving. 
The subscript~$t$ indicates the reference time at which all time-dependent \textit{parameters} are evaluated. 
Consistently replacing \mbox{$\boldsymbol P(t) \rightarrow \boldsymbol P^\mathrm{(i)}(t)+ \boldsymbol P^\mathrm{(a)}(t)+\dots$}
and collecting all terms of the same order in the slow-driving expansion in the kinetic equation leads to 
\begin{subequations}
\label{eq_master_ad}
\begin{align}
 0 &= \big\{\mathcal W \boldsymbol P \big\}_t^\mathrm{(i)},\\
 \partial_t \boldsymbol P^\mathrm{(i)}(t)&= \big\{\mathcal W \boldsymbol P \big\}_t^\mathrm{(a)}.
\end{align}
\end{subequations}
We use the compact curly-bracket notation~\cite{Kashuba12,Riwar13} 
\begin{subequations}
\label{eq_brackets}
 \begin{align}
  \big\{ A B \big\}_t^\mathrm{(i)} &= A^\mathrm{(i)}(t) B^\mathrm{(i)}(t),\\
   \big\{ A B \big\}_t^\mathrm{(a)} &= 
                               \begin{aligned}[t]
                                 & A^\mathrm{(i)}(t) B^\mathrm{(a)}(t) + A^\mathrm{(a)}(t) B^\mathrm{(i)}(t) \\
                                 &+ \partial A^\mathrm{(i)}(t)\, \dot B^\mathrm{(i)}(t),
                               \end{aligned}
 \end{align}
\end{subequations}
for two generic functions, $A(t)$ and $B(t)$. 
Here, in addition to the occupation vector, this involves the Laplace transforms of the kernels, \mbox{$\mathcal{W}_t^{(i/a)}(z)=\int_0^\infty d(t-t_1)e^{-z(t-t_1)}\mathcal{W}_t^{(i/a)}(t-t_1)$}, and derivatives thereof,
in the limit of zero Laplace frequency, abbreviated by
\mbox{$\mathcal{W}_t^{(i/a)}=\lim_{z\to0}\mathcal{W}_t^{(i/a)}(z)$} and \mbox{$\partial \mathcal{W}_t^{(i)}=\lim_{z\to0}(\partial \mathcal{W}_t^{(i)}(z)/\partial z)$} (equivalent notations apply to current and noise kernels used later).
Equation~\eqref{eq_master_ad}, together with the normalization conditions \mbox{$\boldsymbol e^T \cdot \boldsymbol P^\mathrm{(i)}(t) = 1$} and \mbox{$\boldsymbol e^T \cdot \boldsymbol P^\mathrm{(a)}(t) = 0$},
determines the instantaneous and adiabatic-response contributions to the occupation vector .

Furthermore, by applying the same line of arguments to the time-dependent current, Eq.~\eqref{eq_current}, we find
\begin{subequations}
\label{eq_current_ad}
 \begin{align}
  \expval{I(t)}^\mathrm{(i)} &= \frac{\boldsymbol e^T}{2} \big\{\mathcal W_I \boldsymbol P\big\}_t^\mathrm{(i)},\\
  \expval{I(t)}^\mathrm{(a)} &= \frac{\boldsymbol e^T}{2} \big\{\mathcal W_I \boldsymbol P\big\}_t^\mathrm{(a)}.
 \end{align} 
\end{subequations}
Note that $\expval{I(t)}^\mathrm{(i)}$ vanishes for the single-lead quantum dot considered in this paper, while $\expval{I(t)}^\mathrm{(a)}$
describes the non-vanishing adiabatic-response of the current.

An equivalent expansion can be performed for the auxiliary function of the finite-frequency noise, Eq.~\eqref{eq_auxiliarynoise_blocks}. 
This expansion is described in more detail in Apps.~\ref{app_adexp_auxfunc}-\ref{app_adexp_reducedpropagator}, and leads to the results
\begin{subequations}
\label{eq_auxiliarynoise_ia}
\begin{align}
\label{eq_auxiliarynoise_i}
\tilde{S}^\mathrm{(i)}(t;\omega) &= \begin{aligned}[t] &\frac{\boldsymbol e^T}{2} \Big\{\mathcal W_I^{<}\Pibar \mathcal W_I^{>} \boldsymbol P \Big\}_{t;\omega}^\mathrm{(i)} \\&
  +\frac{\boldsymbol e^T}{2} \Big\{ \mathcal W_{II} \boldsymbol P \Big\}_{t;\omega}^\mathrm{(i)} - 2 \Big\{ \tilde{I} I \Big\}_{t;\omega}^\mathrm{(i)}, \end{aligned}
  \\
\label{eq_auxiliarynoise_a}
\tilde{S}^\mathrm{(a)}(t;\omega) &= \begin{aligned}[t] &\frac{\boldsymbol e^T}{2} \Big\{\mathcal W_I^{<}\Pibar \mathcal W_I^{>} \boldsymbol P \Big\}_{t;\omega}^\mathrm{(a)} \\&
  +\frac{\boldsymbol e^T}{2} \Big\{ \mathcal W_{II} \boldsymbol P \Big\}_{t;\omega}^\mathrm{(a)} - 2 \Big\{ \tilde{I} I \Big\}_{t;\omega}^\mathrm{(a)}, \end{aligned}
\end{align}
\end{subequations}
Here, curly brackets with four operators can be obtained by successively applying Eq.~\eqref{eq_brackets}, see also Eq.~\eqref{eq_brackets2}. 
The additional subscript~$\omega$ indicates that the frequency-dependent functions $\mathcal W_I^{<},\mathcal W_I^{>},\Pibar$ and $\tilde{I}$ are evaluated at this frequency. 
The functions  \mbox{$ \tilde{I}^\mathrm{(i)}(t;\omega)= \boldsymbol e^T [\{\mathcal W_I \boldsymbol P\}_t^\mathrm{(i)} - \{\mathcal W_I \boldsymbol P\}_{t;\omega}^\mathrm{(i)}]/2 i\omega$} 
and \mbox{$\tilde{I}^\mathrm{(a)}(t;\omega)= \boldsymbol e^T [ \{\mathcal W_I \boldsymbol P\}_t^\mathrm{(a)} - \{\mathcal W_I \boldsymbol P\}_{t;\omega}^\mathrm{(a)}]/2 i\omega$} 
are derived in App.~\ref{app_adexp_tildeI}, and expressions for the instantaneous and adiabatic-response contributions to the reduced propagator are given in App.~\ref{app_adexp_reducedpropagator}.

Using Eqs.~\eqref{eq_auxiliarynoise_ia} together with Eqs.~\eqref{eq_noise_adexp}, the instantaneous and the adiabatic-response contributions to the time-resolved finite-frequency noise and its harmonics can be evaluated. 
These quantities are the objects of main interest in this paper.
An additional resummation of higher-order contributions in the slow-driving expansion of the noise is only considered in the special case of noise at high frequencies, as presented in Sec.~\ref{sec_results_resummedHF} and indicated by~(s).
A list of superscripts denoting different approximation schemes used in this paper is provided in Tab.~\ref{tab_super}.

\subsection{Expansion in the tunnel-coupling strength}
\label{sec_gammaschemes}

On top of the adiabatic expansion, outlined in Sec.~\ref{sec_adexp}, we perform a perturbative expansion in the tunnel-coupling strength~$\Gamma$. 
Since we are interested in a weakly coupled quantum dot, \mbox{$\beta \Gamma \ll 1$}, we restrict the following discussion to the \lq sequential tunneling limit\rq, 
where first-order tunneling processes govern the dynamics of the driven quantum dot. We expect that second or higher order processes are reasonably suppressed for the 
system of interest.\footnote{For a time-dependently driven, interacting quantum dot, effects of second order in the tunnel coupling have been studied for the 
relaxation dynamics~\cite{Splettstoesser10}, the pumping current~\cite{Splettstoesser06} and the zero-frequency pumping-current noise~\cite{Riwar13}.}

While an  order-by-order expansion in $\Gamma$ (see the end of this section for more details), has been applied for the calculation of the pumping current~\cite{Splettstoesser06} and the zero-frequency noise of these systems~\cite{Riwar13}, 
it is in general not applicable for the calculation of the finite-frequency noise, see e.\,g.~Refs.~\cite{Kack03,Braun06,Droste15}.  
The reason is that the frequency-dependent propagator, Eq.~\eqref{eq_Pibardef}, can in general not be expanded order-by-order in $\Gamma$, 
which can intuitively be understood from the Dyson equation~\eqref{eq_dysontime} and also follows from its determining equations given in Eq.~\eqref{eq_Pibar_dysonlaplace2_ia}.
To evaluate the finite-frequency noise we therefore use a different scheme, which we refer to as the \textit{crossover scheme}. 
It means that only kernels are expanded in the tunnel-coupling strength, while for other objects resulting from them---like the reduced propagator---we keep all orders in~$\Gamma$. 
More specifically, we first derive the adiabatic expansion of Eqs.~\eqref{eq_master}-\eqref{eq_auxiliarynoise_blocks} as outlined in Sec.~\ref{sec_adexp} and Apps.~\ref{app_adexp_auxfunc}-\ref{app_adexp_reducedpropagator}. 
We then keep all terms on the right-hand sides of the resulting equations which include kernels in first order in~$\Gamma$.
For explicit expressions we refer to App.~\ref{app_adexp_kernels}.

While this scheme is in principle applicable for all noise frequencies, it turns out to be in some cases overcomplicated.
Furthermore, for noise-frequencies of the order of the tunnel-coupling, \mbox{$\omega\sim\Gamma$}, care has to be taken to consistently treat higher-order coupling terms~\cite{Kack03}, see App.~\ref{app_adexp_kernels} for details. 
We therefore only employ this full crossover scheme when calculating the finite-frequency noise for intermediate noise frequencies, which is done in Sec.~\ref{sec_results_crossover}. 
For the regime of low noise frequencies \mbox{($\omega\ll\Gamma$)}, as well as for the regime of high noise frequencies \mbox{($\omega\gg\Gamma$)}, simplified schemes for the perturbative approximation can be employed, as we explain in the following.

\subsubsection{\texorpdfstring{Low noise frequencies, $\omega \ll \Gamma$}{Low noise frequencies}}

We find that for low noise frequencies, \mbox{$\omega \ll \Gamma$}, it is reasonable to neglect the frequency dependence of kernels in Eq.~\eqref{eq_auxiliarynoise_ia} and to only keep the frequency dependence of the propagator, see also Ref.~\cite{Braun06}. 
Any correction in~$\omega$ to the zero-frequency kernel would be smaller than the neglected cotunneling terms, as long as \mbox{$\omega \ll \Gamma$}.
More specifically, we keep the frequency dependence of free parts of the contour only, which means that also the kernel in the Dyson equation~\eqref{eq_dysontime} is evaluated at zero frequency.
This is justified because for low noise frequencies the time scales of fluctuations is much larger than the support of the kernels (given by~$\beta$).
The frequency line in the propagator is therefore expected to rather play a role for free parts of the propagation.
Quantities calculated in this regime are indicated by a superscript~(LF).

\subsubsection{\texorpdfstring{High noise frequencies, $\omega \gg \Gamma$}{High noise frequencies}}

In the high noise-frequency regime, \mbox{$\omega \gg \Gamma$} indicated by~(HF), it turns out that the previously mentioned order-by-order scheme can be employed. 
The reason is that the frequency-dependent propagator in Eq.~\eqref{eq_auxiliarynoise_ia} is well described by the frequency-dependent \emph{free} propagator in this regime, see also Eqs.~\eqref{eq_Pibar_dysonlaplace2_ia}.

This scheme is then consistently applied to the kinetic equation~\eqref{eq_master_ad}, to the current in Eq.~\eqref{eq_current_ad}, as well as to the auxiliary function in Eq.~\eqref{eq_auxiliarynoise_ia}, after the adiabatic expansion has been performed.
After expanding all contributions to these equations up to first order in the tunnel coupling, we sort all terms on the left-~and the right-hand sides~of the resulting equations by their order in the tunnel-coupling strength
and keep the leading contributions only.
For the auxiliary function in Eq.~\eqref{eq_auxiliarynoise_blocks}, this turns out to be
\begin{align}
\label{eq_auxiliarynoiseHF}
 \tilde{S}^{(\text{l},\mathrm{HF})}(t;\omega) &= \frac{\boldsymbol e^T}{2} \mathcal W_{II}^\mathrm{(i)}(t;\omega)\, \boldsymbol P^{(\text{l},\mathrm{HF})}(t),
\end{align}
with \mbox{$l=\mathrm{i/a}$} for the instantaneous contribution and the adiabatic response, respectively.
The derivation of Eq.~\eqref{eq_auxiliarynoiseHF} is outlined in App.~\ref{app_HFscheme}, and the equation is applied in Sec.~\ref{sec_results_HF}, where we analyze noise and noise harmonics evaluated at high noise frequencies.

The simplicity of Eq.~\eqref{eq_auxiliarynoiseHF} makes it possible to go beyond the first-order (adiabatic-response) approximation for the slow driving,
by employing the same strategy as discussed up to here. 
In the sequential-tunneling regime (for weak coupling $\Gamma$), the kinetic equation~\eqref{eq_master_ad} and the current formula, Eq.~\eqref{eq_current_ad}, 
can be extended to higher orders in the slow-driving expansion by keeping the instantaneous kernel and taking successively higher-order terms in the slow-driving expansion for the occupation vector, see also Ref.~\cite{Cavaliere09,Riwar16}. 
Importantly, we here find that this extension is also possible for the auxiliary function of Eq.~\eqref{eq_auxiliarynoiseHF}, hence the parameter~$l$, denoting the expansion order.
The case where all higher-order contributions to the slow-driving expansion are resummed is studied in Sec.~\ref{sec_results_resummedHF} and indicated by the superscript~(s).

A list of different orders in the slow-driving expansion and of all employed approximation schemes for the tunnel-coupling expansion is provided in Tab.~\ref{tab_super}.

\section{Noise at high frequencies}
\label{sec_results_HF}

Many results shown in this as well as in the following sections can be understood by comparing the time scales which are present in our system.
On the one hand, the relaxation dynamics of the quantum dot is controlled by the time scale on which the occupation vector varies, i.\,e.,~roughly $\Gamma^{-1}$.
On the other hand, we associate the time scale $\omega^{-1}$ to charge fluctuations with frequency~$\omega$, where a hole or electron excitation is momentarily created in the reservoir, paired with an electron entering or leaving the quantum dot.
Although further time scales are in principle introduced by the applied time-dependent driving scheme, we recall that we consider slow driving in this work; this means that the timescale due to driving is much larger than the relaxation time.

We start by considering high noise frequencies, \mbox{$\omega \gg \Gamma$}, indicated in the following by the superscript (HF). 
Here, the competition between the described time scales means that we can think of fluctuations as temporary processes, which occur much faster than any variations of the occupation vector caused by relaxation dynamics in response to the gate-voltage driving.
This intuitive picture agrees with the explicit high-frequency noise expressions, as we now outline.

\subsection{Noise expressions and fluctuation vector}
\label{sec_results_HF_general}
We first repeat that for the auxiliary function, $\tilde S(t;\omega)$, of the general noise-harmonics expression in Eq.~\eqref{eq_noise2}, we already derived the simple high-frequency form given in Eq.~\eqref{eq_auxiliarynoiseHF}.
Following the procedure of the order-by-order scheme explained in Sec.~\ref{sec_gammaschemes}, we insert this equation in Eqs.~\eqref{eq_noise_adexp} and only keep leading-order terms in $\Gamma$.
We then obtain for the instantaneous noise ($l=0$) and its adiabatic response ($l=1$) the two expressions\footnote{
The third term on the right-hand side~of Eq.~\eqref{eq_noise_adexp_a} does not contribute to Eq.~\eqref{eq_noise_HF_for_flvector_a} in leading order in the order-by-order $\Gamma$ expansion.}
\begin{subequations}
\label{eq_noise_HF_for_flvector}
\begin{align}
\label{eq_noise_HF_for_flvector_i}
\mathcal S^{(\mathrm{i,HF})}(n;\omega) &= \begin{aligned}[t] & \int_0^T \frac{dt}{T} e^{i n\Omega t} \,\boldsymbol F^{\mathrm{(i,HF)}}(t;\omega) \\ & \cdot  \boldsymbol{P}^{\mathrm{(i,HF)}}(t), \end{aligned}
\end{align}
\begin{align}
\label{eq_noise_HF_for_flvector_a}
\mathcal S^{(\mathrm{a,HF})}(n;\omega) &= \begin{aligned}[t] & \int_0^T \frac{dt}{T} e^{i n\Omega t} \,\boldsymbol F^{\mathrm{(i,HF)}}(t;\omega) \\ & \cdot  \boldsymbol{P}^{\mathrm{(a,HF)}}(t), \end{aligned}
\end{align}
\end{subequations}
Equations~\eqref{eq_noise_HF_for_flvector} can be applied to calculate time-averaged noise spectra, \mbox{$n=0$}, as well as noise harmonics, \mbox{$n\neq0$}, of the slowly driven interacting quantum dot.
Both Eqs.~\eqref{eq_noise_HF_for_flvector} contain the vector
\begin{align}
\label{eq_flvector}
\boldsymbol F^{\mathrm{(i,HF)}}(t;\omega) &= \frac{\boldsymbol{e}^T}{2}\Big[\mathcal W_{II}^{\mathrm{(i)}}(t;\omega) + \mathcal W_{II}^{\mathrm{(i)}}(t;-\omega)\Big],
\end{align}
which we denote as the \emph{instantaneous fluctuation vector}. 
As written before, the explicit form of Eqs.~\eqref{eq_noise_HF_for_flvector} can be seen as a consequence of the fact that high-frequency fluctuations occur much faster than any relaxation dynamics of the quantum dot.
Hence, the fluctuation vector appearing in Eqs.~\eqref{eq_noise_HF_for_flvector} is given by the instantaneous one, defined in Eq.~\eqref{eq_flvector}, where the contributing kernels are evaluated with parameters frozen at time~$t$.
The retarded response of the system to the time-dependent driving enters only in terms of the lag of the occupation vector itself:
the instantaneous occupation appears in Eq.~\eqref{eq_noise_HF_for_flvector_i}, while Eq.~\eqref{eq_noise_HF_for_flvector_a} is evaluated with its adiabatic response.

\begin{figure}[b]
\begin{center}
\includegraphics[width=0.98\columnwidth]{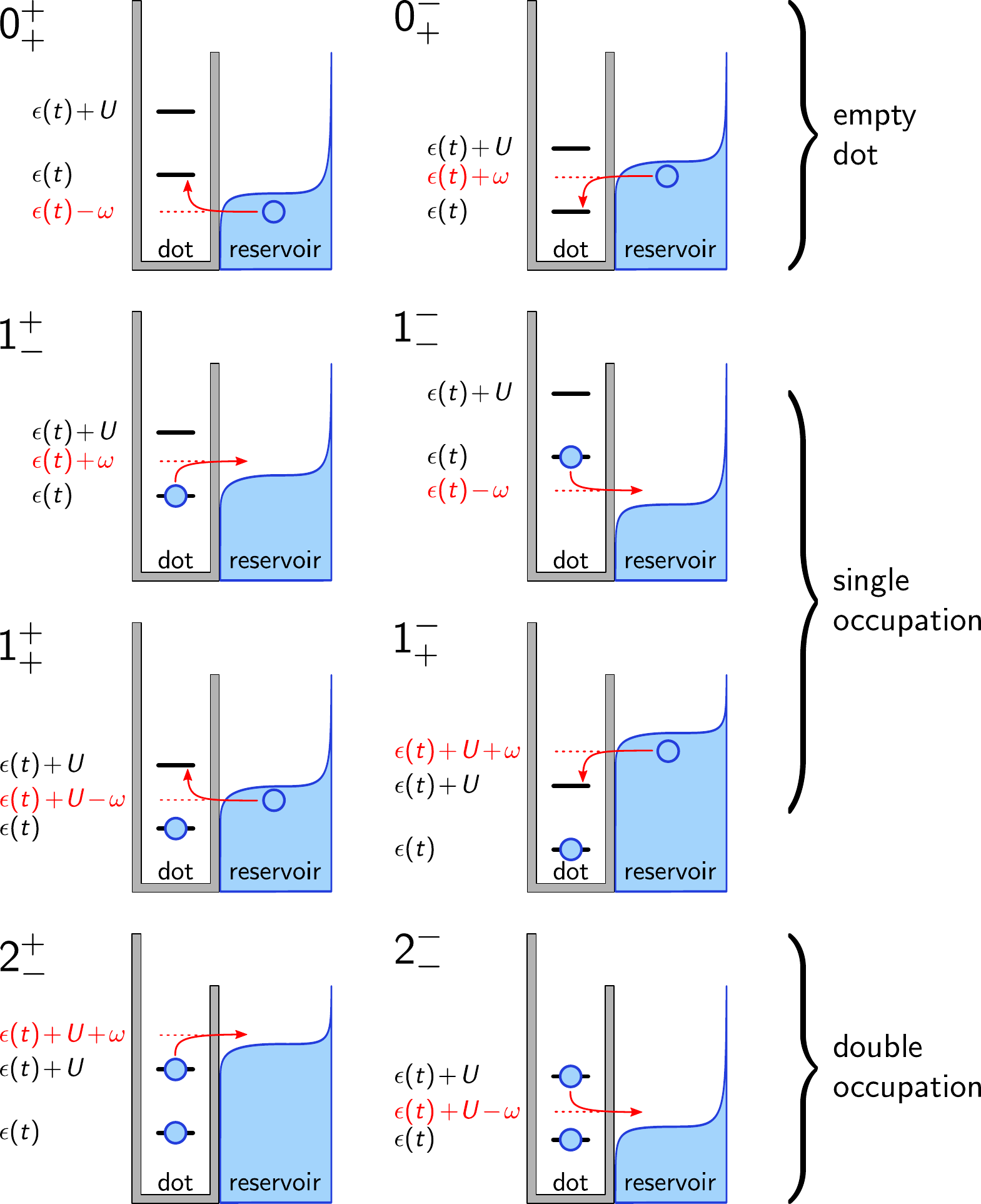}
\end{center}\vspace{-5mm}
	\caption{
	Sketch of fluctuation processes which are included in the instantaneous fluctuation vector in Eq.~\eqref{eq_flvector_example} for the spin-symmetric case, \mbox{$\epsilon_\uparrow=\epsilon_\downarrow$} 
	(`single occupation' refers to mixed states of $\uparrow$ and $\downarrow$).
	The processes are denoted as $0^\pm_+,1^\pm_\pm$ and $2^\pm_-$, where the superscript indicates the temporal absorption or emission of the energy~$\omega$ during the fluctuation, while
	the subscript indicates if the occupation momentarily increases or decreases by one electron.
	Probabilities for individual processes are proportional to the occupation of the reservoir at the relevant energies.
	An additional Zeeman splitting would split each of the shown processes in two (not shown in this figure).
	}
	\label{fig_fluctuationvector}
\end{figure}

We continue by studying the instantaneous fluctuation vector, Eq.~\eqref{eq_flvector}, in more detail, leading us towards an intuitive picture of noise spectra in the high noise-frequency regime.
For the model introduced in Eq.~\eqref{eq_hamiltonian} (and \mbox{$\epsilon_\uparrow = \epsilon_\downarrow$}) we derive
\begin{align}
\label{eq_flvector_example}
\frac{\boldsymbol F^{\mathrm{(i,HF)}} (t;\omega)}{\Gamma} &= \left(\!\begin{matrix}
                        2 f^+\big(\epsilon(t);\omega\big) \\
                        f^-\big(\epsilon(t);\omega\big) + f^+\big(\epsilon(t)+U;\omega\big) \\[0.5mm]
                        f^-\big(\epsilon(t);\omega\big) + f^+\big(\epsilon(t)+U;\omega\big) \\
                        2 f^-\big(\epsilon(t)+U;\omega\big) \\
                       \end{matrix}\!\right),
\end{align}
with \mbox{$f^\pm(x;\omega) = f^\pm(x+\omega)+f^\pm(x-\omega)$} and the Fermi functions \mbox{$f^\pm(x)=(1+e^{\pm \beta x})^{-1}$}.
The extension of Eq.~\eqref{eq_flvector_example} to the spin-split case is provided in Eq.~\eqref{app_eq_flvector_spinsplit}.

The explicit form of the instantaneous fluctuation vector in Eq.~\eqref{eq_flvector_example} can be understood by studying possible fluctuation processes.
As an example, let us consider the first entry of this vector, which in Eqs.~\eqref{eq_noise_HF_for_flvector} is  multiplied by the first entry of the instantaneous (adiabatic-response) occupation vector, 
i.e., the probability of the empty quantum-dot configuration.
A fluctuation originating from an empty dot involves an electron, which momentarily tunnels from the reservoir onto the dot and back.
The first entry on the right-hand side~of Eq.~\eqref{eq_flvector_example} represents two possible scenarios for this fluctuation:
the tunneling electron either absorbs or emits the energy quantum~$\omega$ temporarily for the time span of the fluctuation.
Both fluctuation processes are sketched in Fig.~\ref{fig_fluctuationvector} in the panels indicated by $0^+_+$ and $0^-_+$.
Since we consider an energy-independent tunnel coupling, the probability that one of the two processes occurs is proportional to the occupation of the reservoir at the initial energy of the tunneling electrons.
This occupation is what is described by the first entry of the instantaneous fluctuation vector in Eq.~\eqref{eq_flvector_example}, where the factor 2 stems from the spin degree-of-freedom.
In analogy, fluctuation processes with respect to the other dot occupations can be assigned to the further entries of the fluctuation vector, see the remaining panels in Fig.~\ref{fig_fluctuationvector}.
Fluctuations in which an electron tunnels momentarily from the quantum dot into an empty reservoir state lead to Fermi functions with the superscript~`$-$'.

We now turn to numerically evaluated high-frequency noise spectra of the quantum dot with a harmonically driven gate voltage, with \mbox{$\epsilon_g(t) = \delta \epsilon \cos(\Omega t)$}, 
and discuss features related to the processes shown in Fig.~\ref{fig_fluctuationvector}  as well as to the extended fluctuation-dissipation theorem of Sec.~\ref{sec_generalprop_FDT}. 

\subsection{Time-resolved finite-frequency noise -- noninteracting quantum dot}
\label{sec_results_HF_spectra}

We start by discussing time-resolved finite-frequency-noise spectra, considering the simplest case of a noninteracting spin-symmetric quantum dot as a reference system for the spin-split as well as the interacting dots studied in the next sections. 
The quantum-dot energy level is driven harmonically around the working point \mbox{$\bar \epsilon=0$}, with amplitude \mbox{$\delta \epsilon=10\Gamma$}, 
which results in the periodic emission and absorption of two electrons by the dot during each period of the drive.
We first present results for the instantaneous part of the noise.

\begin{figure}[tb]
\begin{center}
\includegraphics[width=\columnwidth]{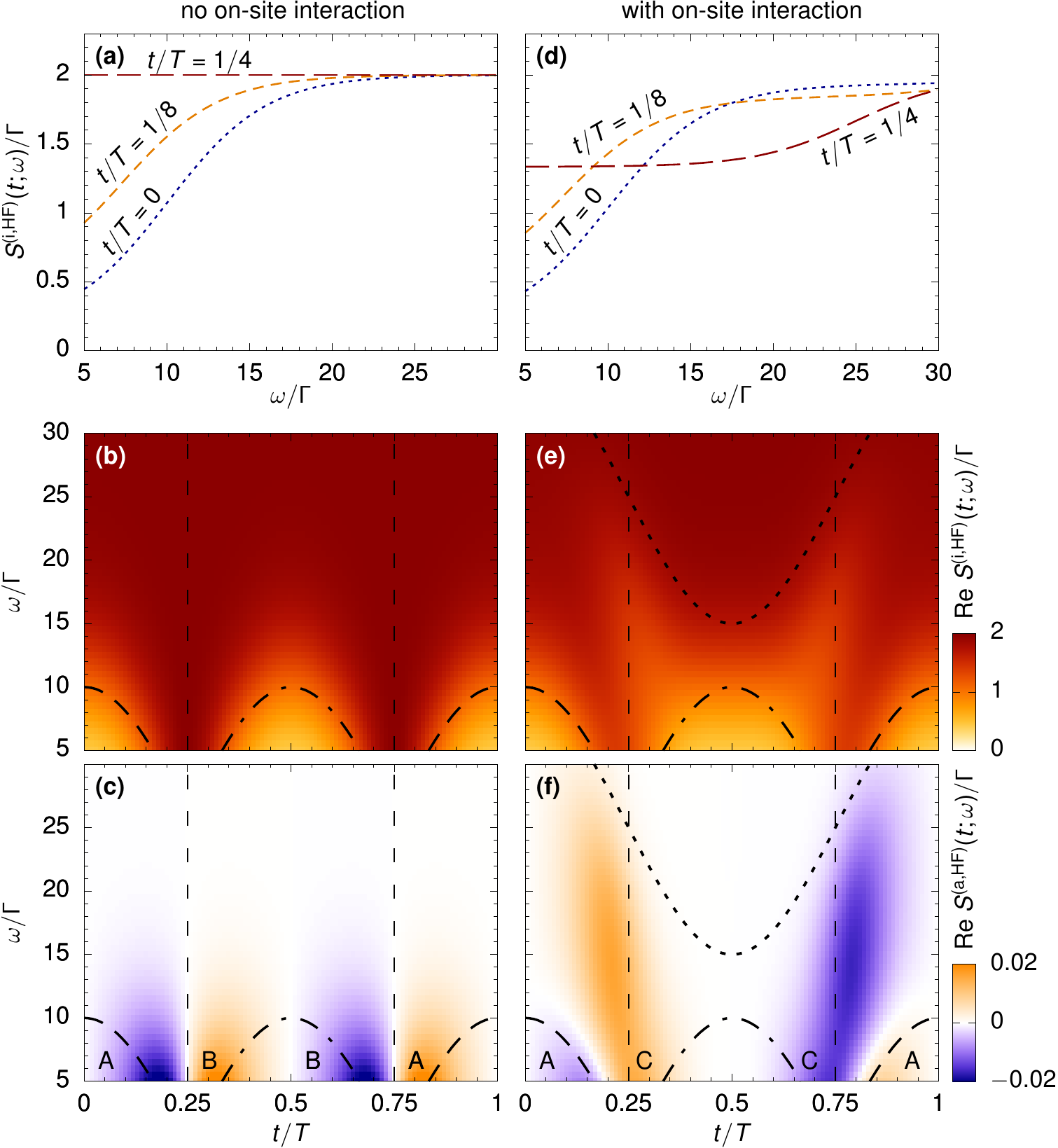}
\end{center}\vspace{-5mm}
	\caption{
	(a)-(b) Instantaneous time-resolved finite-frequency noise (HF regime) for a harmonic gate voltage and \mbox{$U=0\Gamma$};
	panel (a) presents cuts of (b) for times \mbox{$t=0,T/8,T/4$} (dotted, short dashed, long dashed lines).
	(c) Adiabatic response to the time-resolved finite-frequency noise (HF regime) for a harmonic gate voltage and \mbox{$U=0\Gamma$}.
	Additional lines in (b)-(c) show $\epsilon(t)$ and $-\epsilon(t)$ (thick dashed and dashed-dotted) and zero-crossings of~$\epsilon(t)$ (thin dashed).
	(d)-(f) Similar to (a)-(c) with \mbox{$U=25\Gamma$} and $\epsilon(t)+U$ shown by the thick dotted line.
	Further parameters are \mbox{$\beta=1/(3\Gamma)$}, \mbox{$\epsilon_\uparrow=\epsilon_\downarrow$}, \mbox{$\bar \epsilon = 0\Gamma$}, \mbox{$\delta \epsilon = 10\Gamma$} and \mbox{$\Omega=0.03\Gamma$}.
	}
	\label{fig_noise_HFtfr}
\end{figure}

The instantaneous time-resolved finite-frequency noise of this setup 
is plotted in Fig.~\ref{fig_noise_HFtfr}~(a), for three different times $t$, and in Fig.~\ref{fig_noise_HFtfr}~(b) as a function of time and frequency.
In both figures we find a smeared-out step roughly centered at the noise frequency \mbox{$\omega=|\epsilon(t)|$}, see also the thick dashed and thick dashed-dotted lines in~(b). 
This step structure at every instant of time is expected, since the strength of an individual fluctuation process strongly depends on the occupation of the reservoir at a specific energy, see Fig.~\ref{fig_fluctuationvector}.
As a consequence, the appearance of a particular step indicates that a related fluctuation processes (or several processes) become energetically possible or suppressed.

In the line-cuts shown in Fig.~\ref{fig_noise_HFtfr}~(a), this step is clearly visible for the dotted line, showing the noise spectrum at time \mbox{$t=0$}, where the level position is given by~\mbox{$\epsilon(0)=10\Gamma$}.
The instantaneous dot occupation at this point in time is given by the empty configuration, which means that only the two processes $0^\pm_+$ of Fig.~\ref{fig_fluctuationvector} can in principle contribute to the noise.
Since the process $0^-_+$ is suppressed for all noise frequencies whenever the energy level is above the Fermi energy, we find that the fluctuation process $0^+_+$ is responsible for the observed noise spectrum.
However, also this process is suppressed in the case \mbox{$\omega < \epsilon(0)$}, which leads to the visible step.

During the time-dependent drive, the quantum-dot level first moves towards the Fermi energy; consequently, the step approaches lower and lower noise frequencies, 
as shown by the short dashed and long dashed lines in Fig.~\ref{fig_noise_HFtfr}~(a).\footnote{At $t=T/4$ the step is not visible in this plot, which does not continue to zero noise-frequency. 
At zero frequency, \mbox{$\omega=0$}, the instantaneous time-resolved finite-frequency noise always vanishes due to charge conservation.}
After the Fermi-energy crossing at \mbox{$t=T/4$}, the quantum-dot level moves to even lower energies and the dot tends towards double occupation.
The dominating fluctuation process is now given by $2^+_-$, because the second possibility, $2^-_-$, is suppressed by the Fermi function of the reservoir, 
and therefore a step appears for \mbox{$\omega<-\epsilon(t)=|\epsilon(t)|$}, see Fig.~\ref{fig_noise_HFtfr}~(b).

For very high noise frequencies, \mbox{$\omega \gg |\epsilon(t)|$}, the instantaneous time-resolved finite-frequency noise reaches a plateau value of $2\Gamma$.
The technical reason for this is that in this regime the instantaneous fluctuation vector becomes time independent and is well described by $2 \Gamma \boldsymbol e^T$ with \mbox{$\boldsymbol e^T=(1,1,1,1)^T$}.
More intuitively, it is explained by the fact that all fluctuation processes which temporarily absorb the energy~$\omega$, i.\,e. the ones in the left column in Fig.~\ref{fig_fluctuationvector}, can contribute to the noise,
while processes which emit the energy~$\omega$ are suppressed by the vanishing occupation of the reservoir at energies far above the Fermi energy.

The instantaneous noise studied in the previous paragraph does not account for the fact that the occupation of the driven system slightly lags behind the driving signal.
To investigate this, we turn to the adiabatic-response contribution $S^{\mathrm{(a,HF)}}(t;\omega)$.
The adiabatic response is plotted in Fig.~\ref{fig_noise_HFtfr}~(c).
A first observation are the alternating signs going along with the fact that the time average of the adiabatic-response noise vanishes, which is a consequence of the one-parameter driving considered in this paper, see Sec.~\ref{sec_generalprop_vanishing_a}.
More specifically, the sign of the adiabatic-response noise changes whenever the energy level crosses the Fermi energy during the drive (thin dashed lines).
The regions indicated by `A/B' in panel~(c) mark regions where the empty/double configuration of the dot dominates in the quasi-stationary state, respectively.
We find that the adiabatic response reduces the time-resolved noise before the crossing (blue regions) and increases the noise after the crossing (orange regions) 
leading to a slight shift of the total time-resolved noise, which reflects the lag of the occupation vector.
The adiabatic-response contribution to the noise is suppressed when the noise frequency is larger than the drive amplitude,~$\delta \epsilon$, 
i.\,e., when the noise probes the reservoir occupation far away from the Fermi energy, see the Fermi functions in Eq.~\eqref{eq_flvector_example}.
Naturally, at these energies the fluctuations are not sensitive to the time-dependent modulation of the energy level.
In this regime, as explained in the previous paragraph, the instantaneous fluctuation vector is approximately given by $2 \Gamma \boldsymbol e^T$, 
because here all fluctuation processes which absorb the energy $\omega$ are energetically allowed.
This leads to a vanishing integrand in Eq.~\eqref{eq_noise_HF_for_flvector_a}, because the adiabatic response of the occupation vector fulfills the normalization constraint \mbox{$\boldsymbol e^T \cdot \boldsymbol P^{\mathrm{(a)}}(t) = 0$}.

\subsection{Time-resolved finite-frequency noise -- interacting quantum dot}

One of the main questions addressed in this paper is how an on-site interaction $U$ changes finite-frequency noise spectra of the driven quantum dot. 
Since an interacting quantum dot, due to the effect of Coulomb blockade, can be \textit{single occupied} over a large range of parameters, we expect the fluctuation processes $1_\pm^\pm$ of Fig.~\ref{fig_fluctuationvector} to play a major role,
leading to additional steps in our noise spectra.
As before, we begin with the instantaneous time-resolved finite-frequency noise, which is presented in Fig.~\ref{fig_noise_HFtfr}~(d) and (e) for the interacting quantum dot.
We consider the interaction strength to be larger than the driving amplitude, which has the consequence that the quantum dot emits and absorbs only a single electron during one period of the drive. The step at $\omega < |\epsilon(t)|$, which is also seen for the noninteracting dot, is now caused by the fluctuation processes $0^+_+$ and $1^+_-$, instead of $0^+_+$ and $2^+_-$ in the noninteracting case.
In addition,  we find in Fig.~\ref{fig_noise_HFtfr}~(d), as expected, a step centered roughly at \mbox{$\omega\approx|\epsilon(t) + U|$}. It emerges when the quantum dot occupation is mostly non-zero.
Indeed, this new step is visible for the long dashed line, which shows the spectrum evaluated at time $t=T/4$, where \mbox{$\epsilon(T/4) + U=25\Gamma$} 
[the first step with \mbox{$\epsilon(T/4)=0$} is not shown due to the frequency range chosen for this plot].
The generating fluctuation process turns out to be $1_+^+$. Note that for a quantum-dot level far below the Fermi energy, a step appears at \mbox{$\omega=-\big(\epsilon(t) + U\big)$}, generated by the process $2^+_-$. 
In Fig~\ref{fig_noise_HFtfr}~(e), we see how these steps involve in time: both steps are indicated by the thick dashed, dashed-dotted and dotted lines. 

Let us now discuss how a finite on-site Coulomb interaction modifies the adiabatic response of the time-resolved finite-frequency noise, see Fig.~\ref{fig_noise_HFtfr}~(f).
When we gradually change the system from noninteracting to interacting, using the parameters of Fig.~\ref{fig_noise_HFtfr}, we find that
the two regions marked with `A' and `B' in panel~(c) separate and two new colored regions, `C', related to a single-occupied dot, emerge in between, see panel~(f).
Here, we also observe that the regions `B'---associated with a double occupied dot---have disappeared.
See also the additional plots in App.~\ref{app_ABCplots}, where the behavior for different interaction strengths is shown.
This disappearance is due to the strong interaction, which prevents the occupation of the quantum dot with two electrons.
Interestingly, the regions `C' extend to much larger noise frequencies than the regions `A' and `B', because only when the noise frequency exceeds the value \mbox{$|\epsilon(t)+U|$}, 
the noise always probes the reservoir occupation far away from the Fermi energy and the effect of the driving disappears in the noise.
What is more, we find that the boundary between the regions `A' and `C' now shows a frequency-dependent bending, which is linked to the fact that the single-occupied dot is doubly degenerate, in contrast to the empty dot.
The consequence is that, in the case where double occupation is suppressed by the strong on-site interaction,
fluctuations which originate from the empty configuration can in principle contribute stronger to the noise than fluctuations beginning from a single-occupied dot.
For example, at \mbox{$\epsilon(t)=0$} and \mbox{$\omega \lesssim U$}, the first entry of the instantaneous fluctuation vector in Eq.~\eqref{eq_flvector_example} roughly doubles the value of the second (third) entry.
Furthermore, since the adiabatic response of the occupation vector, $\boldsymbol P^\mathrm{(a,HF)}(t)$, is proportional to $(1,-1/2,-1/2,0)$, when evaluated in the vicinity of the energy-level's zero-crossing, 
the boundary between `A' and `C'  reveals the difference between the first and the second (third) entry of the instantaneous fluctuation vector.
The boundary line indicates points where fluctuation processes originating from an empty and a singly occupied dot are of equal magnitude, see also Eq.~\eqref{eq_noise_HF_for_flvector_a}.
If the dot is noninteracting, the additionally allowed fluctuations between single and double occupation result in a cancellation of this frequency dependence, as it is visible in panel~(c).
Note that for strong Coulomb interaction on the quantum dot, the described boundary generally differs from the point, where the \emph{tunnel rates} which change the instantaneous dot occupation from empty to single and vice versa are equal.
The latter leads to the known condition \mbox{$\epsilon(t)=\log(2)/\beta$}, which defines the emission and absorption times of the first electron [the second electron is emitted and absorbed at \mbox{$\epsilon(t)=-\log(2)/\beta-U$}].

\subsection{Noise harmonics -- noninteracting, interacting and spin-split quantum dot}
\label{sec_results_HF_spectra_harmonics}

\begin{figure*}[tb]
\begin{center}
\includegraphics[width=.98\textwidth]{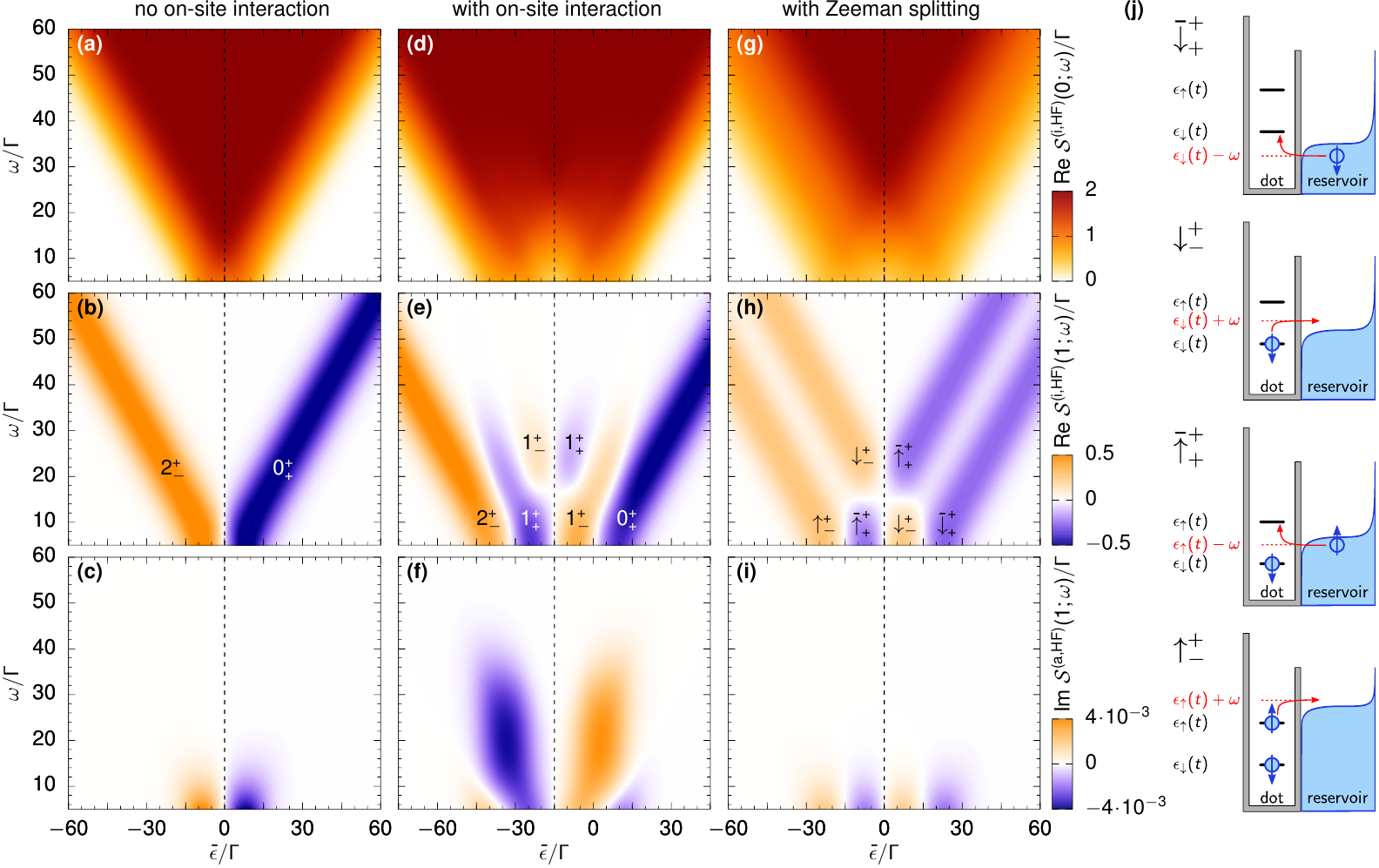}
\end{center}\vspace{-5mm}
	\caption{
	(a)-(i) Noise harmonics (HF regime) for a harmonic gate voltage. 
	Shown is the zeroth harmonic of the instantaneous noise (first row), 
	the first harmonic of the instantaneous noise (second row) and
	the first harmonic of the adiabatic-response noise (third row) 
	for \mbox{$\beta=1/(3\Gamma)$}, \mbox{$\delta \epsilon = 10\Gamma$} and \mbox{$\Omega=0.02\Gamma$} and (a)-(c) \mbox{$U = 0\Gamma$},
	(d)-(f) \mbox{$U = 30\Gamma$}, (g)-(i) \mbox{$U = 0\Gamma$} and \mbox{$\bar \epsilon_\uparrow - \bar \epsilon_\uparrow = 30\Gamma$}. 
	Dotted lines indicate the electron-hole symmetric point.
	(j) Fluctuation processes for a spin-split quantum dot. 
	The \mbox{$\downarrow/\bar \downarrow$} denote processes where an electron with 
	respective spin projection leaves/enters the quantum dot during the fluctuation [similarly for \mbox{$\uparrow/\bar \uparrow$}].
	The sup-/superscripts $\pm$ are defined as in Fig.~\ref{fig_fluctuationvector}.
	}
	\label{fig_noise_HFharmonics}
\end{figure*}
In this work, we promote the study of noise harmonics as a tool to identify particular fluctuation processes in the finite-frequency noise. The aim is to thereby identify interaction physics from a combination with time-dependent driving.
For this purpose, the central object of this paper is the decomposition of the time-resolved finite-frequency current noise, $S(t;\omega)$, in individual current-noise harmonics, $\mathcal S(n;\omega)$, as introduced in Eq.~\eqref{eq_noise}.
In particular, we focus on the first noise harmonic in addition to the more standardly considered zeroth harmonic, i.\,e., the average of the noise over one driving period. The first noise harmonic can, e.\,g., be accessed in experiments by multiplying the noise with the driving signal \cite{Marguerite16}.

In this section, we show that contributions to the first noise harmonic can be linked to individual fluctuation processes. Furthermore, specific patterns appear when Coulomb interaction is present on the quantum dot, which can be clearly distinguished from a sequence of two resonances due to  a Zeeman splitting.
Another benefit of noise harmonics is that any constant background noise cancels for harmonics with $n\geq 1$.
Besides that, we already proved in Sec.~\ref{sec_generalprop_vanishing_a} that the time-averaged adiabatic-response noise of our model system always vanishes, 
and thus it is natural to analyze the first harmonic of the adiabatic response.

In Fig.~\ref{fig_noise_HFharmonics}, we present noise harmonics for three different scenarios: a noninteracting spin-symmetric quantum dot, an interacting spin-symmetric quantum dot and a noninteracting spin-split quantum dot. The comparison of these three cases allows us to unanimously identify effects due to the many-body Coulomb interaction.
All harmonics are plotted as a function of the noise frequency and the working point,~$\bar \epsilon$, of the harmonic gate-voltage drive.
In the first row of Fig.~\ref{fig_noise_HFharmonics} we plot instantaneous parts of \emph{zeroth} noise harmonics.
These represent time averages of time-resolved finite-frequency noise spectra similar to the ones shown in Fig.~\ref{fig_noise_HFtfr}~(b) and~(d).
We see that the zeroth noise harmonic is finite whenever the noise frequency exceeds the distance between the lead's Femi energy and a dot addition energy.
As mentioned before, the related adiabatic-response contributions of the zeroth noise harmonics vanish.

We now turn to \emph{first} noise harmonics, which are shown in the second and third row of Fig.~\ref{fig_noise_HFharmonics}, and which we use to study the effect of time-dependent driving on the noise.

As expected from the extension of the fluctuation-dissipation theorem to the first noise harmonics, Eq.~\eqref{eq_fdt_i1}, the panels in the middle row in Fig.~\ref{fig_noise_HFharmonics} are
closely related to the first derivative, with respect to~$\bar \epsilon$, of the noise shown in the upper row in the same figure.
Interestingly,  we can assign a single fluctuation process, which generates the contribution to the first noise harmonic, namely to each of the colored regions.
Fig.~\ref{fig_noise_HFharmonics}~(b) presents the instantaneous noise contribution of a noninteracting spin-symmetric quantum dot. 
In this figure, we find two straight, broadened lines centered around \mbox{$\omega= \pm \bar \epsilon$}. 
Their widths are given by twice the amplitude of the gate-voltage drive.
The dominant processes are $2^+_-$ and $0^+_+$, as defined in Fig.~\ref{fig_fluctuationvector} and indicated in Fig.~\ref{fig_noise_HFharmonics}~(b).
The reason that the first harmonic clearly differentiates between these processes is that all fluctuations involving reservoir states far away from the Fermi energy cannot contribute: 
the magnitudes of the latter are not sensitive to the gate-voltage drive.
In other words, the first noise harmonic only includes fluctuations, where the fluctuating electron comes from or tunnels into a reservoir state close to the Fermi energy.
From these fluctuations, the occupation vector in Eq.~\eqref{eq_noise_HF_for_flvector_i} then selects the processes $2^+_-$ and $0^+_+$ in the two colored lines in Fig.~\ref{fig_noise_HFharmonics}~(b).
We find that for the related adiabatic-response noise (purely imaginary at high noise frequencies) in Fig.~\ref{fig_noise_HFharmonics}~(c), 
no clear identification of fluctuation processes is possible, because the adiabatic-response of the quantum-dot occupation allows for several processes to contribute with comparable strengths in Eq.~\eqref{eq_noise_HF_for_flvector_a}.
Here, a nonzero first harmonic not only requires the condition that reservoir states in the vicinity of the Fermi energy take part, but also that the quantum-dot occupation itself strongly varies during the drive.
To fulfill the second condition, the quantum-dot level needs to be close to the Fermi energy, additionally requiring \mbox{$|\bar \epsilon| \lesssim \delta \epsilon$} for the working points.

Coulomb interaction on the quantum dot strongly modifies the first noise harmonic.
To study its impact, it is instructive to compare the first column of Fig.~\ref{fig_noise_HFharmonics} with the second column, where a strong on-site interaction~$U$ has been included.
In the instantaneous part of the first noise harmonic in panel~(e), we again find two straight lines with widths set by twice the driving amplitude, but now centered\footnote{We recall that 
\mbox{$\epsilon(t)=\log 2/\beta$} and \mbox{$\epsilon(t)=-U-\log 2/\beta$} set the emission/absorption times of electrons from the interacting quantum dot.} 
around \mbox{$\omega - \bar \epsilon= \log 2/\beta$} and \mbox{$\omega- \bar \epsilon= -U-\log 2/\beta$}.
In the region bounded by these (external) lines, we observe a new pattern, which turns out to be specific to the presence of on-site Coulomb interaction.
The new pattern is anti-symmetric with respect to the electron-hole symmetric point (dotted lines) and it tends to zero, when the noise frequency exceeds the value of~$U$. 
Again, we can assign a dominating fluctuation process to each region in Fig.~\ref{fig_noise_HFharmonics}~(e) as indicated.
An important difference with respect to the noninteracting case is that dominant processes for the first noise harmonic not only depend on the working point, but also on the noise frequency.
This additional dependency occurs in the region where the dot is mostly single occupied, because here fluctuations between empty/single occupation and single/double occupation 
are both possible but contribute at different noise frequencies.

The adiabatic response to the first noise harmonic, Fig.~\ref{fig_noise_HFharmonics}~(f), also shows additional contributions when compared to Fig.~\ref{fig_noise_HFharmonics}~(c).
Here, the strongest contributions appear if the system is driven around the working points \mbox{$\bar \epsilon=\log 2/\beta$} and \mbox{$\bar \epsilon=-U-\log 2/\beta$}, which
are the points where the occupation vector changes most strongly during the drive. 
This first harmonic of the adiabatic-response noise exhibits the  characteristic features of the interplay between time-dependent driving and strong Coulomb interaction, 
which we have previously identified in the time-resolved finite-frequency noise: 
we see the bended line of the sign change as a function of noise frequency and working point as well as non-vanishing noise regions extended to a value set by the Coulomb interaction $U$.

As a third scenario, we analyze the impact of a magnetic field to the noninteracting quantum dot, leading to a spin splitting of the energy level. 
In a measurement limited to currents, this spin-splitting could  be confused with the two split dot resonances due to Coulomb interaction.
Here, we show that the first noise harmonic constitutes an unambiguous way to distinguish the two cases.
The zeroth and first noise harmonics of this system are shown in the third row of Fig.~\ref{fig_noise_HFharmonics}.
By comparing panels~(g)-(i) with panels (a)-(c) in the same figure, we find that the spin splitting doubles the structures which are seen in the noise harmonics of the noninteracting quantum dot.
The dominant fluctuation processes for the instantaneous first harmonic of the spin-split case are indicated in panel~(h) and sketched in Fig.~\ref{fig_noise_HFharmonics}~(j).
It is instructive to analyze the difference between the instantaneous first noise harmonic of the spin-split quantum dot in panel~(h) and the interacting quantum dot in panel~(e).
While the instantaneous contributions are qualitatively similar for noise frequencies with \mbox{$\omega \lesssim \frac{U}{2}$}, they differ strongly for higher noise frequencies.
The main reason is that processes which contain a `1' in Fig.~\ref{fig_noise_HFharmonics}~(e) can only occur when a single electron already occupies the quantum dot.
On the contrary, for the processes `$\bar \downarrow^+_+$' and `$\bar \uparrow^+_+$' in Fig.~\ref{fig_noise_HFharmonics}~(h) only the occupation of either down or up electrons is relevant, respectively,
independent of the occupation with electrons of opposite spin direction.
For the adiabatic-response noise in panel~(i), which doubles the pattern visible in panel~(c), we again find that no clear identification of fluctuation processes is possible, 
because the adiabatic-response occupation vector selects several processes with comparable strengths in Eq.~\eqref{eq_noise_HF_for_flvector_a}.

\subsection{Noise spectra beyond the adiabatic response}
\label{sec_results_resummedHF}
In the previous sections we analyzed either instantaneous contributions to the noise or adiabatic-response contributions, 
the latter describing corrections to the instantaneous noise as a consequence of a small retarded response of the system.
Both contributions rely on a slow driving of the quantum dot. In an experimental realization, such a slow driving might reduce the magnitude of the signal to be detected. 
It is therefore of interest to find out  whether the driving frequency can be increased without modifying the features described in the previous sections.
Interestingly, in the high noise-frequency regime, it turns out that the results of Secs.~\ref{sec_results_HF_general}-\ref{sec_results_HF_spectra_harmonics} are transferable to faster driving schemes, i.\,e.,~beyond the adiabatic response.
This is possible as long as the time scale of fluctuations, $\omega^{-1}$, is smaller than both the scale of quantum-dot relaxation dynamics and any time scales introduced by the driving scheme.
In this case, the auxiliary noise function of Eq.~\eqref{eq_auxiliarynoise} obtains the simple form given in Eq.~\eqref{eq_auxiliarynoiseHF}, see also App.~\ref{app_HFscheme}.
As discussed in Sec.~\ref{sec_gammaschemes}, this equation not only holds for instantaneous, \mbox{$l=0$}, and adiabatic-response contributions, \mbox{$l=1$}, 
but also in all orders in $l$ of the slow-driving expansion.
By summing up this expansion series, we obtain the generalized auxiliary function
\begin{align}
 \label{eq_auxiliarynoiseHF_resummed}
 \tilde{S}^\mathrm{(s,HF)}(t;\omega) &= \frac{\boldsymbol e^T}{2} \mathcal W_{II}^\mathrm{(i)}(t;\omega) \, \boldsymbol P^{\mathrm{(s,HF)}}(t),
\end{align}
marked with the additional superscript (s) for `sum', see Tab.~\ref{tab_super}.
Similarly, we write for the occupation vector~\cite{Cavaliere09}
\begin{align}
\label{eq_master_resum}
\partial_t \boldsymbol P^\mathrm{(s,HF)}(t) &= \mathcal W^\mathrm{(i)}_t \boldsymbol P^\mathrm{(s,HF)}(t).
\end{align}
Consequently, we find the generalized noise formula,
\begin{align}
 \label{eq_noiseHF_resummed}
 \mathcal S^\mathrm{(s,HF)}(n;\omega) &= \begin{aligned}[t] & \int_0^T \frac{dt}{T} e^{i n \Omega t}\, \boldsymbol F^\mathrm{(i,HF)}(t;\omega) \\ & \cdot \boldsymbol P^\mathrm{(s,HF)}(t), \end{aligned}
\end{align}
where the instantaneous fluctuation vector is the one defined in Eq.~\eqref{eq_flvector} and the occupation vector is derived using Eq.~\eqref{eq_master_resum}. See App.~\ref{app_HFscheme} for details. Besides the restrictions to high noise frequencies and weak tunnel coupling, we expect Eq.~\eqref{eq_noiseHF_resummed} to be valid for \mbox{$\Omega \lesssim \Gamma$} \cite{Cavaliere09,Riwar13}.
Therefore, in comparison to the instantaneous and adiabatic-response noise in Eqs.~\eqref{eq_noise_HF_for_flvector},
the generalized noise formula in Eq.~\eqref{eq_noiseHF_resummed} can be applied to study faster driving schemes.
What is more, since the equations share a similar structure, Eq.~\eqref{eq_noiseHF_resummed} generalizes the results outlined in Secs.~\ref{sec_results_HF_general}-\ref{sec_results_HF_spectra_harmonics}.

\begin{figure}[t]
\begin{center}
\includegraphics[width=\columnwidth]{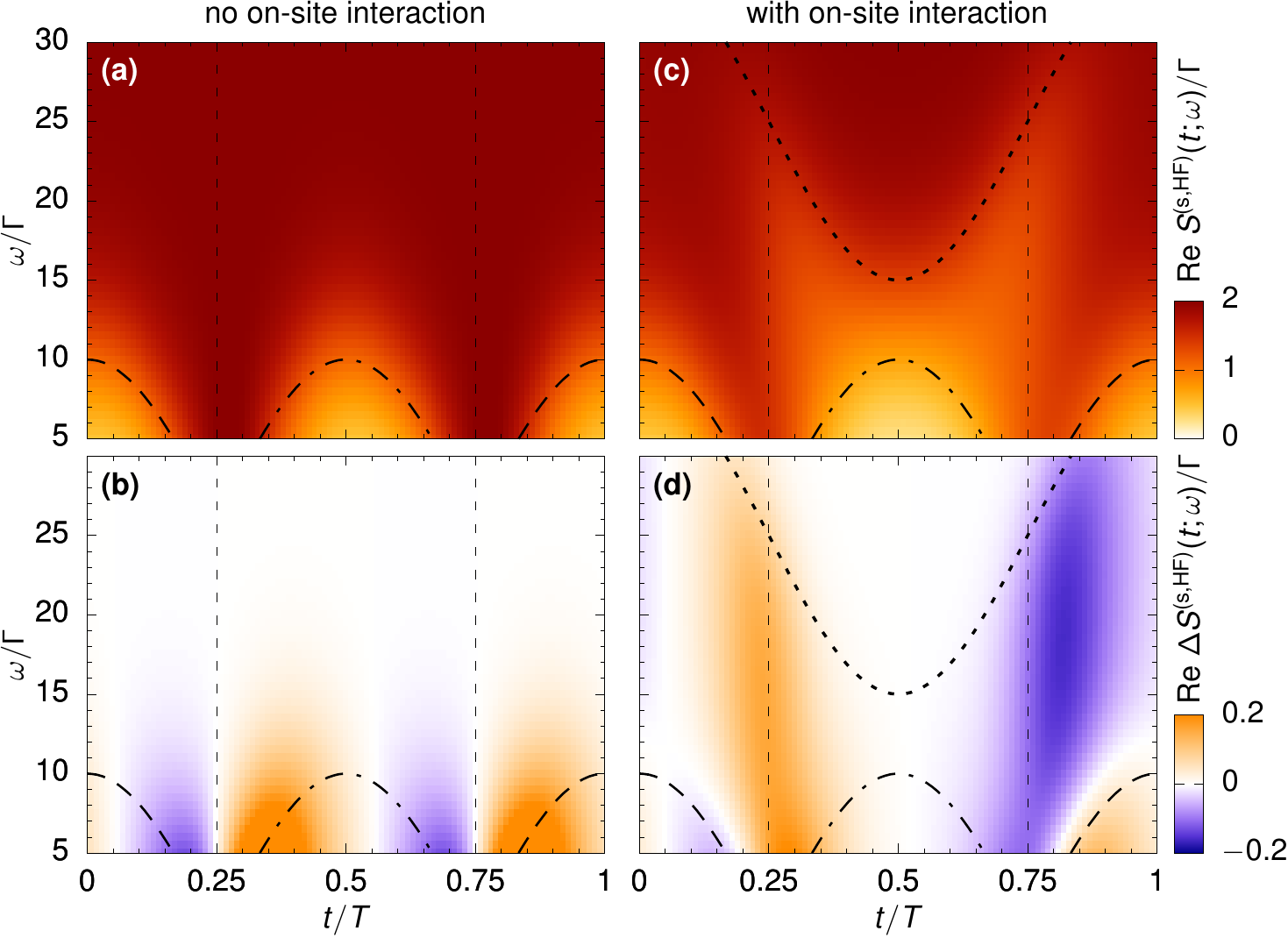}
\end{center}\vspace{-5mm}
	\caption{
	(a) Summed-up time-resolved finite-frequency noise (HF regime) for a harmonic gate voltage and \mbox{$U=0\Gamma$}.
	(b) Same as (a) with instantaneous time-resolved finite-frequency noise subtracted.
	Additional lines in (a)-(b) show $\epsilon(t)$, $-\epsilon(t)$ and $\epsilon(t)+U$ (thick dashed, dashed-dotted and dotted) and zero-crossings of~$\epsilon(t)$ (thin dashed).
	(c)-(d) Similar to (a)-(b) with \mbox{$U=25\Gamma$}.
	Further parameters are \mbox{$\beta=1/(3\Gamma)$}, \mbox{$\epsilon_\uparrow=\epsilon_\downarrow$}, \mbox{$\bar \epsilon = 0\Gamma$}, \mbox{$\delta \epsilon = 10\Gamma$} and \mbox{$\Omega=0.3\Gamma$}.
	}
	\label{fig_noise_HFtfr_resummed}
\end{figure}

In Fig.~\ref{fig_noise_HFtfr_resummed}, we show the time-resolved finite-frequency noise of a noninteracting and an interacting spin-symmetric quantum dot calculated from Eq.~\eqref{eq_noiseHF_resummed}.
The figure shows the summed-up noise in panels (a) and (c) as well as the difference between the latter and the instantaneous noise in panels (b) and (d).
The parameters are similar to the ones used in Fig.~\ref{fig_noise_HFtfr}, except that the driving frequency in Fig.~\ref{fig_noise_HFtfr_resummed} has been increased by an order of magnitude.
As expected, we find that the density plots in both figures show a similar qualitative behavior.
However, what might be of importance for experimental realizations: the differences between the instantaneous and the summed-up noise, shown in Fig.~\ref{fig_noise_HFtfr_resummed}~(b) and~(d), are an order of magnitude larger than the adiabatic-response noise, 
presented in Fig.~\ref{fig_noise_HFtfr}~(c) and~(f), respectively.
We find a similar behavior for zeroth and first noise harmonics (not shown).

\section{Low noise-frequencies}
\label{sec_results_LF}
We now turn to low noise frequencies, \mbox{$\omega \ll \Gamma$}, where the time scale of fluctuations exceeds the time scale of relaxation dynamics of the quantum dot.

\subsection{Noise harmonics -- instantaneous contribution}
\label{sec_results_LF_general}

\begin{figure*}[tb]
\begin{center}
\includegraphics[width=0.8\textwidth]{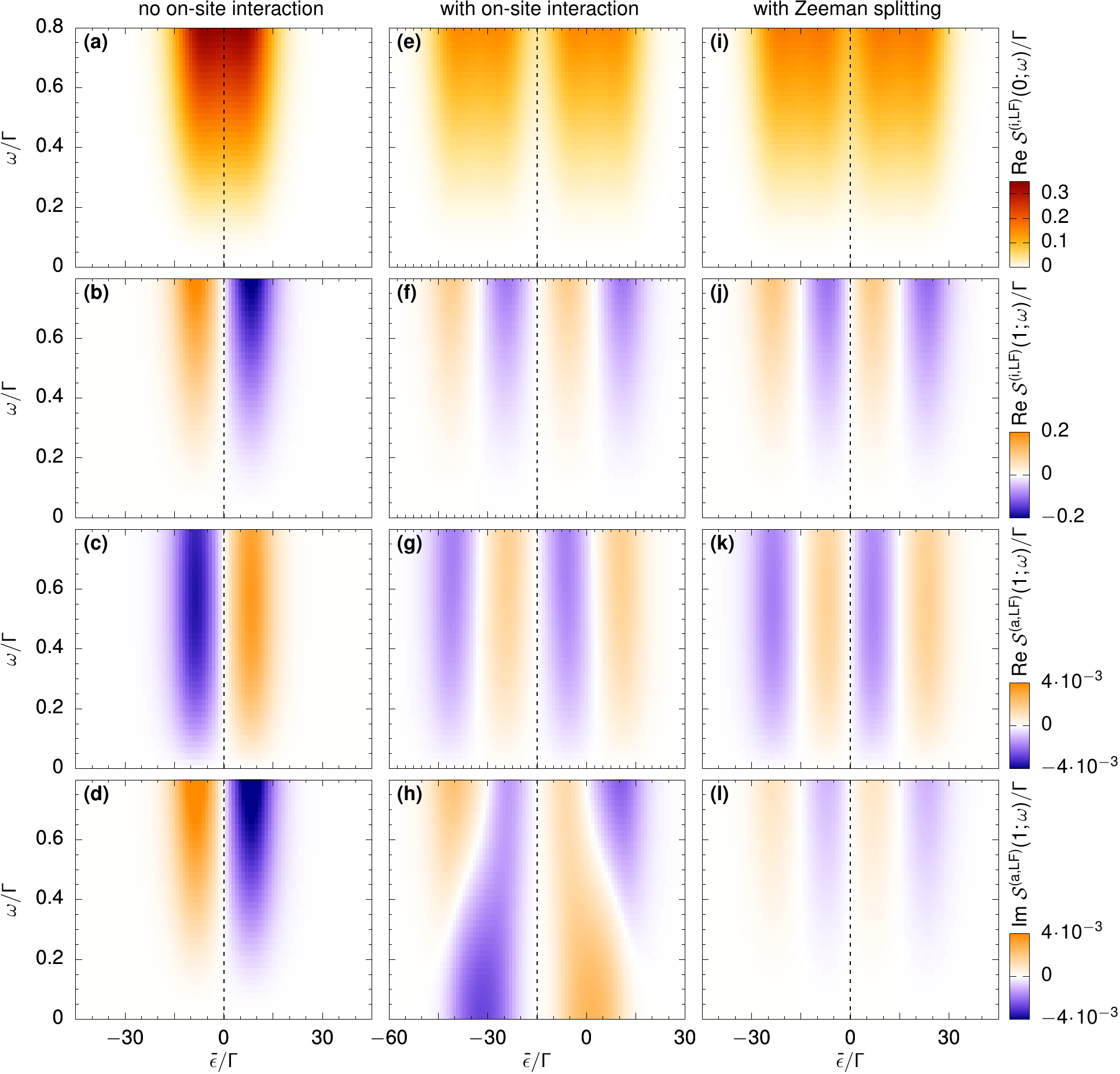}
\end{center}\vspace{-5mm}
	\caption{
	Noise harmonics for a time-dependently driven noninteracting quantum dot for low noise frequencies (LF). Dotted lines indicate the electron-hole symmetric point.
        First row (a), (e), (i): zeroth harmonic of the instantaneous part. Second row (b), (f), (j):  first harmonic of the instantaneous part.
	Third row: (c), (g), (k): real part of the first harmonic of the adiabatic response. Fourth row: (d), (h), (l): imaginary part of the first harmonic of the adiabatic response.
	Parameters are \mbox{$\beta=1/(3\Gamma)$}, \mbox{$\delta \epsilon = 10\Gamma$} and \mbox{$\Omega=0.02\Gamma$}. 
	Furthermore, different values of the interaction strength and the Zeeman splitting are chosen for the different columns:
	first column (a)-(d): \mbox{$U = 0\Gamma$} with \mbox{$\bar \epsilon_\uparrow - \bar \epsilon_\downarrow = 0$};
	second column (e)-(h): \mbox{$U = 30\Gamma$} with \mbox{$\bar \epsilon_\uparrow - \bar \epsilon_\downarrow = 0$}; 
	third column (i)-(l): \mbox{$U = 0\Gamma$} with \mbox{$\bar \epsilon_\uparrow - \bar \epsilon_\downarrow = 30\Gamma$}.
	}
	\label{fig_noise_LFharmonics}
\end{figure*}

We start by analyzing the instantaneous contribution to the low-frequency noise, Eqs.~\eqref{eq_noise_adexp_i} and \eqref{eq_auxiliarynoise_i}.

By employing the crossover scheme of the $\Gamma$ expansion, while additionally neglecting the frequency dependence of kernels, see Secs.~\ref{sec_adexp} and \ref{sec_gammaschemes}, 
we can derive a compact analytical expression for the instantaneous low-frequency noise:
\begin{align}
 \label{eq_noise_LF_for_flvector_i}
 \mathcal S^\mathrm{(i,LF)}(n;\omega) &= \int_0^T \frac{dt}{T} e^{i n\Omega t}\, \boldsymbol F^\mathrm{(i,LF)}(t;\omega) \cdot \boldsymbol{P}^\mathrm{(i)}(t).
\end{align}
The fluctuation vector occurring here is given by the expression
\begin{equation}\label{eq_fluct_low}
\boldsymbol F^\mathrm{(i,LF)}(t;\omega)=\frac{\omega^2}{\lambda_\mathrm{c}(t)^2+\omega^2}\, \boldsymbol F^\mathrm{(i,HF)}(t;0).
\end{equation}
It differs from the one at high noise frequencies, Eq.~\eqref{eq_flvector}, in two ways. 
First, due to the different time scales of the fluctuations considered here, it contributes only at \mbox{$\omega=0$}, and second, it features a factor \mbox{$\omega^2/(\lambda_\mathrm{c}(t)^2+\omega^2)$}. 
This frequency-dependent Lorentzian factor suppresses the noise when the time scale associated with a fluctuation, $\omega^{-1}$, exceeds the time scale $\lambda_\mathrm{c}^{-1}(t)$, 
which equals the physical charge relaxation time for a system with parameters frozen at time $t$ \cite{Cavaliere09,Splettstoesser10},
\begin{align}
 \label{eq_charge_relax}
 \lambda_\mathrm{c}(t) &= \Gamma\left[1+f\big(\epsilon(t)\big)-f\big(\epsilon(t)+U\big)\right].
\end{align}
In the first two rows of panels in Fig.~\ref{fig_noise_LFharmonics}, we show the zeroth and first noise harmonic of this instantaneous contribution to the low-frequency noise for the three different scenarios,
which for high noise frequencies were displayed in Fig.~\ref{fig_noise_HFharmonics} of Sec.~\ref{sec_results_HF_spectra_harmonics}: a non-interacting spin-symmetric dot, an interacting spin-symmetric dot and a non-interacting spin-split dot.
Two important remarks about these figures are at place. 
First, we clearly see the features prescribed by the fluctuation-dissipation theorem, Eqs.~\eqref{eq_fdt_i0} and \eqref{eq_fdt_i1}, for the zeroth and first harmonic of the instantaneous noise:
these noise contributions are directly related to the finite-frequency conductance and its derivative. 
Furthermore, in agreement with these fluctuation-dissipation theorems, both noise harmonics vanish in the limit of \mbox{$\omega\rightarrow 0$}, as dictated by the vanishing zero-frequency conductance of the single-lead quantum dot. 

Importantly, from the noise features displayed in these two rows, no clear distinction is possible between the case of finite interaction and no magnetic field, and vanishing interaction and finite magnetic field. 
This is different when studying the adiabatic-response contribution.

\subsection{Noise harmonics -- adiabatic-response}
\label{sec_results_LF_spectra}

For the adiabatic-response noise at low frequencies, no compact analytical expression for $\mathcal S^\mathrm{(a,LF)}(n;\omega)$ is accessible\footnote{
The reason is that the time-dependent driving causes a retarded response not only of the occupation vector, but of all objects which appear in the auxiliary function in Eq.~\eqref{eq_auxiliarynoise_blocks}.
While these additional terms hardly contribute at high noise frequencies---they can be neglected in the order-by-order scheme---these terms strongly influence the low frequency noise.
The final expression for the adiabatic response related to Eq.~\eqref{eq_noise_LF_for_flvector_i} is too large to be shown here, and we therefore analyze it numerically this section.}
 and it is more insightful to analyze the plots of this function given in the third and fourth rows of Fig.~\ref{fig_noise_LFharmonics}. 
 Since the adiabatic-response of the first harmonic, $\mathcal S^\mathrm{(a,LF)}(1;\omega)$, is a complex quantity, we show its real and imaginary part separately.

For the first harmonic of the noise, the real part of the adiabatic response behaves very similar to the instantaneous part, even though the order of magnitude is much smaller. 
In both cases, the noise vanishes with decreasing frequencies and the overall behavior with alternating signs is equivalent, however, with an opposite overall sign.
In contrast, the imaginary part of the adiabatic response, displayed in the fourth row of panels in Fig.~\ref{fig_noise_LFharmonics}, shows a very different behavior.

For the noninteracting dot, both in the presence and in the absence of a Zeeman field, the difference between real and imaginary part are merely opposite signs and a stronger suppression for low frequencies in the imaginary part. 
However, a key finding of this paper is that the \emph{interacting} quantum dot behaves completely differently: the imaginary part of the adiabatic response of the  first harmonic stays finite even at zero noise frequency, as it is
evident from Fig.~\ref{fig_noise_LFharmonics}~(h).
This effect is unique to strong Coulomb interaction and cannot be mimicked by a Zeeman splitting with equal magnitude. 
Importantly, it is only visible when combined with the time-dependent driving: it is not visible in the time-averaged noise, i.\,e., the zeroth harmonic, where the adiabatic response vanishes, see Sec.~\ref{sec_generalprop_vanishing_a}; 
the described feature is also absent in the zeroth as well as the first harmonic of the instantaneous noise in Fig.~\ref{fig_noise_LFharmonics}~(e)-(f). 
We attribute the signature of non-vanishing noise in the adiabatic response to its sensitivity to the modified charge-relaxation rate, Eq.~\eqref{eq_charge_relax}, 
due to interaction and the resulting difference in degeneracy of the quantum-dot ground states. 
This is also at the origin of deviations from the equilibrium fluctuation-dissipation theorem in the adiabatic response of interacting quantum-dot pumps, see Ref.~\cite{Riwar13}. 

The contribution to $\mathcal S^\mathrm{(a,LF)}(1;\omega)$ for the interacting dot evolves from two features with a sign change to two resonant contributions with a single maximum (or minimum) with decreasing noise frequencies. 
These features are situated around working points in the vicinity of \mbox{$\bar \epsilon=\log 2/\beta$} or \mbox{$\bar \epsilon=-U-\log 2/\beta$}, i.\,e.,
when the dot is driven around energies at which electrons are emitted and absorbed. 
The sign of the contribution, when approaching zero noise frequency, reveals if the quantum dot is driven between the empty configuration and the singly occupied state or between the singly and the doubly occupied states, 
see Fig.~\ref{fig_noise_LFharmonics}~(h), namely whether the ground-state degeneracy increases or decreases with the working-point position, see also Ref.~\cite{Reckermann10}.

We note that for the particular case of vanishing on-site interaction, our noise expressions derived for low frequencies agree with results from scattering-matrix theory.
In order to perform this comparison, we extended calculations valid for low temperatures \cite{Moskalets09,Moskalets13} to the temperature scale relevant for this work.

\section{Noise at intermediate frequencies}
\label{sec_results_crossover}

We finally present results for arbitrary noise frequencies and use them to show how the transition from high to low noise frequencies takes place. 
This is particularly relevant in the range where the time scale of fluctuations is similar to the time scale on which the dot occupation probabilities change, \mbox{$\omega \sim \Gamma$}. 
Our main finding is that noise spectra and noise harmonics derived in this range connect well our high and low noise-frequency results, which have been discussed in Secs.~\ref{sec_results_HF}-\ref{sec_results_LF}.
In this section, we focus on a spin-symmetric quantum dot.

\subsection{Noise harmonics -- instantaneous contribution}
\label{sec_results_crossover_general}

In order to calculate the noise at intermediate noise frequencies, we employ the technically more challenging full crossover scheme of the expansion in the tunnel coupling, as outlined in Sec.~\ref{sec_gammaschemes}.
For the instantaneous noise this leads to the expression
\begin{align}
 \label{eq_noise_crossover_for_flvector_i}
 \mathcal S^\mathrm{(i)}(n;\omega) &= \int_0^T \frac{dt}{T} e^{i n\Omega t}\, \boldsymbol F^\mathrm{(i)}(t;\omega) \cdot \boldsymbol{P}^\mathrm{(i)}(t).
\end{align}
In this equation, \mbox{$\boldsymbol F^\mathrm{(i)}(t;\omega)=\frac{\omega^2}{\lambda_\mathrm{c}(t;\omega)^2+\omega^2}\, \boldsymbol F^\mathrm{(i,HF)}(t;\omega)$} and
\begin{align}
 \label{eq_charge_relax_w}
 \lambda_\mathrm{c}(t;\omega) &= \Gamma\left(1+\frac{f\big(\epsilon(t);\omega\big)-f\big(\epsilon(t)+U;\omega\big)}{2}\right),
\end{align}
with $f(x;\omega)$ defined below Eq.~\eqref{eq_flvector_example}.
Equation~\eqref{eq_noise_crossover_for_flvector_i} combines and extends the instantaneous noise expressions which we derived previously, i.\,e., 
in the high and low noise-frequency regimes, see Eqs.~\eqref{eq_noise_HF_for_flvector_i} and~\eqref{eq_noise_LF_for_flvector_i}.
Again, we find that the instantaneous noise can be expressed in terms of the instantaneous quantum-dot occupation, $\boldsymbol{P}^\mathrm{(i)}(t)$, 
and an instantaneous fluctuation vector, $\boldsymbol F^\mathrm{(i)}(t;\omega)$, as written in Eq.~\eqref{eq_noise_crossover_for_flvector_i}.
The instantaneous fluctuation vector in Eq.~\eqref{eq_noise_crossover_for_flvector_i} differs from its high-frequency limit, $\boldsymbol F^\mathrm{(i,HF)}(t;\omega)$, 
by a frequency-dependent factor, which suppresses the noise for frequencies \mbox{$\omega < \lambda_\mathrm{c}(t;\omega)$}.
The difference compared to the result at low noise frequencies [Eq.~\eqref{eq_charge_relax}] is that the quantity $\lambda_\mathrm{c}(t;\omega)$ in Eq.~\eqref{eq_charge_relax_w} is now frequency dependent itself.
The quantity $\lambda_\mathrm{c}(t;\omega)$ equals the physical charge relaxation rate of the quantum dot \cite{Cavaliere09,Splettstoesser10} only in the limit \mbox{$\omega \rightarrow 0$}.
For general noise frequencies, $\lambda_\mathrm{c}(t;\omega)$ can be expressed as the average of two charge relaxation rates, namely the rates associated with quantum-dot levels frozen at \mbox{$\epsilon(t)\pm\omega$}.
Importantly, for a noninteracting quantum dot, the quantity $\lambda_\mathrm{c}(t;\omega)$ remains frequency independent and the suppression factor in Eq.~\eqref{eq_noise_crossover_for_flvector_i} becomes a Lorentzian, 
as for the low-frequency noise discussed before. 
For the noninteracting dot the transition between the high- and low-frequency noise is therefore expected to be trivial and we focus on the interacting dot, when displaying the results in Fig.~\ref{fig_noise_CRharmonics}. 
The result for the instantaneous contribution to the noise in Fig.~\ref{fig_noise_CRharmonics}~(a) shows the suppression of the noise with decreasing frequencies.

\subsection{Noise harmonics -- adiabatic response}
\label{sec_results_crossover_harmonics}

\begin{figure}[bt]
\begin{center}
\includegraphics[width=\columnwidth]{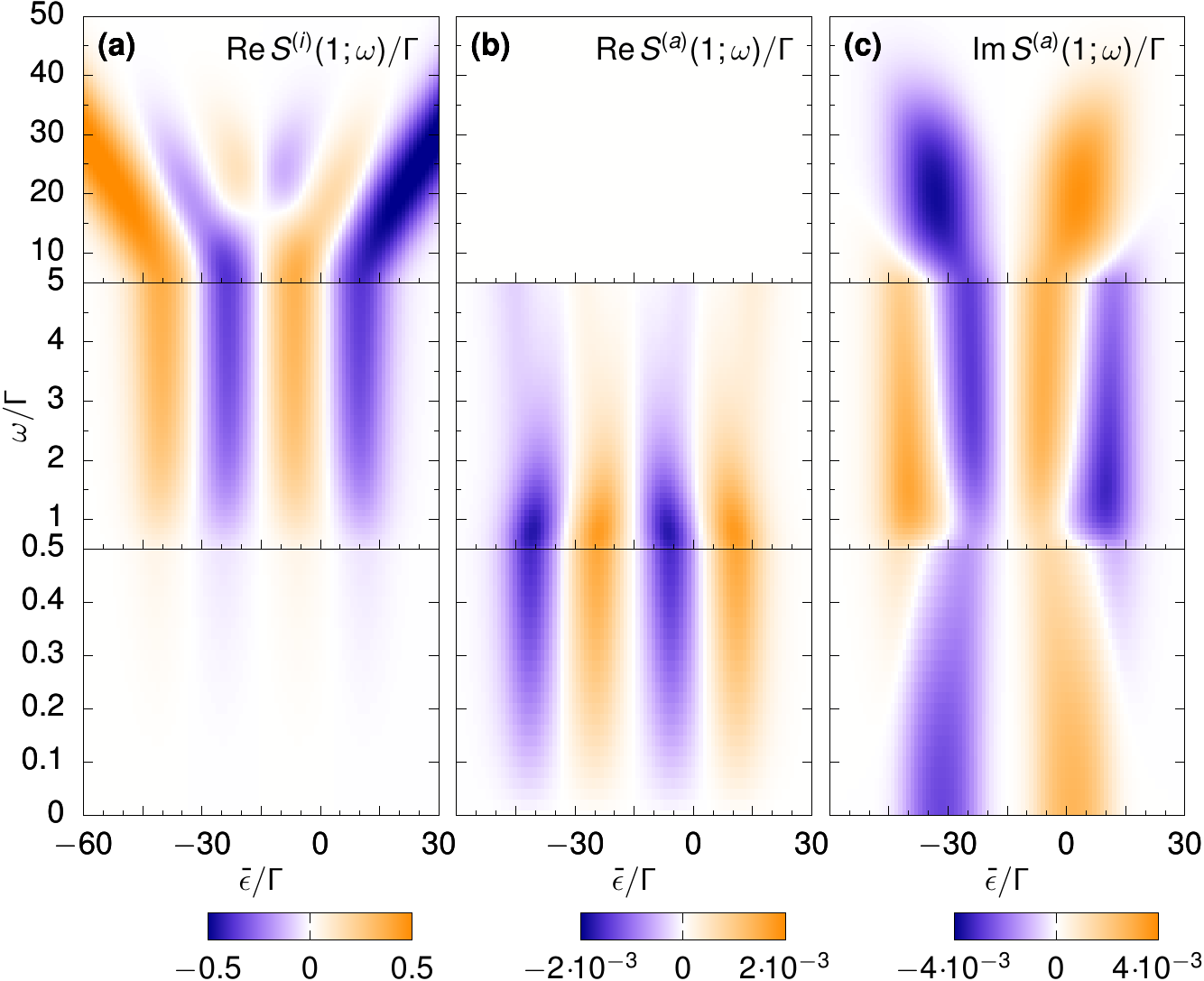}
\end{center}\vspace{-5mm}
	\caption{
	Full frequency dependence of the finite-frequency noise for a strongly interacting quantum dot. 
	We show results for the first noise harmonic displaying (a) the instantaneous contribution and (b)-(c) the real and imaginary part of the adiabatic response. 
	The high- and low-frequency results repeat the ones discussed in Secs.~\ref{sec_results_HF} and \ref{sec_results_LF}, while the middle row is obtained by employing the full crossover scheme required for the regime of intermediate frequencies.
	Parameters are \mbox{$U = 30\Gamma$}, \mbox{$\beta=1/(3\Gamma)$}, \mbox{$\epsilon_\uparrow=\epsilon_\downarrow$}, \mbox{$\delta \epsilon = 10\Gamma$} and \mbox{$\Omega=0.02\Gamma$}.
	}
	\label{fig_noise_CRharmonics}
\end{figure}

The adiabatic response turns out to be more sensitive to the frequency dependence of the noise.
To also investigate the adiabatic response of the noise, we again find that it is more insightful to analyze numerical results. 

The real and imaginary part of the adiabatic response of the first harmonic are shown in Figs.~\ref{fig_noise_CRharmonics}~(b) and (c). 
We again find that the results present a smooth connection between the high- and low-frequency-noise results.
A minor deviation is visible in column~(b), where the crossover scheme gives a small contribution even at higher frequencies, which is not captured by the order-by-order scheme employed in the upper plot of this column.
Note, however, that the non-vanishing contribution visible in the upper part of the middle plot in column~(b) turns out to be further suppressed, if we choose a higher temperature than the one applied here [\mbox{$\beta = 1/(3\Gamma)$}].

The center panels of Figs.~\ref{fig_noise_CRharmonics}~(b) and (c) show a shifting of features (such as maxima and sign changes) as a function of the working-point position depending on the noise frequencies. 
The reason for this is the delicate interplay between time scales given by the time-dependent driving, the time scale of fluctuations, and the charge relaxation time, 
where the latter is only working-point dependent in the case of Coulomb interaction, see Eq.~\eqref{eq_charge_relax}.

\section{Conclusion}
\label{sec_conclusion}

We investigated the finite-frequency current noise of an interacting quantum dot, coupled to a single contact, when the system is subject to a slow harmonic gate-voltage driving.
By extending a perturbative real-time diagrammatic technique, we set up a framework to access instantaneous as well as adiabatic-response contributions to the noise over a large range of noise frequencies.
We then used this approach to analyze time-resolved finite-frequency noise spectra and, importantly, also their decomposition into individual noise harmonics.

In the case of high noise frequencies, where the time scale of fluctuations is much smaller than the time scale of the quantum-dot relaxation dynamics, 
we found simple noise expressions, which allow us to identify dominating fluctuation processes in the instantaneous first noise harmonic.
In the opposite limit of low noise frequencies, a key result is that the combination of strong Coulomb interaction and periodic driving leads to a non-vanishing imaginary part in the adiabatic response of the first noise harmonic.
We emphasize that this contribution provides unambiguous evidence of Coulomb interaction in the low-frequency noise of the driven quantum dot:
for a noninteracting and possibly spin-split quantum dot, both the instantaneous and the adiabatic-response contributions to the first noise harmonic vanish, when the noise-frequency approaches zero.

Our results thus promote the study of noise harmonics not only as a spectroscopic tool to access contributions of individual fluctuation processes, 
but also to identify Coulomb interaction in the noise of the time-dependently driven quantum dot---both of which we expect to be of use in future related experiments.

\section*{Acknowledgements} 
We thank D.~C.~Glattli, G.~Johansson, F.~D.~Parmentier, F.~Portier, R.-P.~Riwar and M.~R.~Wegewijs for stimulating discussions and 
we acknowledge funding from the DFG via RTG1995. J.~S. also acknowledges funding from the Knut and Alice Wallenberg foundation and the Swedish VR.

\appendix

\section{Additional plots for the time-resolved finite-frequency noise of the interacting quantum dot}

\label{app_ABCplots}
In this appendix we provide additional plots which add to the discussion of the time-resolved finite-frequency noise in Sec.~\ref{sec_results_HF_spectra}, in particular of the adiabatic responses shown in Fig.~\ref{fig_noise_HFtfr}~(c) and~(f).
In Fig.~\ref{figapp_noise_HFtfr} we present how the adiabatic-response noise of the time-dependently driven quantum dot changes, when the interaction strength is modified from \mbox{$U=0\Gamma$}~(a) to \mbox{$U=25\Gamma$}~(e).
The Coulomb interaction leads to newly emerging regions `C'---related to single occupation of the dot---in between the regions `A' and `B'. 
If the strong interaction together with the applied driving scheme permits the occupation of the dot with two electrons, the regions `B' as well as one part of the regions `C' disappear, see panel~(e).
\begin{widetext}

\begin{figure*}[h]
\begin{center}
\includegraphics[width=\columnwidth]{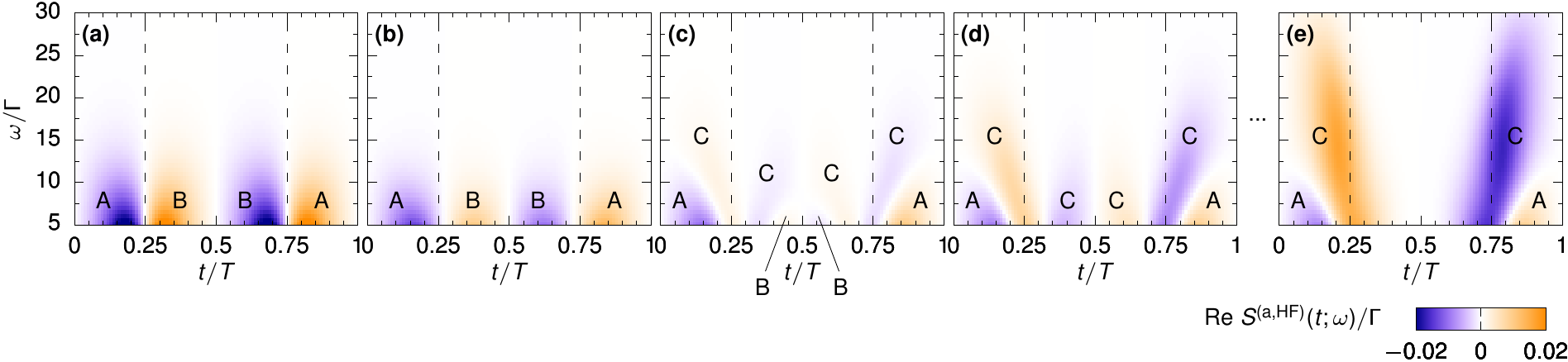}
\end{center}\vspace{-5mm}
	\caption{
	(a)-(e) Adiabatic response of the time-resolved finite-frequency noise (HF regime) for a harmonic gate voltage.
	Thin dashed lines show zero-crossings of~$\epsilon(t)$.
	The interaction strength changes from (a) to (e) as \mbox{$U/\Gamma = 0,3,6,9,25$}.
	Further parameters are \mbox{$\beta=1/(3\Gamma)$}, \mbox{$\epsilon_\uparrow=\epsilon_\downarrow$}, \mbox{$\bar \epsilon = 0\Gamma$}, \mbox{$\delta \epsilon = 10\Gamma$} and \mbox{$\Omega=0.03\Gamma$}.
	}
	\label{figapp_noise_HFtfr}
\end{figure*}

\section{Auxiliary function for diagrammatic noise calculations}
\label{app_method_details}
In this and the following appendices we provide technical details of our noise calculations.

We begin by deriving Eq.~\eqref{eq_noise2} of the main text by rewriting Eq.~\eqref{eq_noise} in terms of the auxiliary function $\tilde{\mathcal{S}}(t;\omega)$ given in Eq.~(\ref{eq_auxiliarynoise}).
First, in Eq.~\eqref{eq_noise}, we split the second integration into two parts and obtain
\begin{align}
\mathcal S(n;\omega) &= \lim_{t_0 \rightarrow -\infty}\Bigg[
\int_0^T \!\frac{dt}{T} \int_{t_0}^0 \!d\tau e^{in\Omega t + i \omega \tau}\mathcal{C}(t,\tau)
+ \int_0^T \!\frac{dt}{T} \int_{t_0}^0 \!d\tau e^{in\Omega t - i \omega \tau}\mathcal{C}(t,-\tau)\Bigg],
\end{align}
with \mbox{$\mathcal{C}(t,\tau) = \expval{\left\{ \delta I(t), \delta I(t+\tau)\right\}}$}.
To treat the second term in the square brackets, we swap its two integrations and---exploiting periodicity---shift the interval of the integration over~$t$ by the amount~$-t'$. 
Incorporating the latter shift into a shift of the variable $t$, swapping the integration order a second time and exploiting the symmetrized form of \mbox{$\mathcal{C}(t,\tau)=\mathcal{C}(t+\tau,-\tau)$}, we derive
\begin{align}
\label{eq_app_noise_intermediatestep}
\mathcal S(n;\omega) &=  \lim_{t_0 \rightarrow -\infty}\Bigg[
\int_0^T \!\frac{dt}{T} \int_{t_0}^0 \!d\tau e^{in\Omega t + i \omega \tau}\mathcal{C}(t,\tau)
+ \int_0^T \!\frac{dt}{T} \int_{t_0}^0 \!d\tau e^{in\Omega \left(t+\tau\right) - i \omega \tau}\mathcal{C}(t,\tau)\Bigg].
\end{align}
Since at finite temperatures considered here, temporal correlations of current fluctuations, given by $\mathcal{C}(t,\tau)$, decay quickly for large values of $\tau$, 
we expect the limit in Eq.~\eqref{eq_app_noise_intermediatestep} to converge separately for both terms in the square brackets.
Therefore, we can replace $t_0$ with $-\infty$ at the integration bounds.
The final step to obtain Eq.~\eqref{eq_noise2} of the main text is then to write the resulting equation in terms of the auxiliary function, which has been defined in Eq.~\eqref{eq_auxiliarynoise}, 
where the time difference $\tau$ has been replaced by \mbox{$t'-t$}.

\section{Expansion of the auxiliary function \texorpdfstring{$\tilde S(t;\omega)$}{S-tilde} for slow driving}
\label{app_adexp_auxfunc}
In Sec.~\ref{sec_model_adexp} we explained that, for slow periodic driving, it is justified to expand the noise expression in Eq.~\eqref{eq_noise2} order-by-order in the small parameter $\delta\epsilon\,\Omega\beta/\Gamma$.
To evaluate the zeroth (first) order of the resulting series [Eqs.~\eqref{eq_noise_adexp}]---referred to as the instantaneous and the adiabatic-response contribution to the noise---we 
first need to derive the respective terms in the slow-driving expansion of the auxiliary function defined in Eq.~\eqref{eq_auxiliarynoise}.
For this expansion, we start from the expression introduced in Sec.~\ref{sec_adexp} and split the frequency-dependent propagator on 
the rhs.~of Eq.~\eqref{eq_auxiliarynoise_blocks} into a decaying and a non-decaying part, as defined in Eq.~\eqref{eq_Pibardef}.
By inserting the reduced propagator into Eq.~\eqref{eq_auxiliarynoise_blocks}, we obtain
\begin{align}
\label{eq_auxiliarynoise_blockspibar}
\tilde{S}(t;\omega) &= 
  \begin{aligned}[t]
    &\lim_{t_0 \rightarrow -\infty} \frac{\boldsymbol e^T}{2} \Bigg[ \int_{t_0}^{t} dt_1 \int_{t_0}^{t_1} dt_2 \int_{t_0}^{t_2} dt_3 
    \mathcal W_I^{<}(t,t_1;\omega)\Pibar(t_1,t_2;\omega) \mathcal W_I^{>}(t_2,t_3;\omega)\boldsymbol P(t_3) \\ &
   + \int_{t_0}^{t} dt_1 \mathcal W_{II}(t,t_1;\omega)\boldsymbol P(t_1) 
   - \int_{t_0}^t dt_1 \int_{t_0}^{t_1} dt_2 \int_{t_0}^{t_1} dt_3 e^{i \omega (t_1-t)} \mathcal W_I(t,t_2) \boldsymbol P(t_2)\otimes\boldsymbol e^T \mathcal W_I(t_1,t_3) \boldsymbol P(t_3) \Bigg].
\end{aligned}
\end{align}
In Sec.~\ref{sec_adexp} we introduced how the integrand of the kinetic equation~\eqref{eq_master} is expanded around the reference time~$t$; 
here we proceed in a similar way, expanding all occupation vectors in the integrands in Eq.~\eqref{eq_auxiliarynoise_blockspibar} around~$t$. 
In addition, also the first time arguments of all kernels and of the reduced propagators are expanded around the reference time. 
These expansions are justified for slow driving, due to the short support times of all kernels, given by the reservoir correlation time~$\beta$, 
and the short support time of the reduced propagator, given by~$\Gamma^{-1}$. 
In addition, again following the same principle as introduced for the kinetic equation in Sec.~\ref{sec_adexp}, all objects $\boldsymbol{P}, \mathcal{W}$ and $\Pibar$ have to be expanded individually order by order in $\delta\epsilon\,\Omega\beta/\Gamma$.
Collecting all terms in zeroth (first) order, we find the instantaneous contribution (adiabatic response) of the auxiliary function as given in Eqs.~\eqref{eq_auxiliarynoise_ia} in the main text,
\begin{subequations}
\label{eq_auxiliarynoise_laplace_ia}
\begin{align}
\label{eq_auxiliarynoise_laplace_i}
\tilde{S}^\mathrm{(i)}(t;\omega) &= \begin{aligned}[t] &\frac{\boldsymbol e^T}{2} \Big\{\mathcal W_I^{<}\Pibar \mathcal W_I^{>} \boldsymbol P \Big\}_{t;\omega}^\mathrm{(i)} 
  +\frac{\boldsymbol e^T}{2} \Big\{ \mathcal W_{II} \boldsymbol P \Big\}_{t;\omega}^\mathrm{(i)} - 2 \Big\{ \tilde{I} I \Big\}_{t;\omega}^\mathrm{(i)}, \end{aligned}
  \\
\label{eq_auxiliarynoise_laplace_a}
\tilde{S}^\mathrm{(a)}(t;\omega) &= \begin{aligned}[t] &\frac{\boldsymbol e^T}{2} \Big\{\mathcal W_I^{<}\Pibar \mathcal W_I^{>} \boldsymbol P \Big\}_{t;\omega}^\mathrm{(a)} 
  +\frac{\boldsymbol e^T}{2} \Big\{ \mathcal W_{II} \boldsymbol P \Big\}_{t;\omega}^\mathrm{(a)} - 2 \Big\{ \tilde{I} I \Big\}_{t;\omega}^\mathrm{(a)}, \end{aligned}
\end{align}
\end{subequations}
where we applied several abbreviations, which we now explain. 
We have given the definition of the curly brackets for an operator product in Eqs.~\eqref{eq_brackets}.
Extended to a four-operator product, this explicitly reads as~\cite{Riwar13}
\begin{subequations}
\label{eq_brackets2}
 \begin{align}
  \big\{ A B C D \big\}^\mathrm{(i)} & = A^\mathrm{(i)} B^\mathrm{(i)} C^\mathrm{(i)} D^\mathrm{(i)},\\
  \big\{ A B C D \big\}^\mathrm{(a)} & = 
  \begin{aligned}[t] &
  A^\mathrm{(i)} B^\mathrm{(i)} C^\mathrm{(i)} D^\mathrm{(a)}
  +A^\mathrm{(i)} B^\mathrm{(i)} C^\mathrm{(a)} D^\mathrm{(i)}
  +A^\mathrm{(i)} B^\mathrm{(a)} C^\mathrm{(i)} D^\mathrm{(i)}
  +A^\mathrm{(a)} B^\mathrm{(i)}(t) C^\mathrm{(i)} D^\mathrm{(i)}\\ 
   & 
   +\partial A^\mathrm{(i)} \, \partial_t \!\Big[B^\mathrm{(i)} C^\mathrm{(i)} D^\mathrm{(i)} \Big]
   +A^\mathrm{(i)} \,\partial B^\mathrm{(i)} \,\partial_t \!\Big[ C^\mathrm{(i)} D^\mathrm{(i)} \Big]+A^\mathrm{(i)} B^\mathrm{(i)} \,\partial C^\mathrm{(i)} \dot D^\mathrm{(i)}.
   \end{aligned}
 \end{align}
\end{subequations}
In Eqs.~\eqref{eq_auxiliarynoise_laplace_ia}, the curly brackets carry a subscript for the reference time~$t$, as well as a frequency argument~$\omega$, which have to be associated to all of objects depending on these parameters. 
The new object $\tilde{I}$ is discussed in App.~\ref{app_adexp_tildeI}.

To proceed with the evaluation of Eqs~\eqref{eq_auxiliarynoise_laplace_ia}, we need to derive the instantaneous contribution and the adiabatic response of all objects appearing in the curly brackets.
These derivations are outlined for the occupation vector and the current in the Sec.~\ref{sec_adexp} in the main text, for the function $\tilde{I}$ in App.~\ref{app_adexp_tildeI} and for the reduced propagator in App.~\ref{app_adexp_reducedpropagator}. 
The adiabatic expansion of kernels is discussed in detail in Refs.~\cite{Splettstoesser06,Riwar13}. 
Explicit expressions in lowest order in the tunnel coupling, needed for the evaluation of the noise in the limits studied in this manuscript, are given in App.~\ref{app_adexp_kernels}. 
The final abbreviation in Eqs.~\eqref{eq_auxiliarynoise_laplace_ia} is that we write $I$ for $\expval{I(t)}$.

\section{The function \texorpdfstring{$\tilde I(t,z;\omega)$}{I-tilde}}
\label{app_adexp_tildeI}

In this appendix, we discuss properties of the function $\tilde{I}$, which appears in Eqs.~\eqref{eq_auxiliarynoise_laplace_ia} and \eqref{eq_auxiliarynoise_ia}. 
We start from the third term on the rhs.~of Eq.~\eqref{eq_auxiliarynoise_blockspibar}, which stems from the product of current operators at different times and a contribution containing the non-decaying part of the propagator $\Pi$. 
The first part of this integral is similar to a current at time $t_1$, but not identical to it, and it also contains a frequency-dependent exponential function. 
It is this term that we want to analyze here and that we abbreviate as
\begin{align}
\label{eq_itilde}
  \tilde{I}(t,t_1;\omega) &= \frac{\boldsymbol e^T}{2} \int_{-\infty}^{t_1} dt_2 e^{i\omega(t_1-t)} \mathcal W_I(t,t_2)\boldsymbol P(t_2).
\end{align}
The Laplace transform of Eq.~\eqref{eq_itilde} with respect to \mbox{$(t_1-1)$} is given by
\begin{align}
  \label{eq_adexp_Itilde}
  \tilde{I}(t,z;\omega) &= \int_{-\infty}^t\! dt_1 e^{z(t_1-t)}  \frac{\boldsymbol e^T}{2} \int_{-\infty}^{t_1}\! dt_2  e^{i\omega(t_1-t)} \mathcal W_I(t,t_2)\boldsymbol P(t_2)\\
  &= \int_{-\infty}^{t}\!dt_2 \int_{t_2}^t \! dt_1 e^{(z+i\omega)(t_1-t)}  \frac{\boldsymbol e^T}{2}\mathcal W_I(t,t_2) \boldsymbol P(t_2)\\
  &= \int_{-\infty}^{t}\!dt_2 \frac{\boldsymbol e^T}{2}\mathcal W_I(t,t_2) \boldsymbol P(t_2) \frac{1-e^{(z+i\omega)(t_2-t)}}{z+i\omega}\nonumber \\
  &= \frac{1}{z+i\omega} \left[\expval{I(t)} - I(t,z+i\omega)\right]\nonumber.
\end{align}
In the last line, we inserted Eq.~\eqref{eq_current} for the expectation value of the current, as well as the additional definition, \mbox{$I(t,z) = \frac{\boldsymbol e^T}{2}\int_{-\infty}^{t} dt_2 e^{z(t_2-t)} \mathcal W_I(t,t_2) \boldsymbol P(t_2)$}.
We now outline the steps to derive the instantaneous part and the adiabatic response of the function $\tilde I(t,z;\omega)$.
Using the form given in  the last line in Eq.~\eqref{eq_adexp_Itilde}, we can expand $\tilde{I}(t,z;\omega)$ in powers of $\delta\epsilon\,\Omega\beta/\Gamma$ following the lines of the current expansion discussed in Sec.~\ref{sec_adexp}.
In the limit \mbox{$z\rightarrow 0^+$}, which is of interest here, we find
\begin{subequations}
  \label{eq_adexp_Itilde_ia}
  \begin{align}
    \tilde{I}^\mathrm{(i)}(t,0^+;\omega) &= \frac{\boldsymbol e^T}{2 i\omega} \left[ \big\{\mathcal W_I \boldsymbol P\big\}_t^\mathrm{(i)} - \big\{\mathcal W_I \boldsymbol P\big\}_{t;\omega}^\mathrm{(i)}\right],\\
    \label{eq_adexp_Itilde_a}
    \tilde{I}^\mathrm{(a)}(t,0^+;\omega) &= \frac{\boldsymbol e^T}{2 i\omega} \left[ \big\{\mathcal W_I \boldsymbol P\big\}_t^\mathrm{(a)} - \big\{\mathcal W_I \boldsymbol P\big\}_{t;\omega}^\mathrm{(a)}\right].
  \end{align}
\end{subequations}
Note that the expression for $\tilde{I}^\mathrm{(a)}(t,0^+;\omega)$ in Eq.~\eqref{eq_adexp_Itilde_a} is only given for completeness and is not needed for the calculations performed here. 
The reason is that in the auxiliary function in Eq.~\eqref{eq_auxiliarynoise_laplace_a} it is always multiplied with $\expval{I(t)}^\mathrm{(i)}$, which vanishes for the single-lead quantum dot considered in this paper.

\section{The reduced propagator \texorpdfstring{$\Pibar(t,z;\omega)$}{Pi-bar}}
\label{app_adexp_reducedpropagator}
Equations~\eqref{eq_auxiliarynoise_laplace_ia} of the auxiliary function also include the instantaneous part and the adiabatic response of the Laplace-transformed reduced propagator, $\Pibar(t,z;\omega)$.
The derivation of these parts is discussed in this appendix.
By combining the definition of the reduced propagator, Eq.~\eqref{eq_Pibardef}, with the Dyson equation of the full propagator, Eq.~\eqref{eq_dysontime}, we find
\begin{align}
 \label{eq_Pibar_dysontime}
 \Pibar(t,t_1;\omega) &=
 \Big[ \boldsymbol1 - \boldsymbol P(t)\otimes \boldsymbol e^T \Big] e^{i \omega (t_1-t)}
    + \int_{t_1}^t \!dt_2 \int_{t_1}^{t_2} \!dt_3 \, e^{i \omega (t_2-t)}  \mathcal W(t_2,t_3;\omega)  
    \Big[\Pibar(t_3,t_1;\omega) 
    + \boldsymbol P(t_3)\otimes \boldsymbol e^T e^{i\omega(t_1-t_3)} \Big].
\end{align}
A key property of $\Pibar(t,t_1;\omega)$ is that this function decays for $|t-t_1| \gg \Gamma^{-1}$, or in other words, when the difference in the time arguments exceeds the relaxation time of the quantum dot.
This property, together with the assumption of slow driving, \mbox{$\delta\epsilon\,\Omega\beta/\Gamma \ll 1$},
justifies the expansion of the $t_2$-dependence of the kernel $\mathcal W(t_2,t_3;\omega)$ in Eq.~\eqref{eq_Pibar_dysontime} around the time~$t$.
Similarly, we expand the $t_3$-dependence of the term in the (second) square brackets in Eq.~\eqref{eq_Pibar_dysontime} around the time~$t$.

At this point, the Laplace transform  $\Pibar(t,z;\omega)$ can be calculated, where we make use of the fact that the Laplace transform of a convolution of three functions, $A(t,t_1),B(t,t_1)$ and $C(t,t_1)$, i.\,e.,
\mbox{$\int_{-\infty}^{t}\!dt_1 e^{z (t_1-t)} \int_{t_1}^{t} dt_2 \int_{t_2}^{t} dt_3 A(t,t_1)B(t_1,t_2)C(t_2,t_3)$},
can be expressed as \mbox{$e^\partial A(t,z) B(t,z) C(t,z)$}, with the abbreviation \mbox{$e^\partial = \exp(\partial_z^A\partial_t^B + \partial_z^A\partial_t^C+\partial_z^B\partial_t^C)$}, see also Ref.~\cite{Riwar13}.
The derivatives in this abbreviation only act on the quantities indicated by their superscripts.
We obtain the equation
\begin{align}
\label{eq_Pibar_dysonlaplace}
\Pibar(t,z;\omega) &=  \frac{\boldsymbol 1-\boldsymbol P(t)\otimes \boldsymbol e^T}{z+i\omega}
    + e^\partial\, \frac{1}{z+i\omega}\mathcal W(t,z;\omega)  \Big[\Pibar(t,z;\omega)+\frac{\boldsymbol P(t)\otimes \boldsymbol e^T}{z+i\omega}\Big],
\end{align}
where the derivatives included in $e^\partial$ act on the three functions \mbox{$A(t,z;\omega)=(z+i\omega)^{-1}$}, \mbox{$B(t,z;\omega)=\mathcal W(t,z;\omega)$} 
and \mbox{$C(t,z;\omega)=\Pibar(t,z;\omega)+\boldsymbol P(t)\otimes \boldsymbol e^T(z+i\omega)^{-1}$}.
We can write Eq.~\eqref{eq_Pibar_dysonlaplace} in a more compact form by using the property that \mbox{$\Pibar(t,z;\omega)=\Pibar(t,z+i\omega,0)=\Pibar(t,z+i\omega)$} and similarly for $W(t,z;\omega)$.
The reason for this is that both objects only contain diagrams in which the frequency line runs over the whole diagram.
The final expression for $\Pibar(t,z)$ is
\begin{align}
\label{eq_Pibar_dysonlaplace2}
\Pibar(t,z) &= 
    \frac{\boldsymbol 1-\boldsymbol P(t)\otimes \boldsymbol e^T}{z} 
    + e^\partial\, \frac{1}{z}\mathcal W(t,z)  \Big[\Pibar(t,z)+\frac{\boldsymbol P(t)\otimes \boldsymbol e^T}{z}\Big].
\end{align}
The limit \mbox{$\lim_{z\rightarrow 0^+}\Pibar(t,z;\omega)$}, which is of interest for our noise calculations, is obtained from Eq.~\eqref{eq_Pibar_dysonlaplace2} by calculating \mbox{$\lim_{z\rightarrow i\omega}\Pibar(t,z)$}.
Importantly, for the finite-frequency noise, this limit can be taken by replacing $z$ with $i\omega$ in Eq.~\eqref{eq_Pibar_dysonlaplace2}, in contrast to the zero-frequency noise \cite{Riwar13}, where the limit must be taken carefully.

We continue by deriving the instantaneous part and the adiabatic response of the function $\Pibar(t,z)$ given in Eq.~\eqref{eq_Pibar_dysonlaplace2}.
To extract these contributions, we proceed as previously and expand the reduced propagator in the small parameter $\delta\epsilon\,\Omega\beta/\Gamma$.
In Eq.~\eqref{eq_Pibar_dysonlaplace2} we replace \mbox{$\Pibar(t,z) \rightarrow \Pibar^{\mathrm(i)}(t,z) + \Pibar^{\mathrm(a)}(t,z) + \dots$} and similarly for $\mathcal W(t,z)$ as well as for $\boldsymbol P(t)$. 
We then collect all contributions in zeroth and first order in $\delta\epsilon\,\Omega\beta/\Gamma$.
The result for the instantaneous part of the reduced propagator is the algebraic equation
\begin{subequations}
\label{eq_Pibar_dysonlaplace2_ia}
\begin{align}
\label{eq_Pibar_dysonlaplace2_i}
 \big[\boldsymbol 1 - \frac{\mathcal W^{(i)}}{z}\big] \,\Pibar^{(i)} &= \frac{\boldsymbol 1-\boldsymbol P^{(i)}\otimes \boldsymbol e^T}{z} + \frac{\mathcal W^{(i)} \boldsymbol P^{(i)}\otimes \boldsymbol e^T}{z^2},
\end{align}	
For readability, we suppress the arguments $(t,z)$ for kernels and the reduced propagator and $(t)$ for the occupation vector in Eq.~\eqref{eq_Pibar_dysonlaplace2_i} and also in Eq.~\eqref{eq_Pibar_dysonlaplace2_a} below.
The adiabatic response of the reduced propagator is calculated by subsequently solving the equation
\begin{align}
\label{eq_Pibar_dysonlaplace2_a}
 \big[\boldsymbol 1 - \frac{\mathcal W^{(i)}}{z}\big] \, \Pibar^{(a)} &= -\frac{\dot{\mathcal W}^{(i)}\,\Pibar^{(i)}}{z^2}-\frac{\mathcal W^{(i)} \, \dot{\Pibar}^{(i)}}{z^2} 
 +\frac{\mathcal W^{(a)} \, \Pibar^{(i)}}{z}  +\frac{\partial \mathcal W^{(i)} \, \dot{\Pibar}^{(i)}}{z}  
 -\frac{\dot{\mathcal W}^{(i)} \, \boldsymbol P^{(i)}\otimes \boldsymbol e^T}{z^3} \\&
  -\frac{\mathcal W^{(i)} \, \dot{\boldsymbol P}^{(i)}\otimes \boldsymbol e^T }{z^3} +\frac{\partial \mathcal W^{(i)} \, \dot{\boldsymbol P}^{(i)}\otimes \boldsymbol e^T}{z^2} 
 +\frac{\mathcal W^{(i)} \, \boldsymbol P^{(a)}\otimes \boldsymbol e^T}{z^2} 
  +\frac{\mathcal W^{(a)}\, \boldsymbol P^{(i)}\otimes \boldsymbol e^T}{z^2}  - \frac{\boldsymbol P^{(a)}\otimes \boldsymbol e^T}{z}.\nonumber
\end{align}
\end{subequations}
The solutions of Eqs.~\eqref{eq_Pibar_dysonlaplace2_ia} are necessary to calculate the instantaneous part and adiabatic response of the auxiliary function in Eqs.~\eqref{eq_auxiliarynoise_laplace_ia} and \eqref{eq_auxiliarynoise_ia}.

A further insight of Eqs.~\eqref{eq_Pibar_dysonlaplace2_ia} is the conclusion that an order-by-order expansion scheme in the tunnel-coupling strength is not generally applicable to the auxiliary noise function.
The reason for this is that the reduced propagator, which is part of the auxiliary noise function, has no well defined order-by-order expansion, 
because the matrix $ \big[\boldsymbol 1 - \frac{\mathcal W^{(i)}}{z}\big]$ on the lhs.~in Eqs.~\eqref{eq_Pibar_dysonlaplace2_ia} mixes the orders: the first term is of order unity, while the second term scales with  $\Gamma/\omega$.
Therefore, only for high noise frequencies, where $\Gamma/\omega\ll1$, the usual order-by-order scheme can be applied, as shown in App.~\ref{app_HFscheme}.
Otherwise, we use a crossover scheme, where all terms in Eqs.~\eqref{eq_Pibar_dysonlaplace2_ia} are kept with kernels derived in leading order in the tunnel-coupling strength.

\section{Explicit kernel expressions from diagrammatic rules}
\label{app_adexp_kernels}
In this work, all $\mathcal W$-kernels are evaluated in the sequential-tunneling regime, namely up to linear order in~$\Gamma$. 
Each kernel is therefore given by a sum over all possible diagrams containing a single tunneling line.
This line connects either two tunnel vertices [$\mathcal W(t,t')$ and $\mathcal W(t,t';\omega)$], a tunnel and a current vertex [$\mathcal W_\mathrm{I}(t,t'), \mathcal W_\mathrm{I}^<(t,t';\omega)$ and $\mathcal W_\mathrm{I}^>(t,t';\omega)$]
or two current vertices [$\mathcal W_\mathrm{II}(t,t';\omega)$].
At zero frequency, \mbox{$\omega=0$}, the diagrammatic rules to calculate instantaneous contributions as well as adiabatic responses of these kernels are outlined in detail in the appendix of Ref.~\cite{Riwar13}.
To derive the diagrams relevant here, we have to take into account that some kernels become frequency dependent due to the exponential factor $e^{i \omega (t'-t)}$ in the auxiliary function defined in Eq.~\eqref{eq_auxiliarynoise}.
As explained in Sec.~\ref{sec_auxnoise}, we include this frequency dependence by adding a line in the respective diagrams, carrying the frequency~$\omega$.
This additional frequency line only leads to a small modification of the diagrammatic rules of Ref.~\cite{Riwar13}, which we now outline.
In Laplace space, a linear-in-$\Gamma$ diagram in instantaneous order is proportional to $1/\Delta E(t)$, where $\Delta E(t)$ is given by the difference of all backward-going minus all forward-going energies.
If we evaluate a diagram which contains an additional frequency line, the only modification is that we have to include this frequency as a forward-going energy in $\Delta E(t)$, hence the frequency line in the diagrammatic picture.
Analogous rules have to be applied for the adiabatic-response diagrams. However, for our system, we can show that these expressions can always be simplified 
to \mbox{$\mathcal W^\mathrm{(a)}(t;z,\omega) = \frac{1}{2} \partial_z \, \dot{\mathcal W}^\mathrm{(i)}(t;z,\omega)$}.

For completeness, we give the instantaneous contribution to all kernels in Laplace representation. 
The kernels are shown for the spin-degenerate case, \mbox{$\epsilon=\epsilon_\uparrow=\epsilon_\downarrow$}, and with finite interaction parameter~$U$, 
since this case is the main focus of this paper. Extensions to spin-split single-particle energies $\epsilon_\uparrow\neq\epsilon_\downarrow$ at vanishing interaction, as discussed in Sec.~\ref{sec_results_HF_spectra} are straightforward.
The instantaneous kernels in linear order in~$\Gamma$ read
\begin{align}
\frac{\mathcal W^\mathrm{(i)}(t,z;\omega)}{\Gamma} &=      \left(\begin{smallmatrix}
                                                                  -F^+(\epsilon,z;\omega) & \frac12 F^-(\epsilon,z;\omega) & \frac12 F^-(\epsilon,z;\omega) & 0\\
                                                                  \frac12 F^+(\epsilon,z;\omega) & -\frac12 F^-(\epsilon,z;\omega) - \frac12 F^+(\epsilon+U,z;\omega) & 0 & \frac12 F^-(\epsilon+U,z;\omega)\\
                                                                  \frac12 F^+(\epsilon,z;\omega) & 0 & -\frac12 F^-(\epsilon,z;\omega) - \frac12 F^+(\epsilon+U,z;\omega) & \frac12 F^-(\epsilon+U,z;\omega)\\
                                                                  0 & \frac12 F^+(\epsilon+U,z;\omega) & \frac12 F^+(\epsilon+U,z;\omega) & -F^-(\epsilon+U,z;\omega)
                                                                 \end{smallmatrix}\right),\\
\frac{\mathcal W^\mathrm{(i)}_\mathrm{I}(t,z)}{\Gamma} &= \left(\begin{smallmatrix}
                                                                  0 & f^-(\epsilon,z;0) & f^-(\epsilon,z;0) & 0\\
                                                                  -f^+(\epsilon,z;0) & 0 & 0 & f^-(\epsilon+U,z;0)\\
                                                                  -f^+(\epsilon,z;0) & 0 & 0 & f^-(\epsilon+U,z;0)\\
                                                                  0 & -f^+(\epsilon+U,z;0) & -f^+(\epsilon+U,z;0) & 0
                                                                 \end{smallmatrix}\right),\\
\frac{\mathcal W^\mathrm{(i)}_\mathrm{II}(t,z;\omega)}{\Gamma} &= \left(\begin{smallmatrix}
                                                                  F^+(\epsilon,z;\omega) & \frac12 F^-(\epsilon,z;\omega) & \frac12 F^-(\epsilon,z;\omega) & 0\\
                                                                  \frac12 F^+(\epsilon,z;\omega) & \frac12 F^-(\epsilon,z;\omega) + \frac12 F^+(\epsilon+U,z;\omega) & 0 & \frac12 F^-(\epsilon+U,z;\omega)\\
                                                                  \frac12 F^+(\epsilon,z;\omega) & 0 & \frac12 F^-(\epsilon,z;\omega) + \frac12 F^+(\epsilon+U,z;\omega) & \frac12 F^-(\epsilon+U,z;\omega)\\
                                                                  0 & \frac12 F^+(\epsilon+U,z;\omega) & \frac12 F^+(\epsilon+U,z;\omega) & F^-(\epsilon+U,z;\omega)
                                                                 \end{smallmatrix}\right),
\end{align}
where the frequency-independent kernel $\mathcal W^\mathrm{(i)}(t,z)$ equals $\mathcal W^\mathrm{(i)}(t,z;0)$. Here, we introduced the abbreviations
\begin{align}
 f^{\pm}(x,z;\omega) &= f^{\pm}(x+\omega-iz) + f^{\pm}(x-\omega+iz),\\
 \psi(x,z;\omega) &=
  \frac{1}{2\pi i} \left[ \tilde{\psi}(\epsilon+\omega-iz) 
    + \tilde{\psi}(-\epsilon-\omega+iz)
    - \tilde{\psi}(\epsilon-\omega+iz) 
    - \tilde{\psi}(-\epsilon+\omega-iz)\right],
   \\
  F^\pm(x,z;\omega) &= f^{\pm}(x,z;\omega) \mp  \psi(x,z;\omega),
\end{align}
where \mbox{$f^\pm(x)=(1+e^{\pm \beta x})^{-1}$} is the Fermi function of the reservoir and $\tilde{\psi}(x)=\psi\left(\frac{1}{2}+\frac{\beta x}{2\pi i}\right) $ with the Digamma function~$\psi$. 
We note that in the limit \mbox{$\omega \rightarrow 0$} and \mbox{$z \rightarrow 0^+$} we find \mbox{$\psi(x,0;0)=0$} and \mbox{$F^\pm(x,0;0) = 2 f^\pm(x)$}.
To write the explicit expressions for the two remaining kernels, $\mathcal W^\mathrm{>,(i)}_\mathrm{I}(t,z;\omega)$ and $\mathcal W^\mathrm{<,(i)}_\mathrm{I}(t,z;\omega)$, we define the additional short-hand notation
\mbox{$g^{\pm\pm}(x,z;\omega) = f^\pm(\epsilon,z;0)\pm F^\pm(\epsilon,z;\omega)$}. 
The first superscript on the left-hand side~refers to the superscripts of the two functions on the right-hand side, while the second superscript defines if the two functions are summed up or subtracted.
Using this abbreviation, the kernels become
\begin{align}
\frac{\mathcal W^\mathrm{>,(i)}_\mathrm{I}(t,z;\omega)}{\Gamma} &= \left(\begin{smallmatrix}
                                                                  -g^{+-}(\epsilon,z;\omega) & \frac12 g^{-+}(\epsilon,z;\omega) & \frac12 g^{-+}(\epsilon,z;\omega) & 0 \\
                                                                  -\frac12 g^{++}(\epsilon,z;\omega) & \frac12 g^{--}(\epsilon,z;\omega)-\frac12 g^{+-}(\epsilon+U,z;\omega) & 0 & \frac12 g^{-+}(\epsilon+U,z;\omega) \\
                                                                  -\frac12 g^{++}(\epsilon,z;\omega) & 0 & \frac12 g^{--}(\epsilon,z;\omega)-\frac12 g^{+-}(\epsilon+U,z;\omega) & \frac12 g^{-+}(\epsilon+U,z;\omega) \\
                                                                  0 & -\frac12 g^{++}(\epsilon+U,z;\omega) & -\frac12 g^{++}(\epsilon+U,z;\omega) &  g^{--}(\epsilon+U,z;\omega) \\
                                                                 \end{smallmatrix}\right),\\
\frac{\mathcal W^\mathrm{<,(i)}_\mathrm{I}(t,z;\omega)}{\Gamma} &= \left(\begin{smallmatrix}
                                                                  g^{+-}(\epsilon,z;\omega) & \frac12 g^{-+}(\epsilon,z;\omega) & \frac12 g^{-+}(\epsilon,z;\omega) & 0 \\
                                                                  -\frac12 g^{++}(\epsilon,z;\omega) & -\frac12 g^{--}(\epsilon,z;\omega)+\frac12 g^{+-}(\epsilon+U,z;\omega) & 0 & \frac12 g^{-+}(\epsilon+U,z;\omega) \\
                                                                  -\frac12 g^{++}(\epsilon,z;\omega) & 0 & -\frac12 g^{--}(\epsilon,z;\omega)+\frac12 g^{+-}(\epsilon+U,z;\omega) & \frac12 g^{-+}(\epsilon+U,z;\omega) \\
                                                                  0 & -\frac12 g^{++}(\epsilon+U,z;\omega) & -\frac12 g^{++}(\epsilon+U,z;\omega) &  -g^{--}(\epsilon+U,z;\omega) \\
                                                                  \end{smallmatrix}\right).
\end{align}
\end{widetext}
Finally, we note that all contributions stemming from Digamma functions, which are included here for completeness, have been neglected in our noise calculations.
The reason is that we denote these contributions to renormalization effects. 
Since tunneling beyond the first order studied here also leads to renormalization of system parameters \cite{Splettstoesser10}, the contributions stemming from the Digamma functions should be excluded for a consistent first-order derivation.
In a calculation in second order in $\Gamma$ (not part of this paper), care must be taken for a proper inclusion of renormalization effects in the finite-frequency noise.

\section{Expressions for high noise frequencies}
\label{app_HFscheme}

In this appendix we derive a simple expression for the reduced propagator, $\Pibar(t,z)$, which is valid for high noise frequencies, $\omega \gg \Gamma$.
This eventually leads to Eq.~\eqref{eq_auxiliarynoiseHF} of the main text.
We begin with Eq.~\eqref{eq_Pibar_dysonlaplace2} for the reduced propagator, which has been derived in App.~\ref{app_adexp_reducedpropagator}.
We remind that the dependence on the noise frequency $\omega$  has been absorbed in the $z$~argument in Eq.~\eqref{eq_Pibar_dysonlaplace2}, which we set to $i\omega$ at the end of the calculation.
The main observation is that the kernel $\mathcal W(t,z)$ in Eq.~\eqref{eq_Pibar_dysonlaplace} has a magnitude of the scale $\Gamma$, while the factor $1/z$ in front turns into a factor $1/\omega$.
We conclude that the second term on the right-hand side~of Eq.~\eqref{eq_Pibar_dysonlaplace2} scales with $\Gamma/\omega$ and is therefore strongly suppressed for high noise frequencies \cite{Kack03}.
This suppression cannot be lifted by the derivatives included in the abbreviation $e^\partial$ in Eq.~\eqref{eq_Pibar_dysonlaplace2}, which only lead to minor corrections for the slow driving considered in this work.
Hence, in the high noise-frequency regime we write
\begin{align}
 \label{eq_app_pibarHF}
 \Pibar^\mathrm{(HF)}(t,z) &= \frac{\boldsymbol 1 - \boldsymbol P(t)\otimes \boldsymbol e^T}{z}.
\end{align}
From this equation we extract the instantaneous part and the adiabatic response of the reduced propagator:
\begin{subequations}
\label{eq_PibarHF}
 \begin{align}
   \Pibar^\mathrm{(i,HF)}(t,z) &= \frac{\boldsymbol 1 - \boldsymbol P^\mathrm{(i)}(t)\otimes \boldsymbol e^T}{z},\\
   \Pibar^\mathrm{(a,HF)}(t,z) &= -\frac{\boldsymbol P^{(a)}(t)\otimes \boldsymbol e^T}{z}.
 \end{align} 
\end{subequations}
Interestingly, the reduced propagator in the high noise-frequency regime, i.e., Eq.~\eqref{eq_app_pibarHF}, has an order-by-order expansion scheme in the tunnel-coupling strength, which means in the small parameter $\Gamma\beta$.
The reason is that this expansion scheme is well defined for the occupation vector on the right-hand side, see also Sec.~\ref{sec_gammaschemes}.

We now derive Eq.~\eqref{eq_auxiliarynoiseHF}, which gives the auxiliary function, $\tilde S^{(l,\mathrm{HF})}(t;\omega)$, in the high-noise frequency regime and in $l$'th order in the slow-driving expansion. 
Importantly, the instantaneous part is included as the case \mbox{$l=0$} and the adiabatic response as \mbox{$l=1$}.
We first remind that for the occupation vector calculated in $l$'th order in the slow-driving expansion, the leading-order term in the additional expansion in the tunnel-coupling strength is given by the $-l$'th order \cite{Splettstoesser06}.
From Eq.~\eqref{eq_app_pibarHF} we conclude that the same is true for leading contributions of the reduced propagator, when the latter is evaluated at high noise frequencies.
Besides that, all kernels begin to contribute in first order in $\Gamma$, irrespective of their order in the slow-driving expansion.
The leading contribution to the current in $l$'th order in the slow-driving expansion is of order  \mbox{$-l+1$} in the expansion in the tunnel-coupling strength.
By only keeping the terms in lowest order in $\Gamma$ for each order in the slow-driving expansion, we arrive at the general Eq.~\eqref{eq_auxiliarynoiseHF} for the auxiliary function calculated in the high-noise frequency regime.

Finally, we also give an explicit expression for the instantaneous fluctuation vector of Eq.~\eqref{eq_flvector} for a spin-split system. At high noise-frequencies, we find the expression
\begin{align}
\label{app_eq_flvector_spinsplit}
\frac{\boldsymbol F^{\mathrm{(i,HF)}} }{\Gamma} &= \left(\begin{matrix}
                        f^+(\epsilon_\uparrow;\omega)+f^+(\epsilon_\downarrow;\omega) \\
                        f^-(\epsilon_\uparrow;\omega) + f^+(\epsilon_\downarrow+U;\omega) \\
                        f^-(\epsilon_\downarrow;\omega) + f^+(\epsilon_\uparrow+U;\omega) \\
                        f^-(\epsilon_\uparrow+U;\omega)+f^-(\epsilon_\downarrow+U;\omega) \\
                       \end{matrix}\right),
\end{align}
where \mbox{$\boldsymbol F^{\mathrm{(i,HF)}} = \boldsymbol F^{\mathrm{(i,HF)}}(t;\omega)$}.


\begin{thebibliography}{47}%
\makeatletter
\providecommand \@ifxundefined [1]{%
 \@ifx{#1\undefined}
}%
\providecommand \@ifnum [1]{%
 \ifnum #1\expandafter \@firstoftwo
 \else \expandafter \@secondoftwo
 \fi
}%
\providecommand \@ifx [1]{%
 \ifx #1\expandafter \@firstoftwo
 \else \expandafter \@secondoftwo
 \fi
}%
\providecommand \natexlab [1]{#1}%
\providecommand \enquote  [1]{``#1''}%
\providecommand \bibnamefont  [1]{#1}%
\providecommand \bibfnamefont [1]{#1}%
\providecommand \citenamefont [1]{#1}%
\providecommand \href@noop [0]{\@secondoftwo}%
\providecommand \href [0]{\begingroup \@sanitize@url \@href}%
\providecommand \@href[1]{\@@startlink{#1}\@@href}%
\providecommand \@@href[1]{\endgroup#1\@@endlink}%
\providecommand \@sanitize@url [0]{\catcode `\\12\catcode `\$12\catcode
  `\&12\catcode `\#12\catcode `\^12\catcode `\_12\catcode `\%12\relax}%
\providecommand \@@startlink[1]{}%
\providecommand \@@endlink[0]{}%
\providecommand \url  [0]{\begingroup\@sanitize@url \@url }%
\providecommand \@url [1]{\endgroup\@href {#1}{\urlprefix }}%
\providecommand \urlprefix  [0]{URL }%
\providecommand \Eprint [0]{\href }%
\providecommand \doibase [0]{http://dx.doi.org/}%
\providecommand \selectlanguage [0]{\@gobble}%
\providecommand \bibinfo  [0]{\@secondoftwo}%
\providecommand \bibfield  [0]{\@secondoftwo}%
\providecommand \translation [1]{[#1]}%
\providecommand \BibitemOpen [0]{}%
\providecommand \bibitemStop [0]{}%
\providecommand \bibitemNoStop [0]{.\EOS\space}%
\providecommand \EOS [0]{\spacefactor3000\relax}%
\providecommand \BibitemShut  [1]{\csname bibitem#1\endcsname}%
\let\auto@bib@innerbib\@empty
\bibitem [{\citenamefont {F{\`e}ve}\ \emph {et~al.}(2007)\citenamefont
  {F{\`e}ve}, \citenamefont {Mah{\'e}}, \citenamefont {Berroir}, \citenamefont
  {Kontos}, \citenamefont {Pla{\c c}ais}, \citenamefont {Glattli},
  \citenamefont {Cavanna}, \citenamefont {Etienne},\ and\ \citenamefont
  {Jin}}]{Feve07}%
  \BibitemOpen
  \bibfield  {author} {\bibinfo {author} {\bibfnamefont {G.}~\bibnamefont
  {F{\`e}ve}}, \bibinfo {author} {\bibfnamefont {A.}~\bibnamefont {Mah{\'e}}},
  \bibinfo {author} {\bibfnamefont {J.-M.}\ \bibnamefont {Berroir}}, \bibinfo
  {author} {\bibfnamefont {T.}~\bibnamefont {Kontos}}, \bibinfo {author}
  {\bibfnamefont {B.}~\bibnamefont {Pla{\c c}ais}}, \bibinfo {author}
  {\bibfnamefont {D.~C.}\ \bibnamefont {Glattli}}, \bibinfo {author}
  {\bibfnamefont {A.}~\bibnamefont {Cavanna}}, \bibinfo {author} {\bibfnamefont
  {B.}~\bibnamefont {Etienne}}, \ and\ \bibinfo {author} {\bibfnamefont
  {Y.}~\bibnamefont {Jin}},\ }\href {\doibase 10.1126/science.1141243}
  {\bibfield  {journal} {\bibinfo  {journal} {Science}\ }\textbf {\bibinfo
  {volume} {316}},\ \bibinfo {pages} {1169} (\bibinfo {year}
  {2007})}\BibitemShut {NoStop}%
\bibitem [{\citenamefont {Blumenthal}\ \emph {et~al.}(2007)\citenamefont
  {Blumenthal}, \citenamefont {Kaestner}, \citenamefont {Li}, \citenamefont
  {Giblin}, \citenamefont {Janssen}, \citenamefont {Pepper}, \citenamefont
  {Anderson}, \citenamefont {Jones},\ and\ \citenamefont
  {Ritchie}}]{Blumenthal07}%
  \BibitemOpen
  \bibfield  {author} {\bibinfo {author} {\bibfnamefont {M.~D.}\ \bibnamefont
  {Blumenthal}}, \bibinfo {author} {\bibfnamefont {B.}~\bibnamefont
  {Kaestner}}, \bibinfo {author} {\bibfnamefont {L.}~\bibnamefont {Li}},
  \bibinfo {author} {\bibfnamefont {S.}~\bibnamefont {Giblin}}, \bibinfo
  {author} {\bibfnamefont {T.~J. B.~M.}\ \bibnamefont {Janssen}}, \bibinfo
  {author} {\bibfnamefont {M.}~\bibnamefont {Pepper}}, \bibinfo {author}
  {\bibfnamefont {D.}~\bibnamefont {Anderson}}, \bibinfo {author}
  {\bibfnamefont {G.}~\bibnamefont {Jones}}, \ and\ \bibinfo {author}
  {\bibfnamefont {D.~A.}\ \bibnamefont {Ritchie}},\ }\href {\doibase
  10.1038/nphys582} {\bibfield  {journal} {\bibinfo  {journal} {Nat Phys}\
  }\textbf {\bibinfo {volume} {3}},\ \bibinfo {pages} {343} (\bibinfo {year}
  {2007})}\BibitemShut {NoStop}%
\bibitem [{\citenamefont {B\"auerle}\ \emph {et~al.}(2018)\citenamefont
  {B\"auerle}, \citenamefont {Glattli}, \citenamefont {Meunier}, \citenamefont
  {Portier}, \citenamefont {Roche}, \citenamefont {Roulleau}, \citenamefont
  {Takada},\ and\ \citenamefont {Waintal}}]{Bauerle18}%
  \BibitemOpen
  \bibfield  {author} {\bibinfo {author} {\bibfnamefont {C.}~\bibnamefont
  {B\"auerle}}, \bibinfo {author} {\bibfnamefont {D.~C.}\ \bibnamefont
  {Glattli}}, \bibinfo {author} {\bibfnamefont {T.}~\bibnamefont {Meunier}},
  \bibinfo {author} {\bibfnamefont {F.}~\bibnamefont {Portier}}, \bibinfo
  {author} {\bibfnamefont {P.}~\bibnamefont {Roche}}, \bibinfo {author}
  {\bibfnamefont {P.}~\bibnamefont {Roulleau}}, \bibinfo {author}
  {\bibfnamefont {S.}~\bibnamefont {Takada}}, \ and\ \bibinfo {author}
  {\bibfnamefont {X.}~\bibnamefont {Waintal}},\ }\href
  {http://stacks.iop.org/0034-4885/81/i=5/a=056503} {\bibfield  {journal}
  {\bibinfo  {journal} {Reports on Progress in Physics}\ }\textbf {\bibinfo
  {volume} {81}},\ \bibinfo {pages} {056503} (\bibinfo {year}
  {2018})}\BibitemShut {NoStop}%
\bibitem [{\citenamefont {Giblin}\ \emph {et~al.}(2012)\citenamefont {Giblin},
  \citenamefont {Kataoka}, \citenamefont {Fletcher}, \citenamefont {See},
  \citenamefont {Janssen}, \citenamefont {Griffiths}, \citenamefont {Jones},
  \citenamefont {Farrer},\ and\ \citenamefont {Ritchie}}]{Giblin12}%
  \BibitemOpen
  \bibfield  {author} {\bibinfo {author} {\bibfnamefont {S.}~\bibnamefont
  {Giblin}}, \bibinfo {author} {\bibfnamefont {M.}~\bibnamefont {Kataoka}},
  \bibinfo {author} {\bibfnamefont {J.}~\bibnamefont {Fletcher}}, \bibinfo
  {author} {\bibfnamefont {P.}~\bibnamefont {See}}, \bibinfo {author}
  {\bibfnamefont {T.}~\bibnamefont {Janssen}}, \bibinfo {author} {\bibfnamefont
  {J.}~\bibnamefont {Griffiths}}, \bibinfo {author} {\bibfnamefont
  {G.}~\bibnamefont {Jones}}, \bibinfo {author} {\bibfnamefont
  {I.}~\bibnamefont {Farrer}}, \ and\ \bibinfo {author} {\bibfnamefont
  {D.}~\bibnamefont {Ritchie}},\ }\href {\doibase 10.1038/ncomms1935}
  {\bibfield  {journal} {\bibinfo  {journal} {Nature Communications}\ }\textbf
  {\bibinfo {volume} {3}},\ \bibinfo {pages} {930} (\bibinfo {year}
  {2012})}\BibitemShut {NoStop}%
\bibitem [{\citenamefont {Pekola}\ \emph {et~al.}(2013)\citenamefont {Pekola},
  \citenamefont {Saira}, \citenamefont {Maisi}, \citenamefont {Kemppinen},
  \citenamefont {M\"ott\"onen}, \citenamefont {Pashkin},\ and\ \citenamefont
  {Averin}}]{Pekola13}%
  \BibitemOpen
  \bibfield  {author} {\bibinfo {author} {\bibfnamefont {J.~P.}\ \bibnamefont
  {Pekola}}, \bibinfo {author} {\bibfnamefont {O.-P.}\ \bibnamefont {Saira}},
  \bibinfo {author} {\bibfnamefont {V.~F.}\ \bibnamefont {Maisi}}, \bibinfo
  {author} {\bibfnamefont {A.}~\bibnamefont {Kemppinen}}, \bibinfo {author}
  {\bibfnamefont {M.}~\bibnamefont {M\"ott\"onen}}, \bibinfo {author}
  {\bibfnamefont {Y.~A.}\ \bibnamefont {Pashkin}}, \ and\ \bibinfo {author}
  {\bibfnamefont {D.~V.}\ \bibnamefont {Averin}},\ }\href {\doibase
  10.1103/RevModPhys.85.1421} {\bibfield  {journal} {\bibinfo  {journal} {Rev.
  Mod. Phys.}\ }\textbf {\bibinfo {volume} {85}},\ \bibinfo {pages} {1421}
  (\bibinfo {year} {2013})}\BibitemShut {NoStop}%
\bibitem [{\citenamefont {Freulon}\ \emph {et~al.}(2015)\citenamefont
  {Freulon}, \citenamefont {Marguerite}, \citenamefont {Berroir}, \citenamefont
  {Pla{\c{c}}ais}, \citenamefont {Cavanna}, \citenamefont {Jin},\ and\
  \citenamefont {F{\`e}ve}}]{Freulon15}%
  \BibitemOpen
  \bibfield  {author} {\bibinfo {author} {\bibfnamefont {V.}~\bibnamefont
  {Freulon}}, \bibinfo {author} {\bibfnamefont {A.}~\bibnamefont {Marguerite}},
  \bibinfo {author} {\bibfnamefont {J.-M.}\ \bibnamefont {Berroir}}, \bibinfo
  {author} {\bibfnamefont {B.}~\bibnamefont {Pla{\c{c}}ais}}, \bibinfo {author}
  {\bibfnamefont {A.}~\bibnamefont {Cavanna}}, \bibinfo {author} {\bibfnamefont
  {Y.}~\bibnamefont {Jin}}, \ and\ \bibinfo {author} {\bibfnamefont
  {G.}~\bibnamefont {F{\`e}ve}},\ }\href {http://dx.doi.org/10.1038/ncomms7854}
  {\bibfield  {journal} {\bibinfo  {journal} {Nature Communications}\ }\textbf
  {\bibinfo {volume} {6}} (\bibinfo {year} {2015})}\BibitemShut {NoStop}%
\bibitem [{\citenamefont {Marguerite}\ \emph {et~al.}(2016)\citenamefont
  {Marguerite}, \citenamefont {Cabart}, \citenamefont {Wahl}, \citenamefont
  {Roussel}, \citenamefont {Freulon}, \citenamefont {Ferraro}, \citenamefont
  {Grenier}, \citenamefont {Berroir}, \citenamefont
  {Pla\ifmmode~\mbox{\c{c}}\else \c{c}\fi{}ais}, \citenamefont {Jonckheere},
  \citenamefont {Rech}, \citenamefont {Martin}, \citenamefont {Degiovanni},
  \citenamefont {Cavanna}, \citenamefont {Jin},\ and\ \citenamefont
  {F\`eve}}]{Marguerite16}%
  \BibitemOpen
  \bibfield  {author} {\bibinfo {author} {\bibfnamefont {A.}~\bibnamefont
  {Marguerite}}, \bibinfo {author} {\bibfnamefont {C.}~\bibnamefont {Cabart}},
  \bibinfo {author} {\bibfnamefont {C.}~\bibnamefont {Wahl}}, \bibinfo {author}
  {\bibfnamefont {B.}~\bibnamefont {Roussel}}, \bibinfo {author} {\bibfnamefont
  {V.}~\bibnamefont {Freulon}}, \bibinfo {author} {\bibfnamefont
  {D.}~\bibnamefont {Ferraro}}, \bibinfo {author} {\bibfnamefont
  {C.}~\bibnamefont {Grenier}}, \bibinfo {author} {\bibfnamefont {J.-M.}\
  \bibnamefont {Berroir}}, \bibinfo {author} {\bibfnamefont {B.}~\bibnamefont
  {Pla\ifmmode~\mbox{\c{c}}\else \c{c}\fi{}ais}}, \bibinfo {author}
  {\bibfnamefont {T.}~\bibnamefont {Jonckheere}}, \bibinfo {author}
  {\bibfnamefont {J.}~\bibnamefont {Rech}}, \bibinfo {author} {\bibfnamefont
  {T.}~\bibnamefont {Martin}}, \bibinfo {author} {\bibfnamefont
  {P.}~\bibnamefont {Degiovanni}}, \bibinfo {author} {\bibfnamefont
  {A.}~\bibnamefont {Cavanna}}, \bibinfo {author} {\bibfnamefont
  {Y.}~\bibnamefont {Jin}}, \ and\ \bibinfo {author} {\bibfnamefont
  {G.}~\bibnamefont {F\`eve}},\ }\href {\doibase 10.1103/PhysRevB.94.115311}
  {\bibfield  {journal} {\bibinfo  {journal} {Phys. Rev. B}\ }\textbf {\bibinfo
  {volume} {94}},\ \bibinfo {pages} {115311} (\bibinfo {year}
  {2016})}\BibitemShut {NoStop}%
\bibitem [{\citenamefont {B{\ifmmode \ddot{u} \else \"{u}\fi}ttiker}\ \emph
  {et~al.}(1993)\citenamefont {B{\ifmmode \ddot{u} \else \"{u}\fi}ttiker},
  \citenamefont {Pr{\ifmmode \hat{e} \else \^{e}\fi}tre},\ and\ \citenamefont
  {Thomas}}]{Buttiker1993Jun}%
  \BibitemOpen
  \bibfield  {author} {\bibinfo {author} {\bibfnamefont {M.}~\bibnamefont
  {B{\ifmmode \ddot{u} \else \"{u}\fi}ttiker}}, \bibinfo {author}
  {\bibfnamefont {A.}~\bibnamefont {Pr{\ifmmode \hat{e} \else \^{e}\fi}tre}}, \
  and\ \bibinfo {author} {\bibfnamefont {H.}~\bibnamefont {Thomas}},\ }\href
  {\doibase 10.1103/PhysRevLett.70.4114} {\bibfield  {journal} {\bibinfo
  {journal} {Phys. Rev. Lett.}\ }\textbf {\bibinfo {volume} {70}},\ \bibinfo
  {pages} {4114} (\bibinfo {year} {1993})}\BibitemShut {NoStop}%
\bibitem [{\citenamefont {Pr{\ifmmode \hat{e} \else \^{e}\fi}tre}\ \emph
  {et~al.}(1996)\citenamefont {Pr{\ifmmode \hat{e} \else \^{e}\fi}tre},
  \citenamefont {Thomas},\ and\ \citenamefont {B{\ifmmode \ddot{u} \else
  \"{u}\fi}ttiker}}]{Pretre1996Sep}%
  \BibitemOpen
  \bibfield  {author} {\bibinfo {author} {\bibfnamefont {A.}~\bibnamefont
  {Pr{\ifmmode \hat{e} \else \^{e}\fi}tre}}, \bibinfo {author} {\bibfnamefont
  {H.}~\bibnamefont {Thomas}}, \ and\ \bibinfo {author} {\bibfnamefont
  {M.}~\bibnamefont {B{\ifmmode \ddot{u} \else \"{u}\fi}ttiker}},\ }\href
  {\doibase 10.1103/PhysRevB.54.8130} {\bibfield  {journal} {\bibinfo
  {journal} {Phys. Rev. B}\ }\textbf {\bibinfo {volume} {54}},\ \bibinfo
  {pages} {8130} (\bibinfo {year} {1996})}\BibitemShut {NoStop}%
\bibitem [{\citenamefont {Blanter}\ and\ \citenamefont
  {B\"uttiker}(2000)}]{Blanter00}%
  \BibitemOpen
  \bibfield  {author} {\bibinfo {author} {\bibfnamefont {Y.}~\bibnamefont
  {Blanter}}\ and\ \bibinfo {author} {\bibfnamefont {M.}~\bibnamefont
  {B\"uttiker}},\ }\href {\doibase
  https://doi.org/10.1016/S0370-1573(99)00123-4} {\bibfield  {journal}
  {\bibinfo  {journal} {Phys. Rep.}\ }\textbf {\bibinfo {volume} {336}},\
  \bibinfo {pages} {1} (\bibinfo {year} {2000})}\BibitemShut {NoStop}%
\bibitem [{\citenamefont {Clerk}\ \emph {et~al.}(2010)\citenamefont {Clerk},
  \citenamefont {Devoret}, \citenamefont {Girvin}, \citenamefont {Marquardt},\
  and\ \citenamefont {Schoelkopf}}]{Clerk10}%
  \BibitemOpen
  \bibfield  {author} {\bibinfo {author} {\bibfnamefont {A.~A.}\ \bibnamefont
  {Clerk}}, \bibinfo {author} {\bibfnamefont {M.~H.}\ \bibnamefont {Devoret}},
  \bibinfo {author} {\bibfnamefont {S.~M.}\ \bibnamefont {Girvin}}, \bibinfo
  {author} {\bibfnamefont {F.}~\bibnamefont {Marquardt}}, \ and\ \bibinfo
  {author} {\bibfnamefont {R.~J.}\ \bibnamefont {Schoelkopf}},\ }\href
  {\doibase 10.1103/RevModPhys.82.1155} {\bibfield  {journal} {\bibinfo
  {journal} {Rev. Mod. Phys.}\ }\textbf {\bibinfo {volume} {82}},\ \bibinfo
  {pages} {1155} (\bibinfo {year} {2010})}\BibitemShut {NoStop}%
\bibitem [{\citenamefont {Landauer}(1998)}]{Landauer98}%
  \BibitemOpen
  \bibfield  {author} {\bibinfo {author} {\bibfnamefont {R.}~\bibnamefont
  {Landauer}},\ }\href {\doibase 10.1038/33551} {\bibfield  {journal} {\bibinfo
   {journal} {Nature}\ }\textbf {\bibinfo {volume} {392}},\ \bibinfo {pages}
  {658} (\bibinfo {year} {1998})}\BibitemShut {NoStop}%
\bibitem [{\citenamefont {Mah{\ifmmode \acute{e} \else \'{e}\fi}}\ \emph
  {et~al.}(2010)\citenamefont {Mah{\ifmmode \acute{e} \else \'{e}\fi}},
  \citenamefont {Parmentier}, \citenamefont {Bocquillon}, \citenamefont
  {Berroir}, \citenamefont {Glattli}, \citenamefont {Kontos}, \citenamefont
  {Pla{\ifmmode \mbox{\c{c}} \else \c{c}\fi}ais}, \citenamefont {F{\ifmmode
  \grave{e} \else \`{e}\fi}ve}, \citenamefont {Cavanna},\ and\ \citenamefont
  {Jin}}]{Mahe10}%
  \BibitemOpen
  \bibfield  {author} {\bibinfo {author} {\bibfnamefont {A.}~\bibnamefont
  {Mah{\ifmmode \acute{e} \else \'{e}\fi}}}, \bibinfo {author} {\bibfnamefont
  {F.~D.}\ \bibnamefont {Parmentier}}, \bibinfo {author} {\bibfnamefont
  {E.}~\bibnamefont {Bocquillon}}, \bibinfo {author} {\bibfnamefont {J.-M.}\
  \bibnamefont {Berroir}}, \bibinfo {author} {\bibfnamefont {D.~C.}\
  \bibnamefont {Glattli}}, \bibinfo {author} {\bibfnamefont {T.}~\bibnamefont
  {Kontos}}, \bibinfo {author} {\bibfnamefont {B.}~\bibnamefont {Pla{\ifmmode
  \mbox{\c{c}} \else \c{c}\fi}ais}}, \bibinfo {author} {\bibfnamefont
  {G.}~\bibnamefont {F{\ifmmode \grave{e} \else \`{e}\fi}ve}}, \bibinfo
  {author} {\bibfnamefont {A.}~\bibnamefont {Cavanna}}, \ and\ \bibinfo
  {author} {\bibfnamefont {Y.}~\bibnamefont {Jin}},\ }\href {\doibase
  10.1103/PhysRevB.82.201309} {\bibfield  {journal} {\bibinfo  {journal} {Phys.
  Rev. B}\ }\textbf {\bibinfo {volume} {82}},\ \bibinfo {pages} {201309}
  (\bibinfo {year} {2010})}\BibitemShut {NoStop}%
\bibitem [{\citenamefont {Parmentier}\ \emph {et~al.}(2012)\citenamefont
  {Parmentier}, \citenamefont {Bocquillon}, \citenamefont {Berroir},
  \citenamefont {Glattli}, \citenamefont {Pla\ifmmode~\mbox{\c{c}}\else
  \c{c}\fi{}ais}, \citenamefont {F\`eve}, \citenamefont {Albert}, \citenamefont
  {Flindt},\ and\ \citenamefont {B\"uttiker}}]{Parmentier12}%
  \BibitemOpen
  \bibfield  {author} {\bibinfo {author} {\bibfnamefont {F.~D.}\ \bibnamefont
  {Parmentier}}, \bibinfo {author} {\bibfnamefont {E.}~\bibnamefont
  {Bocquillon}}, \bibinfo {author} {\bibfnamefont {J.-M.}\ \bibnamefont
  {Berroir}}, \bibinfo {author} {\bibfnamefont {D.~C.}\ \bibnamefont
  {Glattli}}, \bibinfo {author} {\bibfnamefont {B.}~\bibnamefont
  {Pla\ifmmode~\mbox{\c{c}}\else \c{c}\fi{}ais}}, \bibinfo {author}
  {\bibfnamefont {G.}~\bibnamefont {F\`eve}}, \bibinfo {author} {\bibfnamefont
  {M.}~\bibnamefont {Albert}}, \bibinfo {author} {\bibfnamefont
  {C.}~\bibnamefont {Flindt}}, \ and\ \bibinfo {author} {\bibfnamefont
  {M.}~\bibnamefont {B\"uttiker}},\ }\href {\doibase
  10.1103/PhysRevB.85.165438} {\bibfield  {journal} {\bibinfo  {journal} {Phys.
  Rev. B}\ }\textbf {\bibinfo {volume} {85}},\ \bibinfo {pages} {165438}
  (\bibinfo {year} {2012})}\BibitemShut {NoStop}%
\bibitem [{\citenamefont {Onac}\ \emph {et~al.}(2006)\citenamefont {Onac},
  \citenamefont {Balestro}, \citenamefont {Trauzettel}, \citenamefont
  {Lodewijk},\ and\ \citenamefont {Kouwenhoven}}]{Onac06}%
  \BibitemOpen
  \bibfield  {author} {\bibinfo {author} {\bibfnamefont {E.}~\bibnamefont
  {Onac}}, \bibinfo {author} {\bibfnamefont {F.}~\bibnamefont {Balestro}},
  \bibinfo {author} {\bibfnamefont {B.}~\bibnamefont {Trauzettel}}, \bibinfo
  {author} {\bibfnamefont {C.~F.~J.}\ \bibnamefont {Lodewijk}}, \ and\ \bibinfo
  {author} {\bibfnamefont {L.~P.}\ \bibnamefont {Kouwenhoven}},\ }\href
  {\doibase 10.1103/PhysRevLett.96.026803} {\bibfield  {journal} {\bibinfo
  {journal} {Phys. Rev. Lett.}\ }\textbf {\bibinfo {volume} {96}},\ \bibinfo
  {pages} {026803} (\bibinfo {year} {2006})}\BibitemShut {NoStop}%
\bibitem [{\citenamefont {Basset}\ \emph {et~al.}(2012)\citenamefont {Basset},
  \citenamefont {Kasumov}, \citenamefont {Moca}, \citenamefont {Zar\'and},
  \citenamefont {Simon}, \citenamefont {Bouchiat},\ and\ \citenamefont
  {Deblock}}]{Basset12}%
  \BibitemOpen
  \bibfield  {author} {\bibinfo {author} {\bibfnamefont {J.}~\bibnamefont
  {Basset}}, \bibinfo {author} {\bibfnamefont {A.~Y.}\ \bibnamefont {Kasumov}},
  \bibinfo {author} {\bibfnamefont {C.~P.}\ \bibnamefont {Moca}}, \bibinfo
  {author} {\bibfnamefont {G.}~\bibnamefont {Zar\'and}}, \bibinfo {author}
  {\bibfnamefont {P.}~\bibnamefont {Simon}}, \bibinfo {author} {\bibfnamefont
  {H.}~\bibnamefont {Bouchiat}}, \ and\ \bibinfo {author} {\bibfnamefont
  {R.}~\bibnamefont {Deblock}},\ }\href {\doibase
  10.1103/PhysRevLett.108.046802} {\bibfield  {journal} {\bibinfo  {journal}
  {Phys. Rev. Lett.}\ }\textbf {\bibinfo {volume} {108}},\ \bibinfo {pages}
  {046802} (\bibinfo {year} {2012})}\BibitemShut {NoStop}%
\bibitem [{\citenamefont {Ubbelohde}\ \emph {et~al.}(2012)\citenamefont
  {Ubbelohde}, \citenamefont {Fricke}, \citenamefont {Flindt}, \citenamefont
  {Hohls},\ and\ \citenamefont {Haug}}]{Ubbelohde12}%
  \BibitemOpen
  \bibfield  {author} {\bibinfo {author} {\bibfnamefont {N.}~\bibnamefont
  {Ubbelohde}}, \bibinfo {author} {\bibfnamefont {C.}~\bibnamefont {Fricke}},
  \bibinfo {author} {\bibfnamefont {C.}~\bibnamefont {Flindt}}, \bibinfo
  {author} {\bibfnamefont {F.}~\bibnamefont {Hohls}}, \ and\ \bibinfo {author}
  {\bibfnamefont {R.~J.}\ \bibnamefont {Haug}},\ }\href
  {http://dx.doi.org/10.1038/ncomms1620} {\bibfield  {journal} {\bibinfo
  {journal} {Nat. Comm.}\ }\textbf {\bibinfo {volume} {3}},\ \bibinfo {pages}
  {612} (\bibinfo {year} {2012})}\BibitemShut {NoStop}%
\bibitem [{\citenamefont {K\"ack}\ \emph {et~al.}(2003)\citenamefont {K\"ack},
  \citenamefont {Wendin},\ and\ \citenamefont {Johansson}}]{Kack03}%
  \BibitemOpen
  \bibfield  {author} {\bibinfo {author} {\bibfnamefont {A.}~\bibnamefont
  {K\"ack}}, \bibinfo {author} {\bibfnamefont {G.}~\bibnamefont {Wendin}}, \
  and\ \bibinfo {author} {\bibfnamefont {G.}~\bibnamefont {Johansson}},\ }\href
  {\doibase 10.1103/PhysRevB.67.035301} {\bibfield  {journal} {\bibinfo
  {journal} {Phys. Rev. B}\ }\textbf {\bibinfo {volume} {67}},\ \bibinfo
  {pages} {035301} (\bibinfo {year} {2003})}\BibitemShut {NoStop}%
\bibitem [{\citenamefont {Engel}\ and\ \citenamefont {Loss}(2004)}]{Engel04}%
  \BibitemOpen
  \bibfield  {author} {\bibinfo {author} {\bibfnamefont {H.-A.}\ \bibnamefont
  {Engel}}\ and\ \bibinfo {author} {\bibfnamefont {D.}~\bibnamefont {Loss}},\
  }\href {\doibase 10.1103/PhysRevLett.93.136602} {\bibfield  {journal}
  {\bibinfo  {journal} {Phys. Rev. Lett.}\ }\textbf {\bibinfo {volume} {93}},\
  \bibinfo {pages} {136602} (\bibinfo {year} {2004})}\BibitemShut {NoStop}%
\bibitem [{\citenamefont {Braun}\ \emph {et~al.}(2006)\citenamefont {Braun},
  \citenamefont {K\"onig},\ and\ \citenamefont {Martinek}}]{Braun06}%
  \BibitemOpen
  \bibfield  {author} {\bibinfo {author} {\bibfnamefont {M.}~\bibnamefont
  {Braun}}, \bibinfo {author} {\bibfnamefont {J.}~\bibnamefont {K\"onig}}, \
  and\ \bibinfo {author} {\bibfnamefont {J.}~\bibnamefont {Martinek}},\ }\href
  {\doibase 10.1103/PhysRevB.74.075328} {\bibfield  {journal} {\bibinfo
  {journal} {Phys. Rev. B}\ }\textbf {\bibinfo {volume} {74}},\ \bibinfo
  {pages} {075328} (\bibinfo {year} {2006})}\BibitemShut {NoStop}%
\bibitem [{\citenamefont {Rothstein}\ \emph {et~al.}(2009)\citenamefont
  {Rothstein}, \citenamefont {Entin-Wohlman},\ and\ \citenamefont
  {Aharony}}]{Rothstein09}%
  \BibitemOpen
  \bibfield  {author} {\bibinfo {author} {\bibfnamefont {E.~A.}\ \bibnamefont
  {Rothstein}}, \bibinfo {author} {\bibfnamefont {O.}~\bibnamefont
  {Entin-Wohlman}}, \ and\ \bibinfo {author} {\bibfnamefont {A.}~\bibnamefont
  {Aharony}},\ }\href {\doibase 10.1103/PhysRevB.79.075307} {\bibfield
  {journal} {\bibinfo  {journal} {Phys. Rev. B}\ }\textbf {\bibinfo {volume}
  {79}},\ \bibinfo {pages} {075307} (\bibinfo {year} {2009})}\BibitemShut
  {NoStop}%
\bibitem [{\citenamefont {Gabdank}\ \emph {et~al.}(2011)\citenamefont
  {Gabdank}, \citenamefont {Rothstein}, \citenamefont {Entin-Wohlman},\ and\
  \citenamefont {Aharony}}]{Gabdank11}%
  \BibitemOpen
  \bibfield  {author} {\bibinfo {author} {\bibfnamefont {N.}~\bibnamefont
  {Gabdank}}, \bibinfo {author} {\bibfnamefont {E.~A.}\ \bibnamefont
  {Rothstein}}, \bibinfo {author} {\bibfnamefont {O.}~\bibnamefont
  {Entin-Wohlman}}, \ and\ \bibinfo {author} {\bibfnamefont {A.}~\bibnamefont
  {Aharony}},\ }\href {\doibase 10.1103/PhysRevB.84.235435} {\bibfield
  {journal} {\bibinfo  {journal} {Phys. Rev. B}\ }\textbf {\bibinfo {volume}
  {84}},\ \bibinfo {pages} {235435} (\bibinfo {year} {2011})}\BibitemShut
  {NoStop}%
\bibitem [{\citenamefont {Marcos}\ \emph {et~al.}(2011)\citenamefont {Marcos},
  \citenamefont {Emary}, \citenamefont {Brandes},\ and\ \citenamefont
  {Aguado}}]{Marcos11}%
  \BibitemOpen
  \bibfield  {author} {\bibinfo {author} {\bibfnamefont {D.}~\bibnamefont
  {Marcos}}, \bibinfo {author} {\bibfnamefont {C.}~\bibnamefont {Emary}},
  \bibinfo {author} {\bibfnamefont {T.}~\bibnamefont {Brandes}}, \ and\
  \bibinfo {author} {\bibfnamefont {R.}~\bibnamefont {Aguado}},\ }\href
  {\doibase 10.1103/PhysRevB.83.125426} {\bibfield  {journal} {\bibinfo
  {journal} {Phys. Rev. B}\ }\textbf {\bibinfo {volume} {83}},\ \bibinfo
  {pages} {125426} (\bibinfo {year} {2011})}\BibitemShut {NoStop}%
\bibitem [{\citenamefont {Orth}\ \emph {et~al.}(2012)\citenamefont {Orth},
  \citenamefont {Urban},\ and\ \citenamefont {Komnik}}]{Orth12}%
  \BibitemOpen
  \bibfield  {author} {\bibinfo {author} {\bibfnamefont {C.~P.}\ \bibnamefont
  {Orth}}, \bibinfo {author} {\bibfnamefont {D.~F.}\ \bibnamefont {Urban}}, \
  and\ \bibinfo {author} {\bibfnamefont {A.}~\bibnamefont {Komnik}},\ }\href
  {\doibase 10.1103/PhysRevB.86.125324} {\bibfield  {journal} {\bibinfo
  {journal} {Phys. Rev. B}\ }\textbf {\bibinfo {volume} {86}},\ \bibinfo
  {pages} {125324} (\bibinfo {year} {2012})}\BibitemShut {NoStop}%
\bibitem [{\citenamefont {M{\ifmmode \ddot{u} \else \"{u}\fi}ller}\ \emph
  {et~al.}(2013)\citenamefont {M{\ifmmode \ddot{u} \else \"{u}\fi}ller},
  \citenamefont {Pletyukhov}, \citenamefont {Schuricht},\ and\ \citenamefont
  {Andergassen}}]{Muller13}%
  \BibitemOpen
  \bibfield  {author} {\bibinfo {author} {\bibfnamefont {S.~Y.}\ \bibnamefont
  {M{\ifmmode \ddot{u} \else \"{u}\fi}ller}}, \bibinfo {author} {\bibfnamefont
  {M.}~\bibnamefont {Pletyukhov}}, \bibinfo {author} {\bibfnamefont
  {D.}~\bibnamefont {Schuricht}}, \ and\ \bibinfo {author} {\bibfnamefont
  {S.}~\bibnamefont {Andergassen}},\ }\href {\doibase
  10.1103/PhysRevB.87.245115} {\bibfield  {journal} {\bibinfo  {journal} {Phys.
  Rev. B}\ }\textbf {\bibinfo {volume} {87}},\ \bibinfo {pages} {245115}
  (\bibinfo {year} {2013})}\BibitemShut {NoStop}%
\bibitem [{\citenamefont {Moca}\ \emph {et~al.}(2014)\citenamefont {Moca},
  \citenamefont {Simon}, \citenamefont {Chung},\ and\ \citenamefont
  {Zar\'and}}]{Moca14}%
  \BibitemOpen
  \bibfield  {author} {\bibinfo {author} {\bibfnamefont {C.~P.}\ \bibnamefont
  {Moca}}, \bibinfo {author} {\bibfnamefont {P.}~\bibnamefont {Simon}},
  \bibinfo {author} {\bibfnamefont {C.-H.}\ \bibnamefont {Chung}}, \ and\
  \bibinfo {author} {\bibfnamefont {G.}~\bibnamefont {Zar\'and}},\ }\href
  {\doibase 10.1103/PhysRevB.89.155138} {\bibfield  {journal} {\bibinfo
  {journal} {Phys. Rev. B}\ }\textbf {\bibinfo {volume} {89}},\ \bibinfo
  {pages} {155138} (\bibinfo {year} {2014})}\BibitemShut {NoStop}%
\bibitem [{\citenamefont {Jin}\ \emph {et~al.}(2015)\citenamefont {Jin},
  \citenamefont {Wang}, \citenamefont {Zheng},\ and\ \citenamefont
  {Yan}}]{Jin15}%
  \BibitemOpen
  \bibfield  {author} {\bibinfo {author} {\bibfnamefont {J.}~\bibnamefont
  {Jin}}, \bibinfo {author} {\bibfnamefont {S.}~\bibnamefont {Wang}}, \bibinfo
  {author} {\bibfnamefont {X.}~\bibnamefont {Zheng}}, \ and\ \bibinfo {author}
  {\bibfnamefont {Y.}~\bibnamefont {Yan}},\ }\href {\doibase 10.1063/1.4922712}
  {\bibfield  {journal} {\bibinfo  {journal} {J. Chem. Phys.}\ }\textbf
  {\bibinfo {volume} {142}},\ \bibinfo {pages} {234108} (\bibinfo {year}
  {2015})}\BibitemShut {NoStop}%
\bibitem [{\citenamefont {Droste}\ \emph {et~al.}(2015)\citenamefont {Droste},
  \citenamefont {Splettstoesser},\ and\ \citenamefont {Governale}}]{Droste15}%
  \BibitemOpen
  \bibfield  {author} {\bibinfo {author} {\bibfnamefont {S.}~\bibnamefont
  {Droste}}, \bibinfo {author} {\bibfnamefont {J.}~\bibnamefont
  {Splettstoesser}}, \ and\ \bibinfo {author} {\bibfnamefont {M.}~\bibnamefont
  {Governale}},\ }\href {\doibase 10.1103/PhysRevB.91.125401} {\bibfield
  {journal} {\bibinfo  {journal} {Phys. Rev. B}\ }\textbf {\bibinfo {volume}
  {91}},\ \bibinfo {pages} {125401} (\bibinfo {year} {2015})}\BibitemShut
  {NoStop}%
\bibitem [{\citenamefont {Zamoum}\ \emph {et~al.}(2016)\citenamefont {Zamoum},
  \citenamefont {Lavagna},\ and\ \citenamefont {Cr\'epieux}}]{Zamoum16}%
  \BibitemOpen
  \bibfield  {author} {\bibinfo {author} {\bibfnamefont {R.}~\bibnamefont
  {Zamoum}}, \bibinfo {author} {\bibfnamefont {M.}~\bibnamefont {Lavagna}}, \
  and\ \bibinfo {author} {\bibfnamefont {A.}~\bibnamefont {Cr\'epieux}},\
  }\href {\doibase 10.1103/PhysRevB.93.235449} {\bibfield  {journal} {\bibinfo
  {journal} {Phys. Rev. B}\ }\textbf {\bibinfo {volume} {93}},\ \bibinfo
  {pages} {235449} (\bibinfo {year} {2016})}\BibitemShut {NoStop}%
\bibitem [{\citenamefont {Stadler}\ \emph {et~al.}(2017)\citenamefont
  {Stadler}, \citenamefont {Rastelli},\ and\ \citenamefont
  {Belzig}}]{Stadler17}%
  \BibitemOpen
  \bibfield  {author} {\bibinfo {author} {\bibfnamefont {P.}~\bibnamefont
  {Stadler}}, \bibinfo {author} {\bibfnamefont {G.}~\bibnamefont {Rastelli}}, \
  and\ \bibinfo {author} {\bibfnamefont {W.}~\bibnamefont {Belzig}},\ }\href
  {https://arxiv.org/abs/1712.06361} {\bibfield  {journal} {\bibinfo  {journal}
  {arXiv}\ } (\bibinfo {year} {2017})},\ \Eprint
  {http://arxiv.org/abs/1712.06361} {1712.06361} \BibitemShut {NoStop}%
\bibitem [{\citenamefont {Cr\'epieux}\ \emph {et~al.}(2018)\citenamefont
  {Cr\'epieux}, \citenamefont {Sahoo}, \citenamefont {Duong}, \citenamefont
  {Zamoum},\ and\ \citenamefont {Lavagna}}]{Crepieux17}%
  \BibitemOpen
  \bibfield  {author} {\bibinfo {author} {\bibfnamefont {A.}~\bibnamefont
  {Cr\'epieux}}, \bibinfo {author} {\bibfnamefont {S.}~\bibnamefont {Sahoo}},
  \bibinfo {author} {\bibfnamefont {T.~Q.}\ \bibnamefont {Duong}}, \bibinfo
  {author} {\bibfnamefont {R.}~\bibnamefont {Zamoum}}, \ and\ \bibinfo {author}
  {\bibfnamefont {M.}~\bibnamefont {Lavagna}},\ }\href {\doibase
  10.1103/PhysRevLett.120.107702} {\bibfield  {journal} {\bibinfo  {journal}
  {Phys. Rev. Lett.}\ }\textbf {\bibinfo {volume} {120}},\ \bibinfo {pages}
  {107702} (\bibinfo {year} {2018})}\BibitemShut {NoStop}%
\bibitem [{\citenamefont {Moskalets}\ and\ \citenamefont
  {B\"uttiker}(2007)}]{Moskalets07}%
  \BibitemOpen
  \bibfield  {author} {\bibinfo {author} {\bibfnamefont {M.}~\bibnamefont
  {Moskalets}}\ and\ \bibinfo {author} {\bibfnamefont {M.}~\bibnamefont
  {B\"uttiker}},\ }\href {\doibase 10.1103/PhysRevB.75.035315} {\bibfield
  {journal} {\bibinfo  {journal} {Phys. Rev. B}\ }\textbf {\bibinfo {volume}
  {75}},\ \bibinfo {pages} {035315} (\bibinfo {year} {2007})}\BibitemShut
  {NoStop}%
\bibitem [{\citenamefont {Moskalets}\ and\ \citenamefont
  {B\"uttiker}(2009)}]{Moskalets09}%
  \BibitemOpen
  \bibfield  {author} {\bibinfo {author} {\bibfnamefont {M.}~\bibnamefont
  {Moskalets}}\ and\ \bibinfo {author} {\bibfnamefont {M.}~\bibnamefont
  {B\"uttiker}},\ }\href {\doibase 10.1103/PhysRevB.80.081302} {\bibfield
  {journal} {\bibinfo  {journal} {Phys. Rev. B}\ }\textbf {\bibinfo {volume}
  {80}},\ \bibinfo {pages} {081302} (\bibinfo {year} {2009})}\BibitemShut
  {NoStop}%
\bibitem [{\citenamefont {Moskalets}(2013)}]{Moskalets13}%
  \BibitemOpen
  \bibfield  {author} {\bibinfo {author} {\bibfnamefont {M.}~\bibnamefont
  {Moskalets}},\ }\href {\doibase 10.1103/PhysRevB.88.035433} {\bibfield
  {journal} {\bibinfo  {journal} {Phys. Rev. B}\ }\textbf {\bibinfo {volume}
  {88}},\ \bibinfo {pages} {035433} (\bibinfo {year} {2013})}\BibitemShut
  {NoStop}%
\bibitem [{\citenamefont {Zhao}\ and\ \citenamefont {Chen}(2013)}]{Zhao13}%
  \BibitemOpen
  \bibfield  {author} {\bibinfo {author} {\bibfnamefont {H.-K.}\ \bibnamefont
  {Zhao}}\ and\ \bibinfo {author} {\bibfnamefont {Q.}~\bibnamefont {Chen}},\
  }\href {\doibase 10.1016/j.physleta.2013.10.014} {\bibfield  {journal}
  {\bibinfo  {journal} {Phys. Lett. A}\ }\textbf {\bibinfo {volume} {377}},\
  \bibinfo {pages} {3235} (\bibinfo {year} {2013})}\BibitemShut {NoStop}%
\bibitem [{\citenamefont {Marguerite}\ \emph {et~al.}(2017)\citenamefont
  {Marguerite}, \citenamefont {Roussel}, \citenamefont {Bisognin},
  \citenamefont {Cabart}, \citenamefont {Kumar}, \citenamefont {Berroir},
  \citenamefont {Bocquillon}, \citenamefont {Pla{\ifmmode \mbox{\c{c}} \else
  \c{c}\fi}ais}, \citenamefont {Cavanna}, \citenamefont {Gennser},
  \citenamefont {Jin}, \citenamefont {Degiovanni},\ and\ \citenamefont
  {F{\ifmmode \grave{e} \else \`{e}\fi}ve}}]{Marguerite2017Oct}%
  \BibitemOpen
  \bibfield  {author} {\bibinfo {author} {\bibfnamefont {A.}~\bibnamefont
  {Marguerite}}, \bibinfo {author} {\bibfnamefont {B.}~\bibnamefont {Roussel}},
  \bibinfo {author} {\bibfnamefont {R.}~\bibnamefont {Bisognin}}, \bibinfo
  {author} {\bibfnamefont {C.}~\bibnamefont {Cabart}}, \bibinfo {author}
  {\bibfnamefont {M.}~\bibnamefont {Kumar}}, \bibinfo {author} {\bibfnamefont
  {J.~M.}\ \bibnamefont {Berroir}}, \bibinfo {author} {\bibfnamefont
  {E.}~\bibnamefont {Bocquillon}}, \bibinfo {author} {\bibfnamefont
  {B.}~\bibnamefont {Pla{\ifmmode \mbox{\c{c}} \else \c{c}\fi}ais}}, \bibinfo
  {author} {\bibfnamefont {A.}~\bibnamefont {Cavanna}}, \bibinfo {author}
  {\bibfnamefont {U.}~\bibnamefont {Gennser}}, \bibinfo {author} {\bibfnamefont
  {Y.}~\bibnamefont {Jin}}, \bibinfo {author} {\bibfnamefont {P.}~\bibnamefont
  {Degiovanni}}, \ and\ \bibinfo {author} {\bibfnamefont {G.}~\bibnamefont
  {F{\ifmmode \grave{e} \else \`{e}\fi}ve}},\ }\href
  {https://arxiv.org/abs/1710.11181} {\bibfield  {journal} {\bibinfo  {journal}
  {arXiv}\ } (\bibinfo {year} {2017})},\ \Eprint
  {http://arxiv.org/abs/1710.11181} {1710.11181} \BibitemShut {NoStop}%
\bibitem [{\citenamefont {Schoeller}\ and\ \citenamefont
  {Sch\"on}(1994)}]{Schoeller94}%
  \BibitemOpen
  \bibfield  {author} {\bibinfo {author} {\bibfnamefont {H.}~\bibnamefont
  {Schoeller}}\ and\ \bibinfo {author} {\bibfnamefont {G.}~\bibnamefont
  {Sch\"on}},\ }\href {\doibase 10.1103/PhysRevB.50.18436} {\bibfield
  {journal} {\bibinfo  {journal} {Phys. Rev. B}\ }\textbf {\bibinfo {volume}
  {50}},\ \bibinfo {pages} {18436} (\bibinfo {year} {1994})}\BibitemShut
  {NoStop}%
\bibitem [{\citenamefont {K\"onig}\ \emph {et~al.}(1996)\citenamefont
  {K\"onig}, \citenamefont {Schmid}, \citenamefont {Schoeller},\ and\
  \citenamefont {Sch\"on}}]{Koenig96b}%
  \BibitemOpen
  \bibfield  {author} {\bibinfo {author} {\bibfnamefont {J.}~\bibnamefont
  {K\"onig}}, \bibinfo {author} {\bibfnamefont {J.}~\bibnamefont {Schmid}},
  \bibinfo {author} {\bibfnamefont {H.}~\bibnamefont {Schoeller}}, \ and\
  \bibinfo {author} {\bibfnamefont {G.}~\bibnamefont {Sch\"on}},\ }\href
  {\doibase 10.1103/PhysRevB.54.16820} {\bibfield  {journal} {\bibinfo
  {journal} {Phys. Rev. B}\ }\textbf {\bibinfo {volume} {54}},\ \bibinfo
  {pages} {16820} (\bibinfo {year} {1996})}\BibitemShut {NoStop}%
\bibitem [{\citenamefont {Riwar}\ \emph {et~al.}(2013)\citenamefont {Riwar},
  \citenamefont {Splettstoesser},\ and\ \citenamefont {K\"onig}}]{Riwar13}%
  \BibitemOpen
  \bibfield  {author} {\bibinfo {author} {\bibfnamefont {R.-P.}\ \bibnamefont
  {Riwar}}, \bibinfo {author} {\bibfnamefont {J.}~\bibnamefont
  {Splettstoesser}}, \ and\ \bibinfo {author} {\bibfnamefont {J.}~\bibnamefont
  {K\"onig}},\ }\href {\doibase 10.1103/PhysRevB.87.195407} {\bibfield
  {journal} {\bibinfo  {journal} {Phys. Rev. B}\ }\textbf {\bibinfo {volume}
  {87}},\ \bibinfo {pages} {195407} (\bibinfo {year} {2013})}\BibitemShut
  {NoStop}%
\bibitem [{\citenamefont {Splettstoesser}\ \emph {et~al.}(2006)\citenamefont
  {Splettstoesser}, \citenamefont {Governale}, \citenamefont {K\"onig},\ and\
  \citenamefont {Fazio}}]{Splettstoesser06}%
  \BibitemOpen
  \bibfield  {author} {\bibinfo {author} {\bibfnamefont {J.}~\bibnamefont
  {Splettstoesser}}, \bibinfo {author} {\bibfnamefont {M.}~\bibnamefont
  {Governale}}, \bibinfo {author} {\bibfnamefont {J.}~\bibnamefont {K\"onig}},
  \ and\ \bibinfo {author} {\bibfnamefont {R.}~\bibnamefont {Fazio}},\ }\href
  {\doibase 10.1103/PhysRevB.74.085305} {\bibfield  {journal} {\bibinfo
  {journal} {Phys. Rev. B}\ }\textbf {\bibinfo {volume} {74}},\ \bibinfo
  {pages} {085305} (\bibinfo {year} {2006})}\BibitemShut {NoStop}%
\bibitem [{\citenamefont {Callen}\ and\ \citenamefont
  {Welton}(1951)}]{Callen51}%
  \BibitemOpen
  \bibfield  {author} {\bibinfo {author} {\bibfnamefont {H.~B.}\ \bibnamefont
  {Callen}}\ and\ \bibinfo {author} {\bibfnamefont {T.~A.}\ \bibnamefont
  {Welton}},\ }\href {\doibase 10.1103/PhysRev.83.34} {\bibfield  {journal}
  {\bibinfo  {journal} {Phys. Rev.}\ }\textbf {\bibinfo {volume} {83}},\
  \bibinfo {pages} {34} (\bibinfo {year} {1951})}\BibitemShut {NoStop}%
\bibitem [{\citenamefont {Thielmann}\ \emph {et~al.}(2003)\citenamefont
  {Thielmann}, \citenamefont {Hettler}, \citenamefont {K{\ifmmode \ddot{o}
  \else \"{o}\fi}nig},\ and\ \citenamefont {Sch{\ifmmode \ddot{o} \else
  \"{o}\fi}n}}]{Thielmann03}%
  \BibitemOpen
  \bibfield  {author} {\bibinfo {author} {\bibfnamefont {A.}~\bibnamefont
  {Thielmann}}, \bibinfo {author} {\bibfnamefont {M.~H.}\ \bibnamefont
  {Hettler}}, \bibinfo {author} {\bibfnamefont {J.}~\bibnamefont {K{\ifmmode
  \ddot{o} \else \"{o}\fi}nig}}, \ and\ \bibinfo {author} {\bibfnamefont
  {G.}~\bibnamefont {Sch{\ifmmode \ddot{o} \else \"{o}\fi}n}},\ }\href
  {\doibase 10.1103/PhysRevB.68.115105} {\bibfield  {journal} {\bibinfo
  {journal} {Phys. Rev. B}\ }\textbf {\bibinfo {volume} {68}},\ \bibinfo
  {pages} {115105} (\bibinfo {year} {2003})}\BibitemShut {NoStop}%
\bibitem [{\citenamefont {Kashuba}\ \emph {et~al.}(2012)\citenamefont
  {Kashuba}, \citenamefont {Schoeller},\ and\ \citenamefont
  {Splettstoesser}}]{Kashuba12}%
  \BibitemOpen
  \bibfield  {author} {\bibinfo {author} {\bibfnamefont {O.}~\bibnamefont
  {Kashuba}}, \bibinfo {author} {\bibfnamefont {H.}~\bibnamefont {Schoeller}},
  \ and\ \bibinfo {author} {\bibfnamefont {J.}~\bibnamefont {Splettstoesser}},\
  }\href {\doibase 10.1209/0295-5075/98/57003} {\bibfield  {journal} {\bibinfo
  {journal} {EPL}\ }\textbf {\bibinfo {volume} {98}},\ \bibinfo {pages} {57003}
  (\bibinfo {year} {2012})}\BibitemShut {NoStop}%
\bibitem [{\citenamefont {Splettstoesser}\ \emph {et~al.}(2010)\citenamefont
  {Splettstoesser}, \citenamefont {Governale}, \citenamefont {K\"onig},\ and\
  \citenamefont {B\"uttiker}}]{Splettstoesser10}%
  \BibitemOpen
  \bibfield  {author} {\bibinfo {author} {\bibfnamefont {J.}~\bibnamefont
  {Splettstoesser}}, \bibinfo {author} {\bibfnamefont {M.}~\bibnamefont
  {Governale}}, \bibinfo {author} {\bibfnamefont {J.}~\bibnamefont {K\"onig}},
  \ and\ \bibinfo {author} {\bibfnamefont {M.}~\bibnamefont {B\"uttiker}},\
  }\href {\doibase 10.1103/PhysRevB.81.165318} {\bibfield  {journal} {\bibinfo
  {journal} {Phys. Rev. B}\ }\textbf {\bibinfo {volume} {81}},\ \bibinfo
  {pages} {165318} (\bibinfo {year} {2010})}\BibitemShut {NoStop}%
\bibitem [{\citenamefont {Cavaliere}\ \emph {et~al.}(2009)\citenamefont
  {Cavaliere}, \citenamefont {Governale},\ and\ \citenamefont
  {K\"onig}}]{Cavaliere09}%
  \BibitemOpen
  \bibfield  {author} {\bibinfo {author} {\bibfnamefont {F.}~\bibnamefont
  {Cavaliere}}, \bibinfo {author} {\bibfnamefont {M.}~\bibnamefont
  {Governale}}, \ and\ \bibinfo {author} {\bibfnamefont {J.}~\bibnamefont
  {K\"onig}},\ }\href {\doibase 10.1103/PhysRevLett.103.136801} {\bibfield
  {journal} {\bibinfo  {journal} {Phys. Rev. Lett.}\ }\textbf {\bibinfo
  {volume} {103}},\ \bibinfo {pages} {136801} (\bibinfo {year}
  {2009})}\BibitemShut {NoStop}%
\bibitem [{\citenamefont {Riwar}\ \emph {et~al.}(2016)\citenamefont {Riwar},
  \citenamefont {Roche}, \citenamefont {Jehl},\ and\ \citenamefont
  {Splettstoesser}}]{Riwar16}%
  \BibitemOpen
  \bibfield  {author} {\bibinfo {author} {\bibfnamefont {R.-P.}\ \bibnamefont
  {Riwar}}, \bibinfo {author} {\bibfnamefont {B.}~\bibnamefont {Roche}},
  \bibinfo {author} {\bibfnamefont {X.}~\bibnamefont {Jehl}}, \ and\ \bibinfo
  {author} {\bibfnamefont {J.}~\bibnamefont {Splettstoesser}},\ }\href
  {\doibase 10.1103/PhysRevB.93.235401} {\bibfield  {journal} {\bibinfo
  {journal} {Phys. Rev. B}\ }\textbf {\bibinfo {volume} {93}},\ \bibinfo
  {pages} {235401} (\bibinfo {year} {2016})}\BibitemShut {NoStop}%
\bibitem [{\citenamefont {Reckermann}\ \emph {et~al.}(2010)\citenamefont
  {Reckermann}, \citenamefont {Splettstoesser},\ and\ \citenamefont
  {Wegewijs}}]{Reckermann10}%
  \BibitemOpen
  \bibfield  {author} {\bibinfo {author} {\bibfnamefont {F.}~\bibnamefont
  {Reckermann}}, \bibinfo {author} {\bibfnamefont {J.}~\bibnamefont
  {Splettstoesser}}, \ and\ \bibinfo {author} {\bibfnamefont {M.~R.}\
  \bibnamefont {Wegewijs}},\ }\href {\doibase 10.1103/PhysRevLett.104.226803}
  {\bibfield  {journal} {\bibinfo  {journal} {Phys. Rev. Lett.}\ }\textbf
  {\bibinfo {volume} {104}},\ \bibinfo {pages} {226803} (\bibinfo {year}
  {2010})}\BibitemShut {NoStop}%
\end{thebibliography}

%

\end{document}